June 2009

# Operations of and Future Plans for the Pierre Auger Observatory

Presentations for the
31st International Cosmic Ray Conference, Łódź , Poland, July 2009

# PIERRE AUGER COLLABORATION

**J. Abraham[8], P. Abreu[71], M. Aglietta[54], C. Aguirre[12], E.J. Ahn[87], D. Allard[31], I. Allekotte[1], J. Allen[90], J. Alvarez-Muñiz[78], M. Ambrosio[48], L. Anchordoqui[104], S. Andringa[71], A. Anzalone[53], C. Aramo[48], E. Arganda[75], S. Argirò[51], K. Arisaka[95], F. Arneodo[55], F. Arqueros[75], T. Asch[38], H. Asorey[1], P. Assis[71], J. Aublin[33], M. Ave[96], G. Avila[10], T. Bäcker[42], D. Badagnani[6], K.B. Barber[11], A.F. Barbosa[14], S.L.C. Barroso[20], B. Baughman[92], P. Bauleo[85], J.J. Beatty[92], T. Beau[31], B.R. Becker[101], K.H. Becker[36], A. Bellétoile[34], J.A. Bellido[11, 93], S. BenZvi[103], C. Berat[34], P. Bernardini[47], X. Bertou[1], P.L. Biermann[39], P. Billoir[33], O. Blanch-Bigas[33], F. Blanco[75], C. Bleve[47], H. Blümer[41, 37], M. Boháčová[96, 27], D. Boncioli[49], C. Bonifazi[33], R. Bonino[54], N. Borodai[69], J. Brack[85], P. Brogueira[71], W.C. Brown[86], R. Bruijn[81], P. Buchholz[42], A. Bueno[77], R.E. Burton[83], N.G. Busca[31], K.S. Caballero-Mora[41], L. Caramete[39], R. Caruso[50], W. Carvalho[17], A. Castellina[54], O. Catalano[53], L. Cazon[96], R. Cester[51], J. Chauvin[34], A. Chiavassa[54], J.A. Chinellato[18], A. Chou[87, 90], J. Chudoba[27], J. Chye[89d], R.W. Clay[11], E. Colombo[2], R. Conceição[71], B. Connolly[102], F. Contreras[9], J. Coppens[65, 67], A. Cordier[32], U. Cotti[63], S. Coutu[93], C.E. Covault[83], A. Creusot[73], A. Criss[93], J. Cronin[96], A. Curutiu[39], S. Dagoret-Campagne[32], R. Dallier[35], K. Daumiller[37], B.R. Dawson[11], R.M. de Almeida[18], M. De Domenico[50], C. De Donato[46], S.J. de Jong[65], G. De La Vega[8], W.J.M. de Mello Junior[18], J.R.T. de Mello Neto[23], I. De Mitri[47], V. de Souza[16], K.D. de Vries[66], G. Decerprit[31], L. del Peral[76], O. Deligny[30], A. Della Selva[48], C. Delle Fratte[49], H. Dembinski[40], C. Di Giulio[49], J.C. Diaz[89], P.N. Diep[105], C. Dobrigkeit [18], J.C. D'Olivo[64], P.N. Dong[105], A. Dorofeev[88], J.C. dos Anjos[14], M.T. Dova[6], D. D'Urso[48], I. Dutan[39], M.A. DuVernois[98], R. Engel[37], M. Erdmann[40], C.O. Escobar[18], A. Etchegoyen[2], P. Facal San Luis[96, 78], H. Falcke[65, 68], G. Farrar[90], A.C. Fauth[18], N. Fazzini[87], F. Ferrer[83], A. Ferrero[2], B. Fick[89], A. Filevich[2], A. Filipčič[72, 73], I. Fleck[42], S. Fliescher[40], C.E. Fracchiolla[85], E.D. Fraenkel[66], W. Fulgione[54], R.F. Gamarra[2], S. Gambetta[44], B. García[8], D. García Gámez[77], D. Garcia-Pinto[75], X. Garrido[37, 32], G. Gelmini[95], H. Gemmeke[38], P.L. Ghia[30, 54], U. Giaccari[47], M. Giller[70], H. Glass[87], L.M. Goggin[104], M.S. Gold[101], G. Golup[1], F. Gomez Albarracin[6], M. Gómez Berisso[1], P. Gonçalves[71], M. Gonçalves do Amaral[24], D. Gonzalez[41], J.G. Gonzalez[77, 88], D. Góra[41, 69], A. Gorgi[54], P. Gouffon[17], S.R. Gozzini[81], E. Grashorn[92], S. Grebe[65], M. Grigat[40], A.F. Grillo[55], Y. Guardincerri[4], F. Guarino[48], G.P. Guedes[19], J. Gutiérrez[76], J.D. Hague[101], V. Halenka[28], P. Hansen[6], D. Harari[1], S. Harmsma[66, 67], J.L. Harton[85], A. Haungs[37], M.D. Healy[95], T. Hebbeker[40], G. Hebrero[76], D. Heck[37], V.C. Holmes[11], P. Homola[69], J.R. Hörandel[65], A. Horneffer[65], M. Hrabovský[28, 27], T. Huege[37], M. Hussain[73], M. Iarlori[45], A. Insolia[50], F. Ionita[96], A. Italiano[50], S. Jiraskova[65], M. Kaducak[87], K.H. Kampert[36], T. Karova[27], P. Kasper[87], B. Kégl[32], B. Keilhauer[37], E. Kemp[18], R.M. Kieckhafer[89], H.O. Klages[37], M. Kleifges[38], J. Kleinfeller[37], R. Knapik[85], J. Knapp[81], D.-H. Koang[34], A. Krieger[2], O. Krömer[38], D. Kruppke-Hansen[36], F. Kuehn[87], D. Kuempel[36], N. Kunka[38], A. Kusenko[95], G. La Rosa[53], C. Lachaud[31], B.L. Lago[23], P. Lautridou[35], M.S.A.B. Leão[22], D. Lebrun[34], P. Lebrun[87], J. Lee[95], M.A. Leigui de Oliveira[22], A. Lemiere[30], A. Letessier-Selvon[33], M. Leuthold[40], I. Lhenry-Yvon[30], R. López[59], A. Lopez Agüera[78], K. Louedec[32], J. Lozano Bahilo[77], A. Lucero[54], H. Lyberis[30], M.C. Maccarone[53], C. Macolino[45], S. Maldera[54], D. Mandat[27], P. Mantsch[87], A.G. Mariazzi[6], I.C. Maris[41], H.R. Marquez Falcon[63], D. Martello[47], O. Martínez Bravo[59], H.J. Mathes[37], J. Matthews[88, 94], J.A.J. Matthews[101], G. Matthiae[49], D. Maurizio[51], P.O. Mazur[87], M. McEwen[76], R.R. McNeil[88], G. Medina-Tanco[64], M. Melissas[41], D. Melo[51], E. Menichetti[51], A. Menshikov[38], R. Meyhandan[14], M.I. Micheletti[2], G. Miele[48], W. Miller[101], L. Miramonti[46], S. Mollerach[1], M. Monasor[75], D. Monnier Ragaigne[32], F. Montanet[34], B. Morales[64], C. Morello[54], J.C. Moreno[6], C. Morris[92], M. Mostafá[85], C.A. Moura[48], S. Mueller[37], M.A. Muller[18], R. Mussa[51], G. Navarra[54], J.L. Navarro[77], S. Navas[77], P. Necesal[27], L. Nellen[64], C. Newman-Holmes[87], D. Newton[81], P.T. Nhung[105], N. Nierstenhoefer[36], D. Nitz[89], D. Nosek[26], L. Nožka[27], M. Nyklicek[27], J. Oehlschläger[37], A. Olinto[96], P. Oliva[36], V.M. Olmos-Gilbaja[78], M. Ortiz[75], N. Pacheco[76], D. Pakk Selmi-Dei[18], M. Palatka[27], J. Pallotta[3], G. Parente[78], E. Parizot[31], S. Parlati[55], S. Pastor[74], M. Patel[81], T. Paul[91], V. Pavlidou[96c], K. Payet[34], M. Pech[27], J. Pękala[69], I.M. Pepe[21], L. Perrone[52], R. Pesce[44], E. Petermann[100], S. Petrera[45], P. Petrinca[49], A. Petrolini[44], Y. Petrov[85], J. Petrovic[67], C. Pfendner[103], R. Piegaia[4], T. Pierog[37], M. Pimenta[71], T. Pinto[74], V. Pirronello[50], O. Pisanti[48], M. Platino[2], J. Pochon[1], V.H. Ponce[1], M. Pontz[42], P. Privitera[96], M. Prouza[27], E.J. Quel[3], J. Rautenberg[36], O. Ravel[35], D. Ravignani[2],**

**A. Redondo[76], B. Revenu[35], F.A.S. Rezende[14], J. Ridky[27], S. Riggi[50], M. Risse[36], C. Rivière[34], V. Rizi[45], C. Robledo[59], G. Rodriguez[49], J. Rodriguez Martino[50], J. Rodriguez Rojo[9], I. Rodriguez-Cabo[78], M.D. Rodríguez-Frías[76], G. Ros[75, 76], J. Rosado[75], T. Rossler[28], M. Roth[37], B. Rouillé-d'Orfeuil[31], E. Roulet[1], A.C. Rovero[7], F. Salamida[45], H. Salazar[59b], G. Salina[49], F. Sánchez[64], M. Santander[9], C.E. Santo[71], E.M. Santos[23], F. Sarazin[84], S. Sarkar[79], R. Sato[9], N. Scharf[40], V. Scherini[36], H. Schieler[37], P. Schiffer[40], A. Schmidt[38], F. Schmidt[96], T. Schmidt[41], O. Scholten[66], H. Schoorlemmer[65], J. Schovancova[27], P. Schovánek[27], F. Schroeder[37], S. Schulte[40], F. Schüssler[37], D. Schuster[84], S.J. Sciutto[6], M. Scuderi[50], A. Segreto[53], D. Semikoz[31], M. Settimo[47], R.C. Shellard[14, 15], I. Sidelnik[2], B.B. Siffert[23], A. Śmiałkowski[70], R. Šmída[27], B.E. Smith[81], G.R. Snow[100], P. Sommers[93], J. Sorokin[11], H. Spinka[82, 87], R. Squartini[9], E. Strazzeri[32], A. Stutz[34], F. Suarez[2], T. Suomijärvi[30], A.D. Supanitsky[64], M.S. Sutherland[92], J. Swain[91], Z. Szadkowski[70], A. Tamashiro[7], A. Tamburro[41], T. Tarutina[6], O. Taşcău[36], R. Tcaciuc[42], D. Tcherniakhovski[38], D. Tegolo[58], N.T. Thao[105], D. Thomas[85], R. Ticona[13], J. Tiffenberg[4], C. Timmermans[67, 65], W. Tkaczyk[70], C.J. Todero Peixoto[22], B. Tomé[71], A. Tonachini[51], I. Torres[59], P. Travnicek[27], D.B. Tridapalli[17], G. Tristram[31], E. Trovato[50], M. Tueros[6], R. Ulrich[37], M. Unger[37], M. Urban[32], J.F. Valdés Galicia[64], I. Valiño[37], L. Valore[48], A.M. van den Berg[66], J.R. Vázquez[75], R.A. Vázquez[78], D. Veberič[73, 72], A. Velarde[13], T. Venters[96], V. Verzi[49], M. Videla[8], L. Villaseñor[63], S. Vorobiov[73], L. Voyvodic[87‡], H. Wahlberg[6], P. Wahrlich[11], O. Wainberg[2], D. Warner[85], A.A. Watson[81], S. Westerhoff[103], B.J. Whelan[11], G. Wieczorek[70], L. Wiencke[84], B. Wilczyńska[69], H. Wilczyński[69], C. Wileman[81], M.G. Winnick[11], H. Wu[32], B. Wundheiler[2], T. Yamamoto[96a], P. Younk[85], G. Yuan[88], A. Yushkov[48], E. Zas[78], D. Zavrtanik[73, 72], M. Zavrtanik[72, 73], I. Zaw[90], A. Zepeda[60b], M. Ziolkowski[42]**

*[1] Centro Atómico Bariloche and Instituto Balseiro (CNEA-UNCuyo-CONICET), San Carlos de Bariloche, Argentina*
*[2] Centro Atómico Constituyentes (Comisión Nacional de Energía Atómica/CONICET/UTN- FRBA), Buenos Aires, Argentina*
*[3] Centro de Investigaciones en Láseres y Aplicaciones, CITEFA and CONICET, Argentina*
*[4] Departamento de Física, FCEyN, Universidad de Buenos Aires y CONICET, Argentina*
*[6] IFLP, Universidad Nacional de La Plata and CONICET, La Plata, Argentina*
*[7] Instituto de Astronomía y Física del Espacio (CONICET), Buenos Aires, Argentina*
*[8] National Technological University, Faculty Mendoza (CONICET/CNEA), Mendoza, Argentina*
*[9] Pierre Auger Southern Observatory, Malargüe, Argentina*
*[10] Pierre Auger Southern Observatory and Comisión Nacional de Energía Atómica, Malargüe, Argentina*
*[11] University of Adelaide, Adelaide, S.A., Australia*
*[12] Universidad Catolica de Bolivia, La Paz, Bolivia*
*[13] Universidad Mayor de San Andrés, Bolivia*
*[14] Centro Brasileiro de Pesquisas Fisicas, Rio de Janeiro, RJ, Brazil*
*[15] Pontifícia Universidade Católica, Rio de Janeiro, RJ, Brazil*
*[16] Universidade de São Paulo, Instituto de Física, São Carlos, SP, Brazil*
*[17] Universidade de São Paulo, Instituto de Física, São Paulo, SP, Brazil*
*[18] Universidade Estadual de Campinas, IFGW, Campinas, SP, Brazil*
*[19] Universidade Estadual de Feira de Santana, Brazil*
*[20] Universidade Estadual do Sudoeste da Bahia, Vitoria da Conquista, BA, Brazil*
*[21] Universidade Federal da Bahia, Salvador, BA, Brazil*
*[22] Universidade Federal do ABC, Santo André, SP, Brazil*
*[23] Universidade Federal do Rio de Janeiro, Instituto de Física, Rio de Janeiro, RJ, Brazil*
*[24] Universidade Federal Fluminense, Instituto de Fisica, Niterói, RJ, Brazil*
*[26] Charles University, Faculty of Mathematics and Physics, Institute of Particle and Nuclear Physics, Prague, Czech Republic*
*[27] Institute of Physics of the Academy of Sciences of the Czech Republic, Prague, Czech Republic*
*[28] Palacký University, Olomouc, Czech Republic*
*[30] Institut de Physique Nucléaire d'Orsay (IPNO), Université Paris 11, CNRS-IN2P3, Orsay, France*
*[31] Laboratoire AstroParticule et Cosmologie (APC), Université Paris 7, CNRS-IN2P3, Paris, France*
*[32] Laboratoire de l'Accélérateur Linéaire (LAL), Université Paris 11, CNRS-IN2P3, Orsay, France*
*[33] Laboratoire de Physique Nucléaire et de Hautes Energies (LPNHE), Universités Paris 6 et Paris 7, Paris Cedex 05, France*

[34] Laboratoire de Physique Subatomique et de Cosmologie (LPSC), Université Joseph Fourier, INPG, CNRS-IN2P3, Grenoble, France
[35] SUBATECH, Nantes, France
[36] Bergische Universität Wuppertal, Wuppertal, Germany
[37] Forschungszentrum Karlsruhe, Institut für Kernphysik, Karlsruhe, Germany
[38] Forschungszentrum Karlsruhe, Institut für Prozessdatenverarbeitung und Elektronik, Karlsruhe, Germany
[39] Max-Planck-Institut für Radioastronomie, Bonn, Germany
[40] RWTH Aachen University, III. Physikalisches Institut A, Aachen, Germany
[41] Universität Karlsruhe (TH), Institut für Experimentelle Kernphysik (IEKP), Karlsruhe, Germany
[42] Universität Siegen, Siegen, Germany
[44] Dipartimento di Fisica dell'Università and INFN, Genova, Italy
[45] Università dell'Aquila and INFN, L'Aquila, Italy
[46] Università di Milano and Sezione INFN, Milan, Italy
[47] Dipartimento di Fisica dell'Università del Salento and Sezione INFN, Lecce, Italy
[48] Università di Napoli "Federico II" and Sezione INFN, Napoli, Italy
[49] Università di Roma II "Tor Vergata" and Sezione INFN, Roma, Italy
[50] Università di Catania and Sezione INFN, Catania, Italy
[51] Università di Torino and Sezione INFN, Torino, Italy
[52] Dipartimento di Ingegneria dell'Innovazione dell'Università del Salento and Sezione INFN, Lecce, Italy
[53] Istituto di Astrofisica Spaziale e Fisica Cosmica di Palermo (INAF), Palermo, Italy
[54] Istituto di Fisica dello Spazio Interplanetario (INAF), Università di Torino and Sezione INFN, Torino, Italy
[55] INFN, Laboratori Nazionali del Gran Sasso, Assergi (L'Aquila), Italy
[58] Università di Palermo and Sezione INFN, Catania, Italy
[59] Benemérita Universidad Autónoma de Puebla, Puebla, Mexico
[60] Centro de Investigación y de Estudios Avanzados del IPN (CINVESTAV), México, D.F., Mexico
[61] Instituto Nacional de Astrofisica, Optica y Electronica, Tonantzintla, Puebla, Mexico
[63] Universidad Michoacana de San Nicolas de Hidalgo, Morelia, Michoacan, Mexico
[64] Universidad Nacional Autonoma de Mexico, Mexico, D.F., Mexico
[65] IMAPP, Radboud University, Nijmegen, Netherlands
[66] Kernfysisch Versneller Instituut, University of Groningen, Groningen, Netherlands
[67] NIKHEF, Amsterdam, Netherlands
[68] ASTRON, Dwingeloo, Netherlands
[69] Institute of Nuclear Physics PAN, Krakow, Poland
[70] University of Łódź, Łódź, Poland
[71] LIP and Instituto Superior Técnico, Lisboa, Portugal
[72] J. Stefan Institute, Ljubljana, Slovenia
[73] Laboratory for Astroparticle Physics, University of Nova Gorica, Slovenia
[74] Instituto de Física Corpuscular, CSIC-Universitat de València, Valencia, Spain
[75] Universidad Complutense de Madrid, Madrid, Spain
[76] Universidad de Alcalá, Alcalá de Henares (Madrid), Spain
[77] Universidad de Granada & C.A.F.P.E., Granada, Spain
[78] Universidad de Santiago de Compostela, Spain
[79] Rudolf Peierls Centre for Theoretical Physics, University of Oxford, Oxford, United Kingdom
[81] School of Physics and Astronomy, University of Leeds, United Kingdom
[82] Argonne National Laboratory, Argonne, IL, USA
[83] Case Western Reserve University, Cleveland, OH, USA
[84] Colorado School of Mines, Golden, CO, USA
[85] Colorado State University, Fort Collins, CO, USA
[86] Colorado State University, Pueblo, CO, USA
[87] Fermilab, Batavia, IL, USA
[88] Louisiana State University, Baton Rouge, LA, USA
[89] Michigan Technological University, Houghton, MI, USA
[90] New York University, New York, NY, USA
[91] Northeastern University, Boston, MA, USA
[92] Ohio State University, Columbus, OH, USA
[93] Pennsylvania State University, University Park, PA, USA
[94] Southern University, Baton Rouge, LA, USA
[95] University of California, Los Angeles, CA, USA

*[96] University of Chicago, Enrico Fermi Institute, Chicago, IL, USA*
*[98] University of Hawaii, Honolulu, HI, USA*
*[100] University of Nebraska, Lincoln, NE, USA*
*[101] University of New Mexico, Albuquerque, NM, USA*
*[102] University of Pennsylvania, Philadelphia, PA, USA*
*[103] University of Wisconsin, Madison, WI, USA*
*[104] University of Wisconsin, Milwaukee, WI, USA*
*[105] Institute for Nuclear Science and Technology (INST), Hanoi, Vietnam*
*[‡] Deceased*
*[a] at Konan University, Kobe, Japan*
*[b] On leave of absence at the Instituto Nacional de Astrofisica, Optica y Electronica*
*[c] at Caltech, Pasadena, USA*
*[d] at Hawaii Pacific University*

*Note added: An additional author, C. Hojvat, Fermilab, Batavia, IL, USA, should be added to papers 2,3,4,5,6,7,8,9 in this collection*

# Performance and operation of the Surface Detector of the Pierre Auger Observatory

**T. Suomijärvi * for the Pierre Auger Collaboration †**

*Institut de Physique Nucleaire, Université Paris-Sud, IN2P3-CNRS, Orsay, France
†Observatorio Pierre Auger, Av. San Martín Norte 304, (5613) Malargüe, Argentina

***Abstract*. The Surface Array of the Pierre Auger Observatory consists of 1660 water Cherenkov detectors that sample at the ground the charged particles and photons of air showers initiated by energetic cosmic rays. The construction of the array in Malargüe, Argentina is now complete. A large fraction of the detectors have been operational for more than five years. Each detector records data locally with timing obtained from GPS units and power from solar panels and batteries. In this paper, the performance and the operation of the array are discussed. We emphasise the accuracy of the signal measurement, the stability of the triggering, the performance of the solar power system and other hardware, and the long-term purity of the water.**

## I. INTRODUCTION

The Surface Detector (SD) of the Pierre Auger Observatory is composed of Water Cherenkov Detectors (WCD) extending over an area of 3000 km$^2$ with 1500 m spacing between detectors. In addition to the detectors in the regular array, some locations of the array were equipped with two and three nearby detectors, placed at $\sim$10 meters from each other. These ”twins” and ”triplets” provide a very useful testbench for studies of signal fluctuation, timing resolution and energy and angular reconstruction precision. Combined with the HEAT telescopes and the AMIGA muon detector array, a denser array of WCD with detector spacing of 750 m has also been deployed. The total number of detector stations is 1660. The hardware of the surface detector is described extensively in [1], [2].

Installation of detectors started in 2002 and the Observatory has been collecting stable data since January 2004. The construction was completed in June 2008. Figure 1 shows the current status of the array.

The Observatory has been running now with its full configuration for nearly one year and its commissioning is completed. The failure rates of various components have been assessed and the Surface Detector is now entering into a regular long term operation and maintenance phase. Some detectors have been operational already for more than 8 years which permits the study of their long term performance. In this paper, after a short description of the Surface Detector, the detector response and uniformity, its acceptance and long-term performance, and finally its operation and maintenance are discussed.

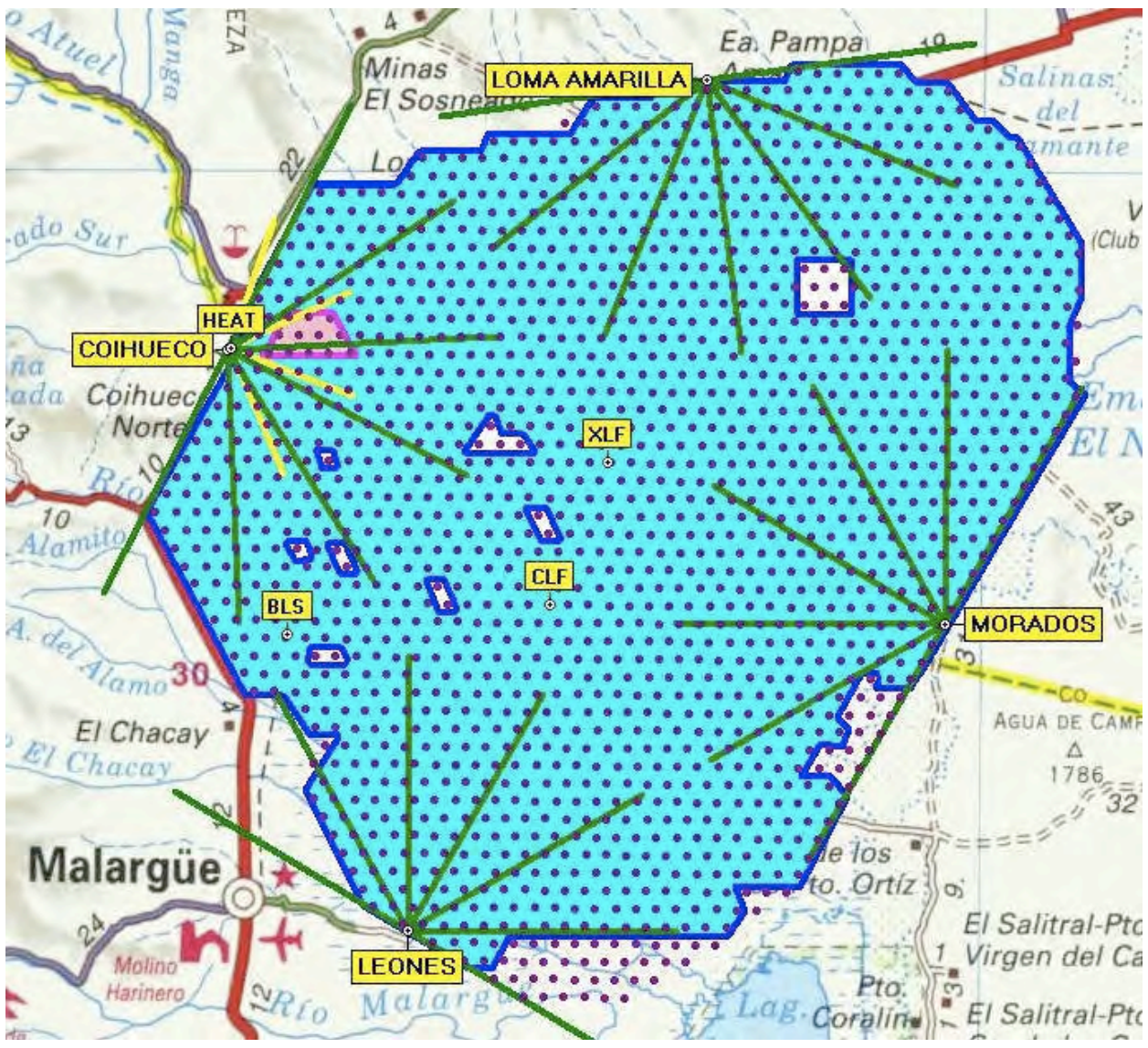

Fig. 1: Current deployment status of the array. Tanks within the shaded area are filled with water and in operation.

## II. DESCRIPTION OF THE SURFACE DETECTOR

Each WCD consists of a 3.6 m diameter water tank containing a Tyvek® liner for uniform reflection of the Cherenkov light. The liner contains 12,000 l of ultra-high purity water with resitivity typically higher than 5 MΩ.cm. Three nine-inch-diameter photomultiplier tubes (PMTs) are symmetrically distributed at a distance of 1.20 m from the center of the tank and look downwards through windows of clear polyethylene into the water to collect the Cherenkov light produced by the passage of relativistic charged particles through the water. The water height of 1.2 m makes it also sensitive to high energy photons, which convert in the water volume. A solar power system provides an average of 10 W for the PMTs and the electronics package consisting of a processor, GPS receiver, radio transceiver and power controller.

The signals produced by the Cherenkov light are read out by three large 9” XP1805 Photonis photomultipliers. The PMTs are equipped with a resistive divider base having two outputs: anode and amplified last dynode [3]. This provides a large dynamic range, totaling 15 bits, extending from a few to about $10^5$ photoelectrons. The high voltage is provided locally. The nominal operating gain of the PMTs is $2{\times}10^5$ and can be extended to $10^6$. The base, together with the HV module, is protected against humidity by silicone potting.

The signals from anode and dynode are filtered and digitised at 40 MHz using 10 bit Flash Analog-Digital Converters (FADC). Two shower triggers are used: threshold trigger (ThT) and time-over-threshold (ToT) trigger. The first one is a simple majority trigger with a threshold at 3.2 VEM (Vertical Equivalent Muon). The ToT trigger requires 12 FADC bins with signals larger than 0.2 VEM in a sliding window of 3 $\mu$s. The time-over-threshold trigger efficiently triggers on the shower particles far away from the shower core. In addition, a muon trigger allows for recording of continuous calibration data. The third level trigger, T3, initiates the data acquisition of the array. It is formed at the Central Data Acquisition System (CDAS), and it is based on the spatial and temporal combination of local station triggers. Once a T3 is formed, all FADC traces from stations passing the local trigger are sent to the CDAS.

A common time base is established for different detector stations by using the GPS system. Each tank is equipped with a commercial GPS receiver (Motorola OnCore UT) providing a one pulse per second output and software corrections. This signal is used to synchronise a 100 MHz clock which serves to timetag the trigger. Each detector station has an IBM 403 PowerPC micro-controller for local data acquisition, software trigger and detector monitoring, and memory for data storage. The station electronics is implemented in a single module called the Unified Board, and mounted in an aluminum enclosure. The electronics package is mounted on top of the hatch cover of one of the PMTs and protected against rain and dust by an aluminum dome.

The detector calibration is inferred from background muons. The average number of photoelectrons per muon collected by one PMT is 95. The measurement of the muon charge spectrum allows us to deduce the charge value for the Vertical Equivalent Muon, $Q_{VEM}$, from which the calibration is inferred for the whole dynamic range. The cross calibration between the two channels, anode and dynode outputs, is performed by using small shower signals in the overlap region of the two channels [4].

The decay constant of the muon signal is related to the absorption length of the light produced. This depends on various parameters such as the Tyvek® reflectivity and the purity of the water. The signal decay constant correlates with the so called area-to-peak (A/P) ratio of the signal:

$$\mathrm{A/P} = \frac{Q_{VEM}}{I_{VEM}} \tag{1}$$

where $I_{VEM}$ is the maximum current of the muon signal. This area-to-peak ratio is a routine monitoring quantity that is directly available from the local station software.

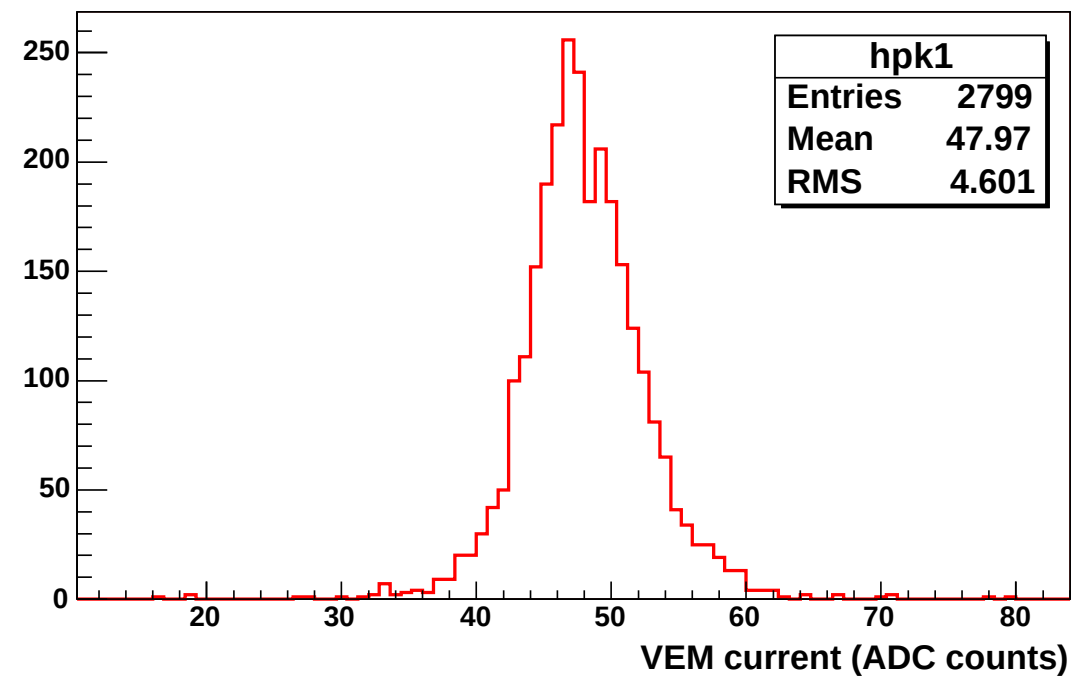

Fig. 2: VEM measured for 2799 PMTs.

## III. Detector response and uniformity

Stable data taking with the Surface Detector started in January 2004 and various parameters are continuously monitored to ensure the good performance of the detectors. The noise levels are very low. For both the anode and dynode channels, the mean value of the pedestal fluctuation RMS is below 0.5 FADC channels corresponding to about 0.01 VEM. The intrinsic resolution of the GPS time tagging system is about 8 ns requiring a good precision for the station location. An accuracy better than 1 meter is obtained for the tank position by measuring the positions with differential GPS.

Figure 2 shows the muon peak current ($I_{VEM}$) values for a large number of PMTs. The mean value of the muon peak ($I_{VEM}$) is at channel 48 with an RMS of 4.6 showing a very good uniformity of the detector response. Trigger rates are also remarkably uniform over all detector stations, also implying good calibration and baseline determination. The mean value of the threshold trigger rate is 22 Hz with dispersion less than 2%. The time-over threshold trigger is about 1 Hz with a larger dispersion. This is due to the fact that this trigger is sensitive to the pulse shape and thus is more sensitive to the characteristics of the detector. It is observed that the new detectors often have ToT values which are higher and then stabilise after a few months to about 1 Hz. The trigger studies and the studies on the muon response show that the detectors have, after a few month stabilisation, a good uniformity.

Day-night atmospheric temperature variations can be larger than 20 °C. In each tank, temperature is measured on the PMT bases, on the electronics board, and on the batteries which allows to correlate various monitoring data with the temperature. Typical day-night variations are of the order of 2 ADC channels for the muon peak. This is mainly due to the sensitivity of the PMTs to temperature. These temperature variations also slightly affect the ToT-trigger. The muon calibration is made online every minute. This continuous calibration allows to correct for the day-night temperature effects.

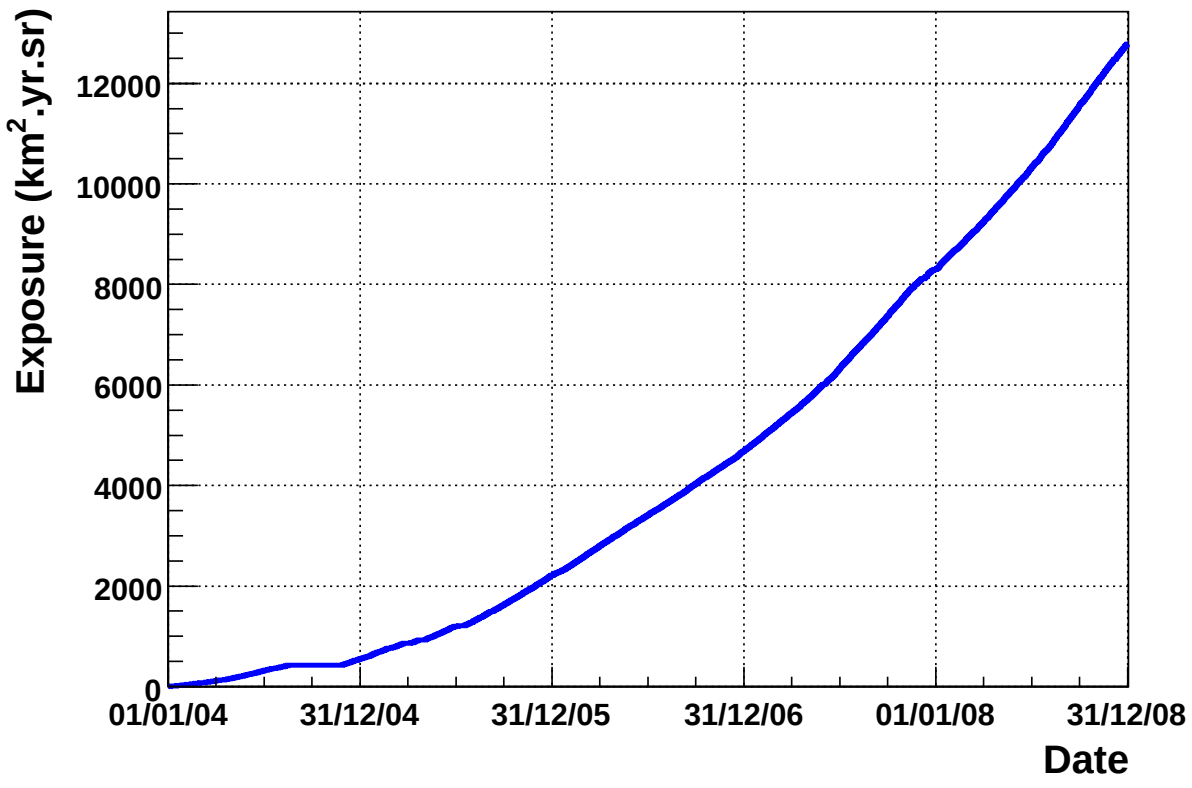

Fig. 3: Evolution of the exposure between January 1st, 2004 and December 31st, 2008.

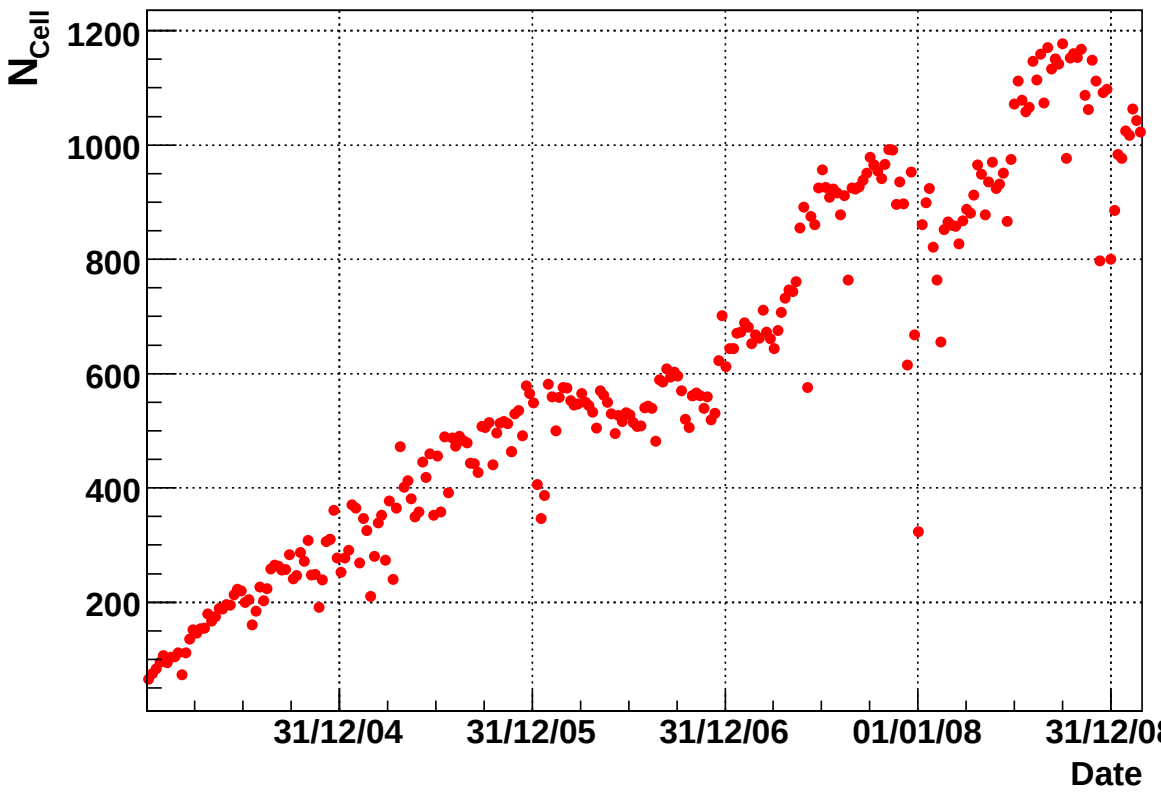

Fig. 4: Evolution of the number of hexagon cells as a function of time between January 1st, 2004 and December 31st, 2008.

## IV. Detector acceptance

To ensure good data quality for physics analysis there are two additional off-line triggers. The physics trigger, T4, is needed to select real showers from the set of stored T3 data. This trigger is mainly based on coincidence between adjacent detector stations within the propagation time of the shower front. In addition, there is a so-called fiducial trigger, T5, which excludes events where a part of the shower may be missing. The full efficiency of the SD trigger and event selection is reached at 3 $10^{18}$ eV. Above this energy, the calculation of the exposure is based on the determination of the geometrical aperture and of the observation time. Figure 3 shows the evolution of the integrated exposure between January 1st, 2004 and December 31st, 2008.

The fiducial trigger is based on hexagons allowing to exploit the regularity of the array. The aperture of the array is obtained as a multiple of the aperture of the elemental hexagon cell. In practice, any active station with six active neighbors contributes exactly to the elementary hexagon aperture, a cell. The number of cells, Ncell(t), is not constant over time due to possible temporary problems in the stations (e. g. failures of electronics, power supply, communication system, etc.). Ncell(t) is constantly monitored. In Fig. 4 the evolution of the number of hexagonal cells with time is shown.

The very precise monitoring of the array configurations has allowed us to exploit data during the construction phase. The evolution as a function of time of the number of trigger cells is globally similar to the evolution of the acceptance. Some differences however can be seen. For example, the pronounced dip seen in Fig. 4 in January 2008 corresponds to a large storm that affected the communication system.

## V. Long-term performance

Figure 5 shows the area-to-peak ratio for a typical PMT channel as a function of time. Two main effects are observed: a slight global decrease and small seasonal variations. The two curves reflect the maximum day-night variations (see figure caption).

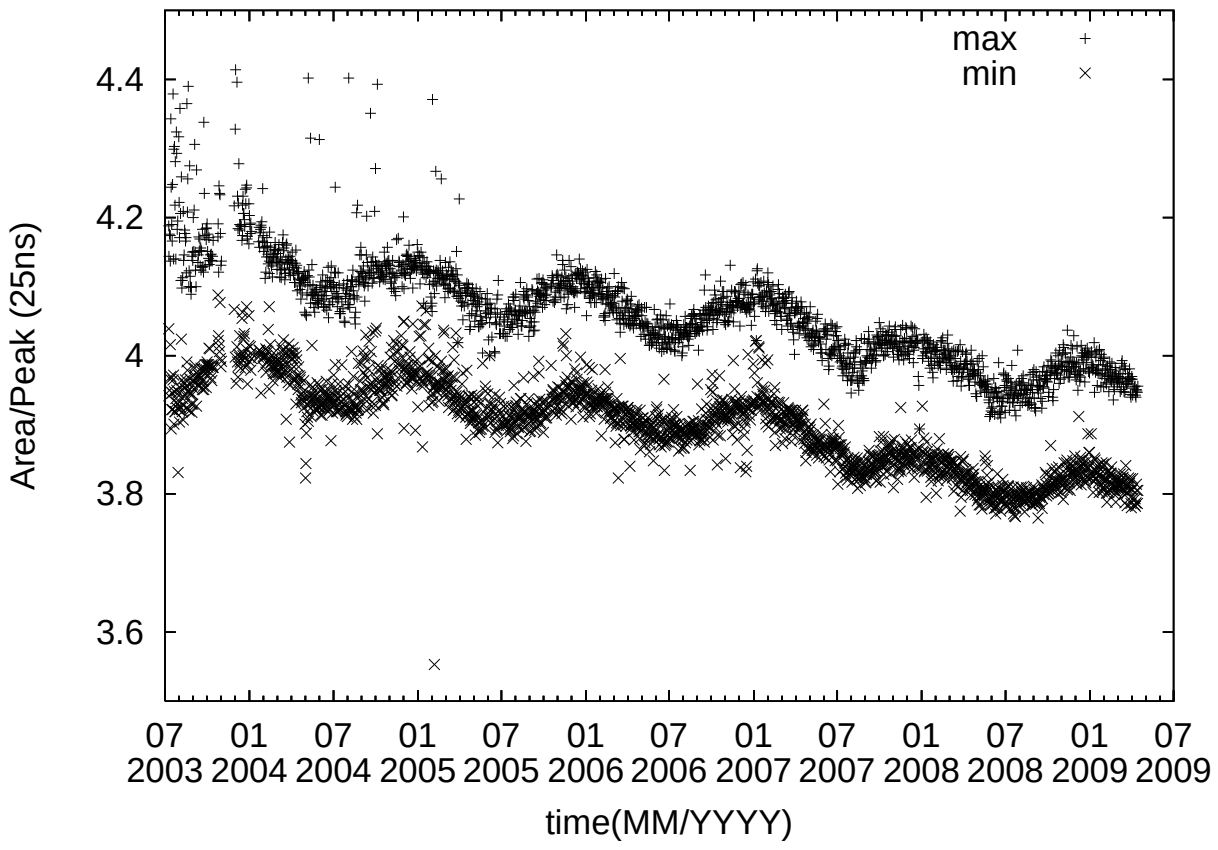

Fig. 5: Area-to-peak ratio as a function of time for a typical PMT channel. The upper curve corresponds to the maximum value and the lower curve to the minimum value during 24 hours reflecting the day-night variations.

The maximum day-night variations are due to outside temperature variations. As explained above, the muon calibration is made on-line every minute which allows to correct for this effect. Seasonal amplitudes of variaions in the area-to-peak ratio are $\sim$1%. The seasonal effects are also mostly due to outside temperature changes. The global decrease of the signal as a function of time could be due to changes in the liner reflectivity or in the water quality. From the current studies, the expected fractional signal loss in 10 years is less than 10% which gives confidence in a very good long-term performance of the Surface Detector.

## VI. Operation and maintenance

Currently more than 1600 surface detector stations are operational. Concerning the WCD itself, only very few failures have been detected. Only a few liners were observed to leak shortly after installation. In this case, which constitutes the worst failure mode, the tank is emptied and brought back to the Assembly Building for replacement of the interior components. Similarly, only a few solar panels have been damaged or were missing. Solar power system parameters are recorded and analyzed using the central data acquisition system. The average battery lifetime is 4 years, and batteries are changed during regular maintenance trips.

The PMTs and electronic boards are the most critical elements of the Surface Detector stations. They are subject to very severe environmental conditions: temperature variations, humidity, salinity and dust. The failure rates of the PMTs are about 20 per year (about 0.5%). Some HV module and base problems have been detected as well as some problems due to bad connections. All other failures except those concerning the PMTs (such as broken photocathode) can be repaired on site. It is currently estimated that the number of spare PMTs is sufficient for about 10-15 more years of operation. The failure rate of electronic boards is about 1% per year. Some of the problems are repaired simply by reflashing the software. Most of the electronic problems can also be repaired on site. All the spare parts are stored on site.

The operation of the array is monitored online and alarms are set on various parameters. The maintenance goal is to have no more that 10 detector stations out of operation at any time. It is currently estimated that the long-term maintenance (including the battery change) requires about 3 field trips per week. This maintenance rate is within the original expectations.The maintenance is organized by the Science Operation Coordinator and performed by local technicians. The Surface Detector does not require a permanent presence of physicists from other laboratories on site. However, remote shifts for the data quality monitoring will be implemented.

## VII. Conclusions

The construction of the Southern site of the Pierre Auger Observatory in Malargüe, Argentina, was completed in June 2008 and the Observatory has been running now with its full configuration for nearly one year. The operation of the Surface Detector array is monitored online and alarms are set on various parameters. The design is robust to withstand the adverse field conditions. The failure rates of various components are low and most of the failures can be repaired on site. The maintenance is performed by local technicians.

The characteristics of the Surface Detector stations are very uniform and the noise levels and resolutions exceed original requirements. The acceptance of the array is constantly monitored. This has allowed to take reliable physics data also in the construction phase.

Some detectors have been operational already for more than 8 years which has allowed to study the long-term performance of the Surface Detector. In particular, these studies have shown that the water quality remains excellent over several years.

Intensive and automatic monitoring, low failure rates, and local maintenance capabilities give confidence in a very stable long-term operation of the Pierre Auger Surface Detector.

# Extension of the Pierre Auger Observatory using high-elevation fluorescence telescopes (HEAT)

**Matthias Kleifges*** **for the Pierre Auger Collaboration**

*Forschungszentrum Karlsruhe, Institut für Prozessdatenverarbeitung und Elektronik, Postfach 3640, 76021 Karlsruhe, Germany*

***Abstract*. The original fluorescence telescopes of the southern Pierre Auger Observatory have a field of view from about 1.5° to 30° in elevation. The construction of three additional telescopes (High Elevation Auger Telescopes HEAT) is nearing completion and measurements with one telescope have started. A second telescope will be operational by the time of the conference. These new instruments have been designed to improve the quality of reconstruction of showers down to energies of $10^{17}$ eV. The extra telescopes are pivot-mounted for operation with a field of view from 30° to 58°. The design is optimised to record nearby showers in combination with the existing telescopes at one of the telescope sites, as well as to take data in hybrid mode using the measurements of surface detectors from a more compact array and additional muon detectors (AMIGA). The design, expected performance, status of construction, and first measurements are presented.**

## I. Introduction

The Pierre Auger Observatory has been designed to measure the energy, arrival direction and composition of cosmic rays from about $10^{18}$ eV to the highest energies with high precision and statistical significance. The construction of the southern site near Malargüe, Province of Mendoza, Argentina is completed since mid 2008 and the analysis of the recorded data has provided first results with respect to the energy spectrum [1], the distribution of arrival directions [2], the composition, and upper limits on the gamma ray and neutrino flux [3], [4]. The measured cosmic ray observables at the highest energies are suitable to tackle open questions like flux suppression due to the GZK cut-off, to discriminate between bottom-up and top-down models and to locate possible extragalactic point sources.

However, for best discrimination between astrophysical models, the knowledge of the evolution of the cosmic ray composition in the transition region from galactic to extragalactic cosmic rays in the range $10^{17}$ eV to $10^{19}$ eV is required. Tests of models for the acceleration and transport of galactic and extragalactic cosmic rays are sensitive to the composition and its energy dependence in the transition region where the current observatory has low efficiency.

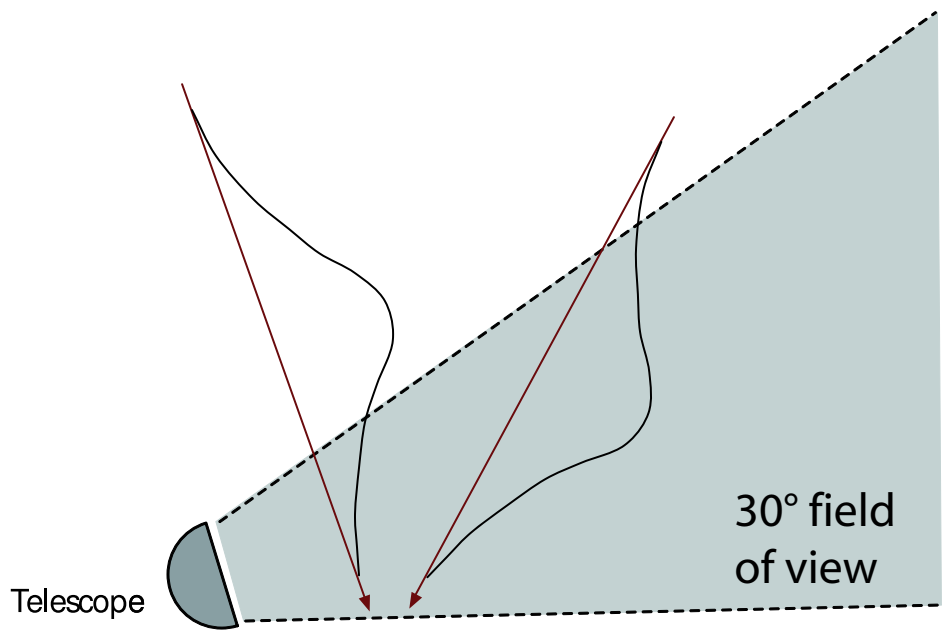

Fig. 1. Effect of limited field of view on reconstruction: Showers approaching the telescope have much higher reconstruction probability than those departing.

The fluorescence technique is best suited to determine the cosmic ray composition by a measurement of the depth of shower maximum. However, it is difficult to lower the energy threshold with the original design of the fluorescence telescopes. As the fluorescence light signal is roughly proportional to the primary particle energy, low energy showers can be detected only at close distance to a telescope. The field of view of the existing Auger fluorescence telescope (FD) is limited to 30° above the horizon (see figure 1). At close distances only the lowest few kilometres of the atmosphere are within the field of view. However, low energy showers reach their maximum of development at higher altitudes. Thus, the crucial region around the shower maximum is generally not observable. The small fraction of the shower development, which falls within the field of view, is mostly very dim and is insufficient to determine the depth of shower maximum $X_{\max}$. In addition, this cut-off effect also depends on primary mass and shower direction. A plain reconstruction of the shower profile using raw data would yield biased results with respect to zenith angle and mass composition. Cuts on the data to remove this bias (anti-bias cuts) are not useful as only very few showers would be left for the $X_{\max}$ determination.

From these arguments it is clear that an effective and unbiased detection of cosmic rays of lower energies requires the extension of the field of view to larger elevations. From the data collected since 2004, we know that the quality of reconstruction is improved considerably if showers are recorded by a hybrid trigger. These hybrid events provide information on the shower profile from the FD telescopes, but in addition at least one surface detector has detected secondary shower particles simultaneously. The data from the SD system restricts the time and the location of the shower impact point on ground. This improves the reconstruction of the shower geometry significantly [5]. An accurate geometry reconstruction with an uncertainty of about 0.5° is the necessary basis for energy and composition determination. But recording of hybrid data needs also adequate trigger efficiency for the individual surface detectors at lowest energies. Therefore, an enlarged energy range down to $10^{17}$ eV with high-quality hybrid events requires an extended field of view for FD telescopes in combination with a surface detector array of higher density in a small fraction of the observatory.

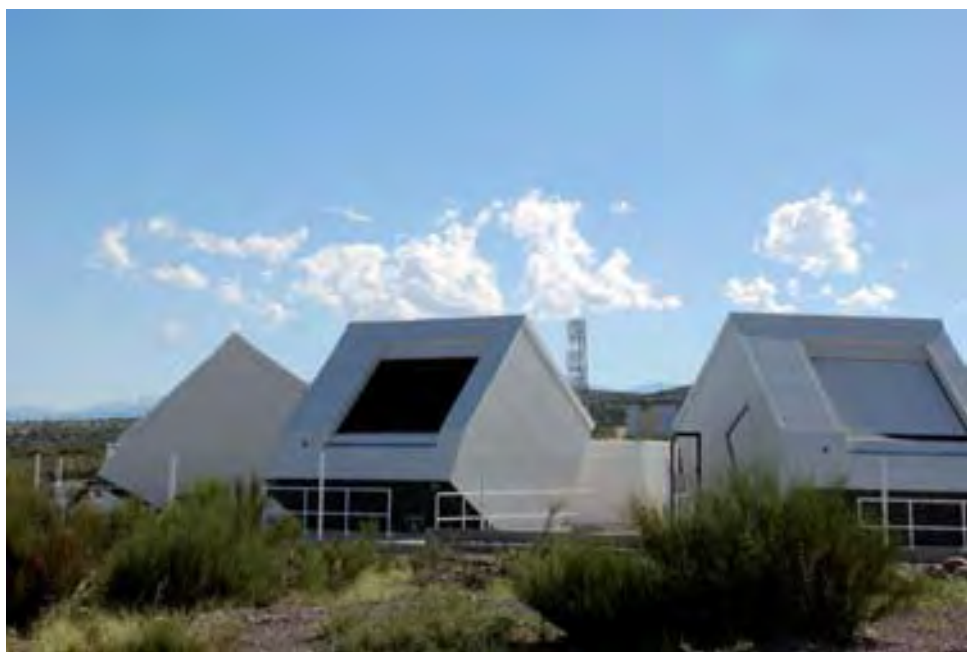

Fig. 2. Photo of the 3 HEAT telescopes tilted upward, end of January 2009. In the background the telecommunication tower of Coihueco is visible.

## II. Design and properties of HEAT

In 2006 the Auger Collaboration decided to extend the original fluorescence detector, a system consisting of 24 telescopes located at four sites, by three High Elevation Auger Telescopes (HEAT). These telescopes have now been constructed, and they are located 180 m north-east of the Coihueco FD building. At the same time, the collaboration deployed extra surface detector stations as an infill array of 25 $km^2$ close to and in the field of HEAT. Additional large area muon detectors (AMIGA) [6] will determine the muon content of the shower and further improve the determination of the composition of the primary cosmic ray particles.

The design of HEAT is very similar to the original FD system, except for the possibility to tilt the telescopes upwards by 29°. In both cases a large field of view of about 30°× 30° is obtained using a Schmidt optics. Fluorescence light entering the aperture is focused by a spherical mirror onto a camera containing 440 hexagonal PMTs. An UV transmitting filter mounted at the entrance window reduces background light from stars effectively. An annular corrector ring assures a spot size of about 0.6° despite the large effective aperture of 3 $m^2$. The high sensitivity of the Auger FD telescopes enables the detection of high energy showers up to 40 km distance. A slow control system for remote operation from Malargüe allows safe handling.

Differences between the conventional FD telescopes and HEAT are caused by the tilting mechanism. While the original 24 FD telescopes are housed in four solid concrete buildings, the 3 HEAT telescopes are installed in individual, pivot-mounted enclosures (see figure 2). Each telescope shelter is made out of lightweight insulated walls coupled to a steel structure. It rests on a strong steel frame filled with concrete. An electric motor can tilt this heavy platform though a commercial hydraulic drive by 29° within two minutes. The whole design is very rigid and can stand large wind and snow loads as required by legal regulations. All optical components are connected to the heavy-weight ground plate to avoid wind induced vibrations and to keep the geometry fixed.

Mirror and camera are adjusted in horizontal position. However, by tilting the telescope the varying gravitational force on camera body and mirror can change their relative position. Supplemental fixing bolts and an improved support structures are foreseen to keep the alignment of the optical system stable, which is essential for telescope pointing and optical resolution. Sensors for inclination are mounted at the mirror top, camera top and bottom, and at the aperture box. Distance sensors monitor the critical distance between camera and several points at the mirror system. These sensors are readout frequently for monitoring purposes.

Another design change for HEAT is the use of an improved DAQ electronics. The concept of the new electronics is the same as before, but as several electronic circuits have become obsolete, every front-end board had to be redesigned. Like the conventional FD electronics, the DAQ of one HEAT telescope contains 20 Analog Boards (AB) for analog signal processing, 20 First Level Trigger (FLT) boards for signal digitizing and storage, and one Second Level Trigger (SLT) board for the recognition of fluorescence light tracks and the initiation of data readout.

Along with faster FPGA logic the sampling rate was increased from 10 MHz to 20 MHz. The cut-off frequency of the anti-aliasing filters on the AB was adapted to about 7 MHz, but the other functions of the board remain the same. The redesigned FLT board implements all functions in FPGA of the Cyclone II FPGA family. A new custom-designed backplane provides dedicated point-to-point links between the FLT and SLT which lead to a factor 40 higher readout speed compared to the previous design. The usage of state-of-the-art FPGA in combination with the higher speed also establishes new fields of application for the DAQ system. The HEAT DAQ system is also the baseline design and the prototype for the Auger North FD electronics.

## III. Operation of HEAT

The horizontal ('down') position is the only position in which a person can physically enter the enclosure. This configuration is used for installation, commissioning, and maintenance of the hardware. The absolute calibration of the telescopes will be performed in this position as well. As the field of view of the existing Coihueco telescopes overlaps with HEAT in down position, it is possible to record air showers or laser tracks simultaneously. By comparing the reconstruction results from both installations one can directly determine the telescope resolution in energy and $X_{\max}$. We also want to reserve part of the time at one HEAT telescope for prototype studies for Auger North. Recording the same event in Coihueco and with the Auger North prototype will allow a direct comparison of the trigger and reconstruction efficiencies.

The tilted ('up') position is the default HEAT state. Telescopes are moved into this position at the beginning of a measuring run and stay that way untill the end of the run. From the trigger point of view the telescopes operate like a fifth FD building. Data of of the different installations (HEAT, diffenent FD sites, infill and Amiga, surface detector) are merged offline only, but the exchange of triggers in real time makes the recording of hybrid showers possible. The combined data will improve the accuracy of shower energy and $X_{\max}$ determination at all energies, but especially at the lower end down to $10^{17}$ eV.

## IV. First Measurements

First measurements were performed with HEAT telescope #2 at the end of January 2009. From January, 30$^{th}$ to February, 1$^{st}$ the telescope was operated for two nights in up and down position. At first, the camera was illuminated with a short light pulse from a blue LED located at the center of the mirror. The High Voltage for the PMTs and the individual electronic gains were adjusted to achieve uniform light response in every pixel. Subsequent measurements with the LED pulser were performed at different tilting angles, but with the same settings as found in in down position. No indications were found for a gain change due to changed orientation of the PMTs in the Earth's magnetic field.

In the next step, the mechanical stability of the optical system was verified. The telescope was tilted several times from down to up position and back. The readings of the inclination and distance sensors were recorded during the movement. The analysis of the distance between camera and center of the mirror showed damped oscillations of low amplitude which stopped within seconds after the movement terminated. At rest the distance change between up and down position is less than 0.5 mm which is neglible for the telescope's optical properties.

After these cross-checks several showers were recorded with the telescope tilted in up position and in coincidence with Coihueco telescopes #4 or #5. One of the recorded events is shown in figure 3. The event data of both telescopes match very well in time (colour of the pixels in figure 3). The reconstruction yields a shower distance of $2.83 \pm 0.06$ km from Coihueco and an energy of the primary particle of $(2.0 \pm 0.2) \cdot 10^{17}$ eV.

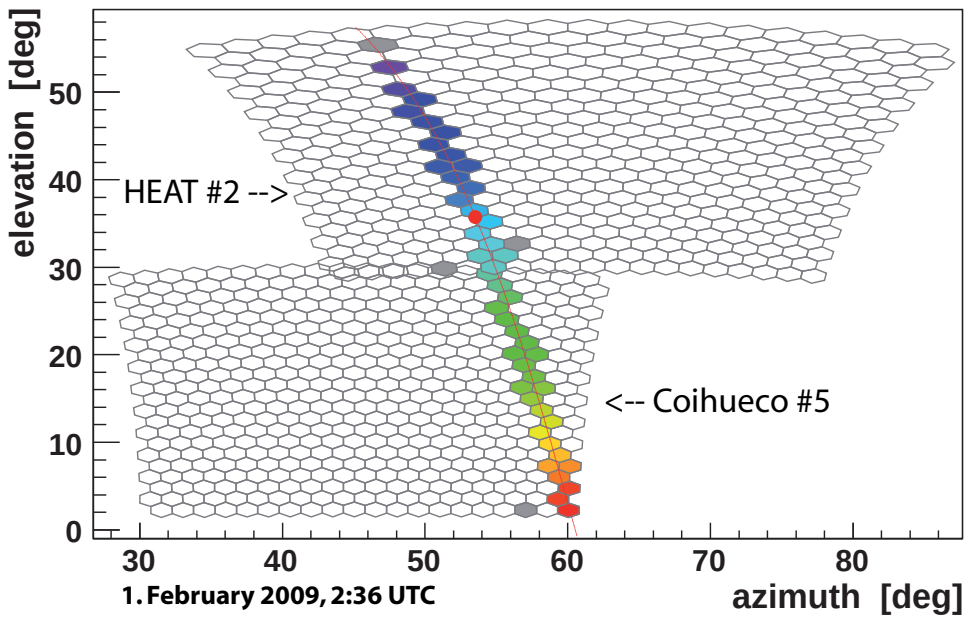

Fig. 3. This shower was recorded by HEAT telescope #2 and Coihueco telescope #5.The relative arrival time of the fluorescence light is coded in the colour of the pixel. The solid line is a fit of the shower detector plane.

In figure 4, the reconstructed longitudinal shower profile is shown together with a fit to a Gaisser-Hillas function. The fit yields a value of $(657 \pm 12)$ g cm$^{-2}$ for $X_{\max}$. The plot also accentuates the need for HEAT telescopes for an accurate reconstruction: Using only the data point above a slant depth of 700 g cm$^{-2}$ (Coihueco data) it would not have been possible to fit the profile and find a precise maximum.

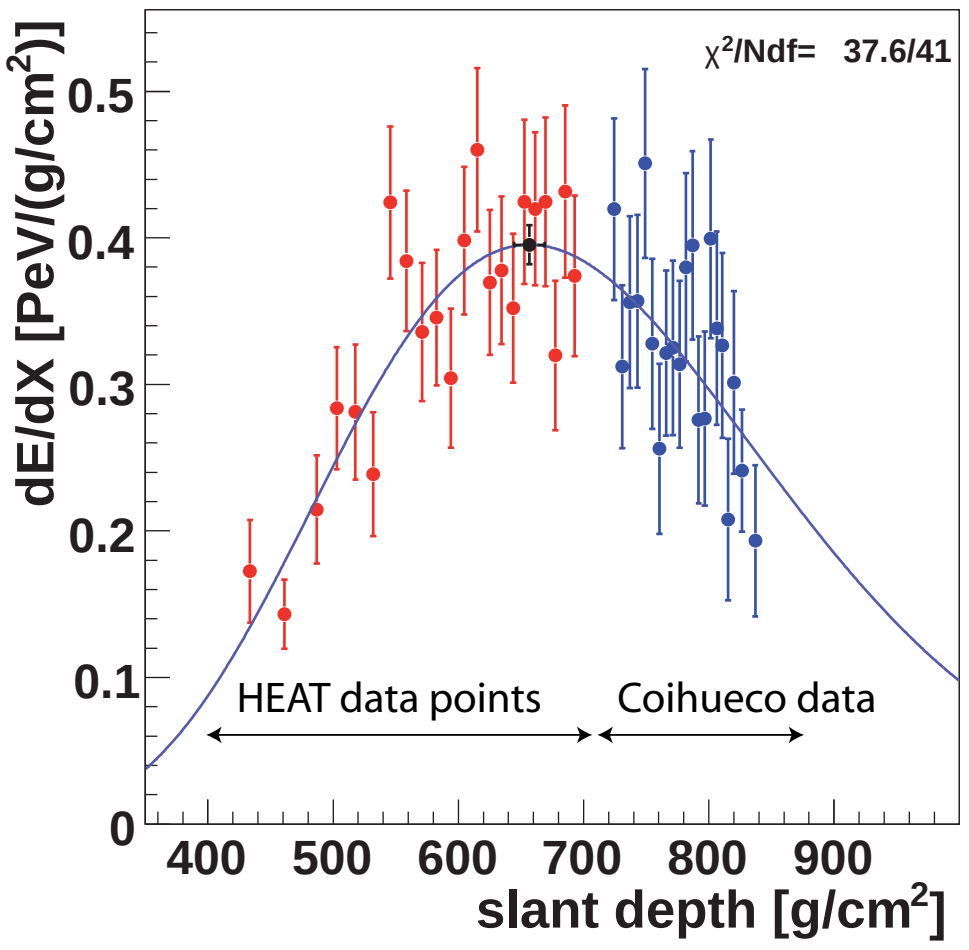

Fig. 4. Longitudinal shower profile of event in figure 3 together with Gaisser-Hillas-fit. Only a fit using both HEAT (left) and Coihueco (right) data points results in a reasonable $X_{\max}$ value.

## V. Conclusion and Outlook

First measurements with a single telescope have demonstrated that HEAT will improve the energy threshold of the Pierre Auger Observatory at the Southern site down to about $10^{17}$ eV. The HEAT design satisfies all requirements with respect to stability and ease of operation. It is expected that all three HEAT telescopes are fully operational in September 2009. They will provide interesting data in the transition region from galactic to extragalactic sources and allow important prototype tests for the design of the Auger North FD system.

# AMIGA - Auger Muons and Infill for the Ground Array of the Pierre Auger Observatory

**Manuel Platino*** **for the Pierre Auger Collaboration**†

*Centro Atómico Constituyentes (Comisión Nacional de Energía Atómica/CONICET/UTN-FRBA)
†Observatorio Pierre Auger, Av. San Martín Norte 304 (5613) Malargüe, Prov. Mendoza, Argentina

***Abstract*. AMIGA consists of 85 detector pairs each one composed of a surface water-Cherenkov detector and a buried muon counter. Each muon counter has an area of 30 m$^2$ and is made of scintillator strips with doped optical fibers glued to them, which guide the light to 64 pixel photomultiplier tubes. The detector pairs are arranged at 433 and 750 m array spacings in order to perform a detailed study of the $10^{17}$ eV to $10^{19}$ eV spectrum region. Design parameters and performance requisites are outlined. Construction of the first muon detectors, associated software and hardware, and the results of laboratory tests are described. Some preliminary results on the performance of the 750 m array of surface detectors are presented.**

## I. Introduction

The Pierre Auger Observatory was built to detect the highest energy cosmic rays known in nature with two distinctive design features, a large size and a hybrid detection system in an effort to observe a large number of events per year with minimum systematic uncertainties. The Southern component of the Auger Observatory is located in the Province of Mendoza, Argentina, and it spans an area of 3000 km$^2$ covered with over 1600 water Cherenkov surface detectors (SDs) deployed on a 1500 m triangular grid with 24 fluorescence detector (FD) telescopes grouped in units of 6 at four sites on the array periphery, each one with a $30^\circ \times 30^\circ$ elevation and azimuth field of view [1], [2]. With such a geometry the Auger Observatory has been able to cast light on two spectral features at the highest energies, the ankle and the GZK-cutoff [3], [4]. At lower energies the electromagnetic fields deflect the particle trajectories rendering impossible to identify the sources, but still composition studies should help discriminate whether the sources are galactic or extragalactic, and where the transition occurs. Besides the mentioned ankle and GZK-cutoff, the cosmic ray spectrum has two other features where the spectral index abruptly changes: the knee ($\sim 4 \times 10^{15}$ eV) and the second knee ($\sim 4 \times 10^{17}$ eV). The transition from galactic to extragalactic sources is assumed to occur either at the second knee or along the ankle (see [5] and references therein) and the way to identify it would be a change in the cosmic ray composition from dominant heavy primaries to either a mixed or light-dominated composition.

## II. AMIGA Baseline Design

Within the Auger baseline design described above, the surface array is fully efficient above $\sim 3 \times 10^{18}$ eV and in the hybrid mode this range is extended down to $\sim 10^{18}$ eV which does not suffice to study the second knee - ankle region. For this purpose [6], Auger has two enhancements: AMIGA ("Auger Muons and Infill for the Ground Array") [5], [7], [8] and HEAT ("High Elevation Auger Telescopes") [9]. Only a small area is needed since the cosmic ray flux increases rapidly with decreasing particle total energy. On the other hand, the detectors have to be deployed at shorter distances among each other in a denser array since lower energies imply a smaller air shower footprint. AMIGA consists of a group of detector pairs deployed in a denser array comprising 61 detectors spaced 750 m apart and 24 extra detectors spaced 433 m apart. All 85 pairs are placed within de 1500 m SD array. This group of SDs is referred in this paper as the Graded Infill Array or simply the infill. Currently the infill has 35 surface detectors, which have been operational since September 2008, accounting for 40% of its planned full aperture. Figure 1 shows the deployment progress of the infill in the Southern Auger Observatory at the time of this paper.

Each AMIGA detector pair consists of a water Cherenkov SD and a neighboring buried muon counter (MC). The MC consists of three scintillator modules each one lodged in a PVC casing. The scintillators are highly segmented in strips 400 cm long $\times$ 4.1 cm wide $\times$ 1.0 cm high made of extruded polystyrene doped with fluor and co-extruded with a $TiO_2$ reflecting coating. They have a groove on the top side where a wave-length-shifter optical fiber is glued and covered with reflective foil. The fibers end at an optical connector matched and aligned to a 64 channel multi anode Hamamatsu H8804 photomultiplier tube (PMT) of 2 mm $\times$ 2 mm pixels. The coupling of the fibers to the PMT requires a fine polishing of the surface that touches the glass on top of the photocathode. For this purpose we use a fly-cutter to polish the surface of the 64 fibers attached to the optical connector as a single unit in order to make the fibers-connector surface as even as possible. The contact of the fiber connector to the PMT glass is lubricated with optical grease and the alignment system chosen for the PMT to fibers coupling is the same one as used by the OPERA project (see Figure 47 in [10]). Each MC would then be composed of 192 independent channels buried alongside a SD at a depth to be experimentally determined [11].

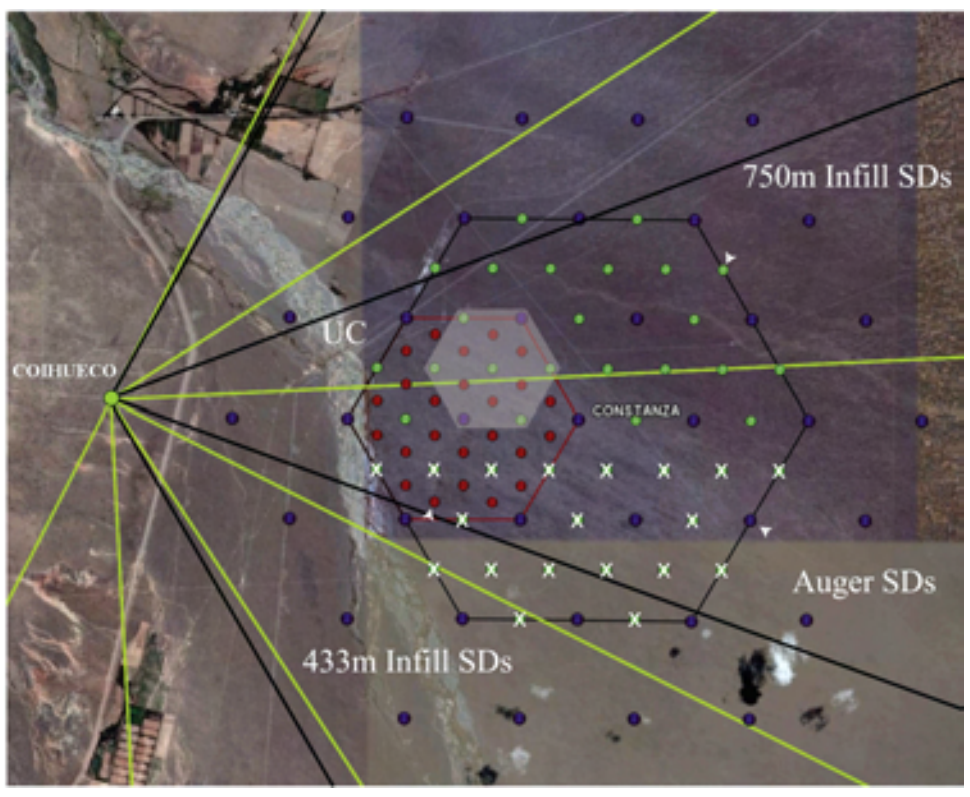

Fig. 1: The Graded Infill Array of the Southern Auger Observatory. The bigger hexagon highlights the 750m spaced SDs. The next smaller hexagon highlights the denser 433m spaced SDs, not deployed yet. The smallest hexagon highlights the UC. The crossed SDs are part of the 750 m array but still not deployed on site at the time of this paper. The darkest spots indicate the SDs that are part of the Auger original array.

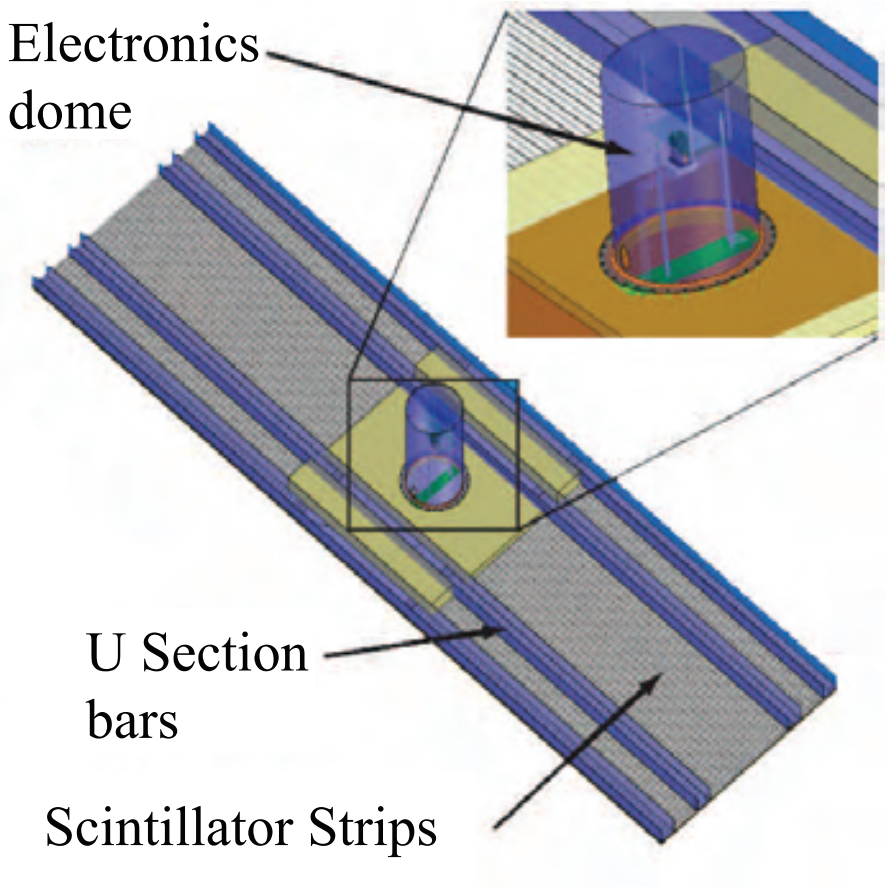

Fig. 2: AMIGA scintillator module. Top PVC panels covering the scintillator strips are not shown. Highlighted are the strips, the PVC U bars, the PMT and electronics dome, zoomed in the top right insert.

To test the baseline design, an engineering array also called Unitary Cell (UC) is being built at 7 of the 61 surface detector stations locations in the 750 m infill array, forming an hexagon around the "Los Piojos" central SD. In this array, each MC is composed of 4 modules rather than 3 since a module is halved into two with 200 cm long scintillator bars, providing 256 independent channels in order to assess muon pile-up close to the airshower core. Each MC of the UC will be buried 2.25 m deep.

Figure 2 shows a diagram of the scintillator module. The 64 400 cm strips are placed in two groups of 32 at each side of a central dome where the electronics is lodged. The module casing is built with PVC panels (transparent in the figure to reveal the strips) onto which the strips with fibers are glued. Structural stability is provided by four PVC U section bars placed at both sides of the central dome all along the module. The UC MCs are equipped with 0.8 mm diameter St. Gobbin and 1.0 mm diameter Kuraray fibers for the 200 cm and 400 cm strips, respectively.

AMIGA electronics have both an underground and a surface component powered by solar panels, which provide a total of 20W, 24V DC supply to the buried MC on each SD pair. Each scintillator module has a PMT to which a set of printed circuit boards (PCBs) are connected to provide PMT pulse filtering and amplification, digitization, sampling and storing in an internal memory (see [12] for more details on the PCBs). Each PMT signal is filtered by an inverter amplifier of gain $\sim 3.8$ and digitized, whose output is sampled and transformed into digital data by a Cyclone III EP3C25 Field Programmable Gate Array (FPGA) from Altera every 3.125 ns [12], [13]. The discrimination level can be controlled independently for each of the 64 channels using 8 units of 8-channel programmable 12 bit digital-to-analog converters (TLV5630 by Texas Instruments) which are commanded at any given time from the FPGA. After laboratory experiments, the threshold was set at $\sim 30\%$ of the pixel mean single photo electron (SPE) amplitude ($\sim$ 15 mV after amplification). The underground electronics is regulated by a microcontroller (TMS470 16-bit Risc by Texas Instruments) which performs the data transfer to the surface and the slow control (temperature, humidity and power supplies monitoring).

The digitized pulses are continuously stored in a circular buffer in a static RAM memory, task performed by the FPGA, and upon reception of an SD trigger signal they are adequately channeled through a CAN bus by the microcontroller to the surface detector [13]. On the surface a single board microcomputer receives the data and upon a request from the central data acquisition system at the Auger campus, transmits both SD and MC data by a radio link using the IEEE 802.15.4 standard, as proposed for the AMIGA network. The topology used for this network is point-to-multipoint, in which every SD of the infill connects to a hub in the nearby Coihueco hill. Ultra low power consumption, industrial grade XBee pro radios working over IP are used for this purpose. The Linux TUN driver running on the single board microcomputer of the surface detector is used to tunnel the IP network through the serial port and the XBee radio. Relatively low bit rate (effective $\sim$ 125 kbps) and 100 bytes frames requires header compression for IP traffic. The first successful tests with the 802.15.4 installed on the "Tierra del Fuego" SD showed no data loss and a reliable communication to the Auger campus.

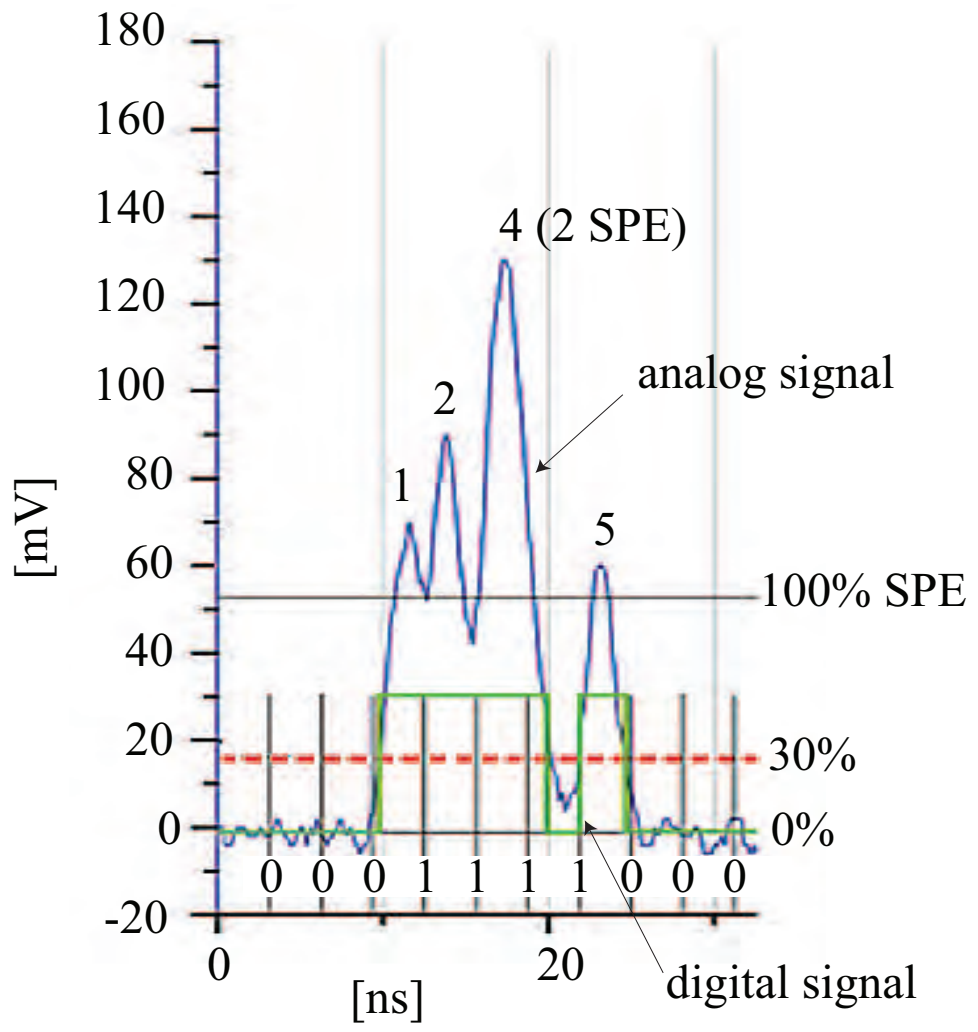

Fig. 3: Analog train of pulses produced by an impinging background muon on a scintillator strip and schematics of data digitalization.

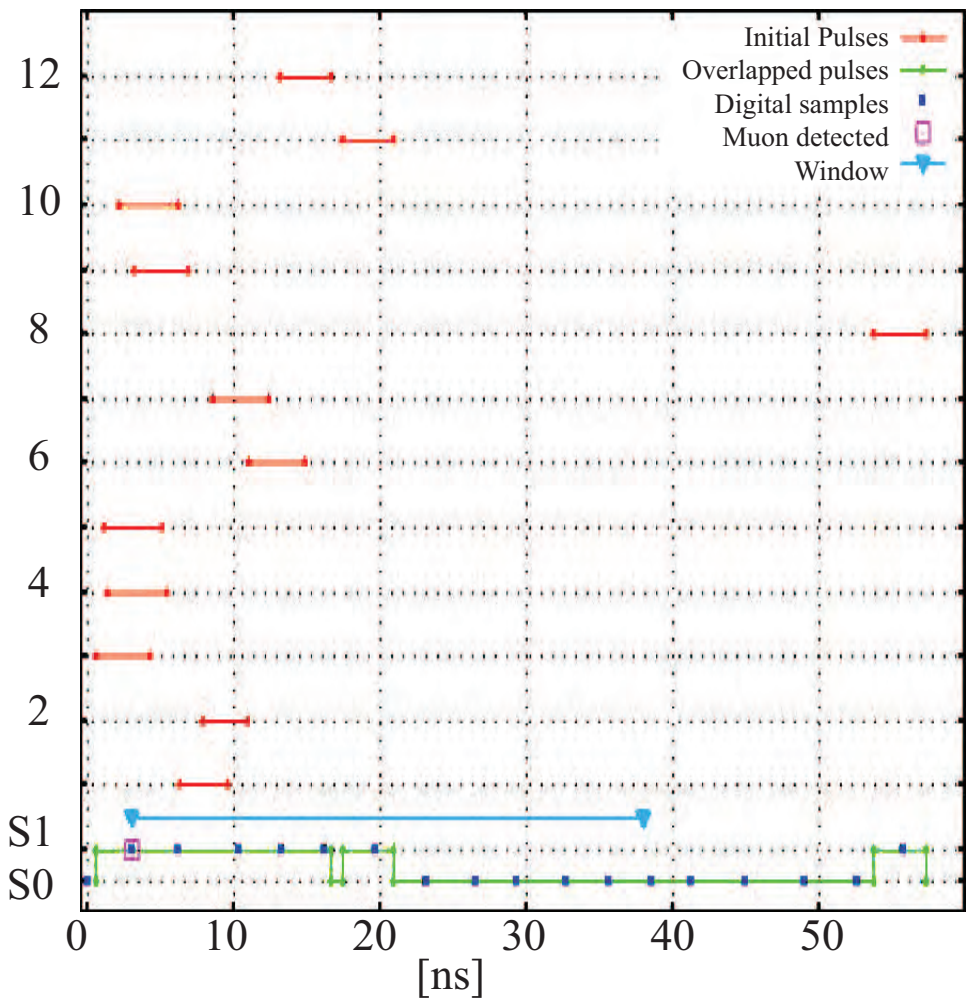

Fig. 4: Diagram of the one-bit sampling system used by the AMIGA front end. The Y axis indicate the number of each arriving pulse at a given channel of the PMT from a muon excited scintillator bar. The X axis indicates time.

A smaller 16 channel version of the scintillator module was built in 2007 and buried for three months in Centro Atómico Constituyentes (CAC) in Buenos Aires. We measured the amplitude and charge histograms for a Hamamatsu H8711-06 PMT both on the surface and underground. The results showed no significant change on the performance in this first successful data taken from a buried prototype [14].

A typical amplified train of pulses produced by a background muon is shown in Figure 3, labeled "analog signal". We see it comprises of four distinctive pulses, the fourth one is an isolated SPE pulse while the third one is a pile up of more than one SPE. Indicated are the equivalent voltage values for the mean SPE amplitude (labeled 100% SPE) and the discrimination level at 30%.

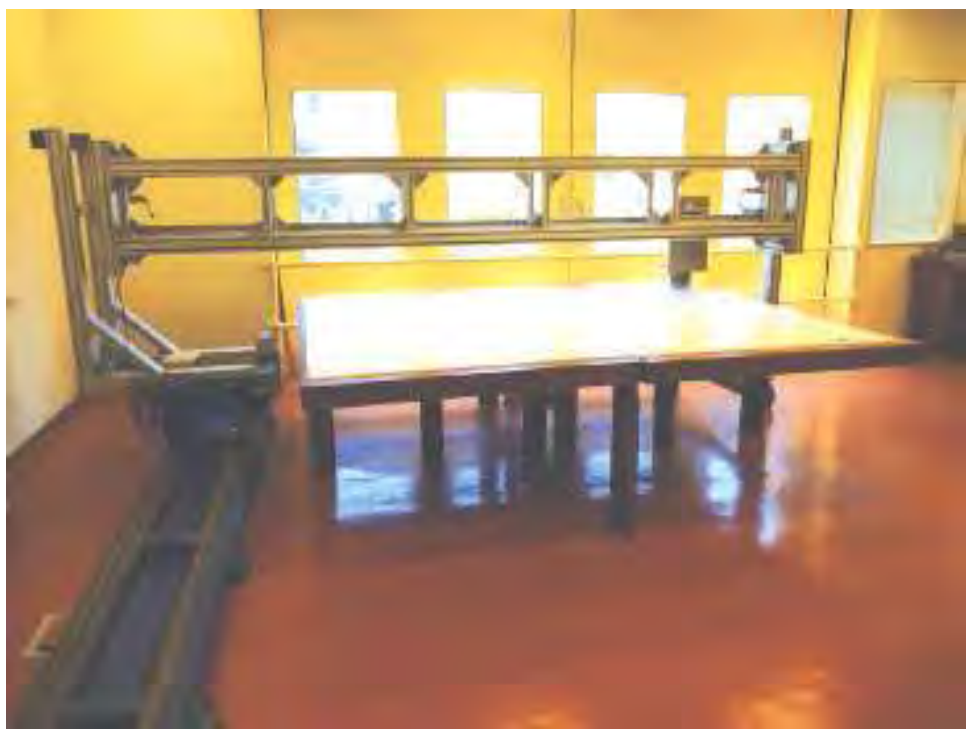

Fig. 5: AMIGA Scanner built at Centro Atómico Constituyentes

The digits on top of each pulse indicate the deduced cumulative number of SPEs that arrive at the PMT. The expected discrimination of this signal is labeled in Figure 3 as "digital signal" and the $0-1$ string illustrates the output of the four possible 3.125 ns FPGA signals (see [12] for prototype measurements of the sampling process). Results of the simulated sampling and muon counting process are shown in Figure 4, as performed by the analog front end and the digital board under a one-bit system. This plot shows the arrival time of 12 SPE pulses (labeled "initial pulses") generated in a given channel of the scintillator module by an impinging background muon and their corresponding duration (indicated by the length in the X direction of the 12 lines distributed in the Y axis from 1 to 12). The arrival time of each "initial pulse" is simulated based on results obtained from the characterization of the scintillator-fiber system, and their duration in time is obtained from measurements of the analog front end characterization. The "overlapped pulses" is the resulting digitized signal obtained at the front end after discrimination, comprised of the overlapping of all the analog "initial pulses" in time. The "digital samples" are taken at locations with a mean 3.125 ns separation with fluctuations which take into account the jitter introduced during the sampling by the FPGA. The "window" between inverted triangles denotes the time frame within which no further muon counts are allowed. The AMIGA trigger counts a muon if three or more "digital samples" of value S1 are within this time window. If exactly two S1s are within the time window, they will be counted as a muon if and only if there is one or more S0s between them. Note that the late coming SPE is not overcounted as a muon. AMIGA digital electronics uses a digital 0 for S1 and a digital 1 for S0. These values are switched in these simulations for clarity purposes.

PMTs will be tested in a dark box built in the AMIGA laboratory at CAC before transportation to the Auger Observatory. The relevant parameters to be checked for each PMT are: relative quantum efficiency, gain uniformity between pixels, crosstalk, overall dark-rate (sum of 64 channels) and the SPE mean amplitude.

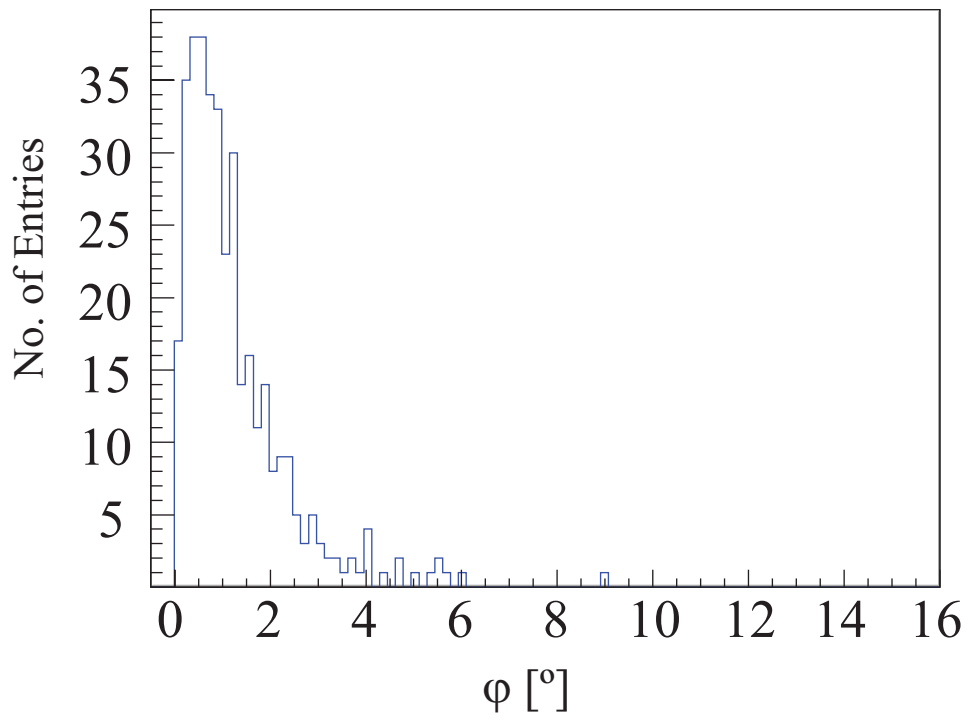

Fig. 6: Reconstruction of arrival direction with and without infill SDs histogram. $\varphi$ represents the difference between angles of arrival of the reconstructed shower with minus without the infill measurements.

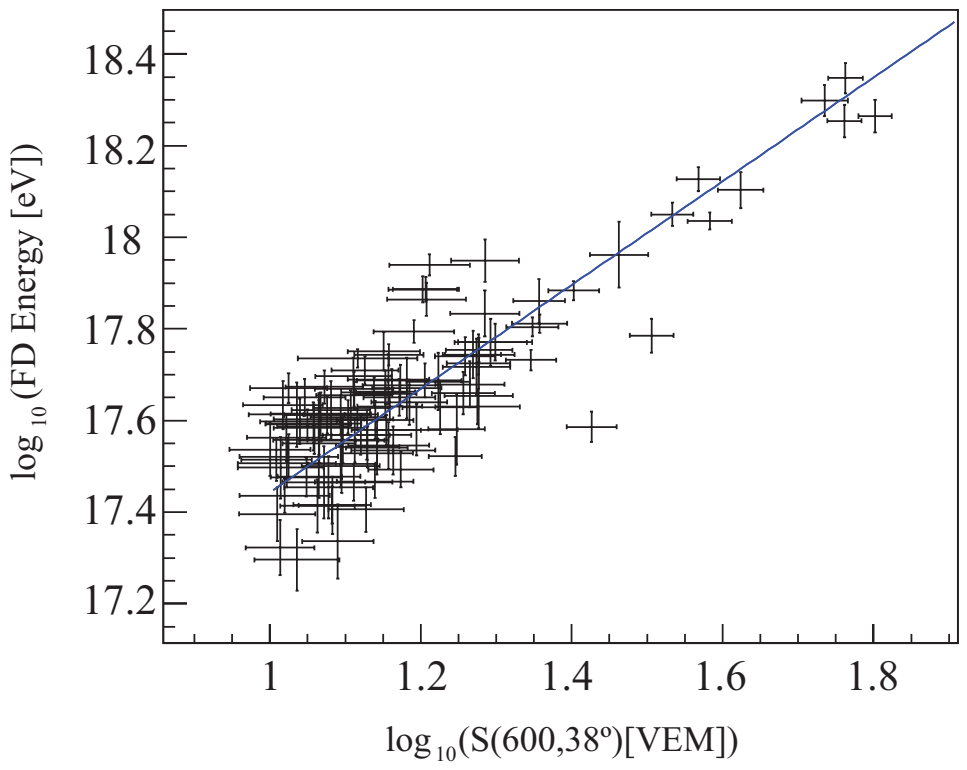

Fig. 7: Correlation between $S(600, 38°)$ and the FD energy

The response of each scintillator strip will be characterized with a 5 mCu $^{137}$Cs radioactive source mounted on an X-Y positioning system or "scanner" designed for this purpose. Figure 5 shows the AMIGA scanner as mounted today in the scintillator module assembly workshop at CAC. The X-Y positioning system has two 1 mm precision linear guides, one for each dimension moved by step-by-step motors. The positioning system has a total effective displacement of 3.75 m in the X axis and 4 m in the Y axis, expandable up to 9 m. Data collected on the scintillator modules tests will be stored on the AMIGA detectors database to keep track of the status of all the MCs during and after the deployment, and for simulation purposes.

## III. Infill Measurements

AMIGA data acquisition has now started with the SD 750 m infilled area and preliminary analyzes are presented with a data set restricted to events with zenith angle $\theta < 60°$ and well constrained within the infill (i.e. with the six SDs of the hexagon enclosing the highest signal SD in its center active). The graded infill was envisaged to both have saturated efficiency down to $10^{17}$ eV and significantly improve and test on the main array reconstruction uncertainties [8]. The latter was tested by reconstructing events with and without the infill and by imposing further data cuts of $\geq 7$ and $\geq 3$ SDs, respectively. The distribution of the space angle differences between the mentioned reconstructions are compared in Figure 6. The 68% of such differences results to be inside $1.4°$, which shows the good compatibility of the two arrays, and is mostly related to the uncertainty of the Auger regular reconstruction due to the expected better infill accuracy being a denser array.

We also focused on a study of the preliminary energy calibration (see [15] for the method details as applied to the 1500 m SD array) based on events simultaneously detected by the infill and the FD. Note that although the infill and main array calibrations are both extracted from FD energy measurements, they will differ since they have to be optimized for different energy ranges, infill signals are more attenuated with zenith angle, and the infill shower ground parameter is taken at 600 m instead of 1000 m.

Figure 7 shows the correlation between the constant intensity cut corrected shower ground parameter $S(600, 38°)$ and the FD assigned energy fitted with an exponential function.

## IV. Summary

The AMIGA baseline design is finalized and outlined. Final design details are being adjusted in the electronics and the mechanical mounting at the time of this paper. The laboratory tests on the scintillator module prototype and the electronics show positive results. A muon detector module was buried and successfully operated. Field tests on the communications also show system reliability. Infill data acquisition has already began and analysis led to a preliminary calibration.

# Radio detection of cosmic rays at the southern Auger Observatory

**A.M. van den Berg*, for the Pierre Auger Collaboration†**

**Kernfysisch Versneller Instituut, University of Groningen, NL-9747 AA, Groningen, The Netherlands*
*†Av. San Martín Norte 304, (5613) Malargüe, Prov. de Mendoza, Argentina*

***Abstract*. An integrated approach has been developed to study radio signals induced by cosmic rays entering the Earth's atmosphere. An engineering array will be co-located with the infill array of the Pierre Auger Observatory. Our R&D effort includes the physics processes leading to the development of radio signals, end-to-end simulations of realistic hardware configurations, and tests of various systems on site, where coincidences with the other detector systems of the Observatory are used to benchmark the systems under development.**

## I. INTRODUCTION

Results from the Southern Pierre Auger Observatory as well as the baseline design of the Northern Observatory [1] point to the need of very large aperture detection systems for ultra-high energy cosmic rays (UHE CRs); see, e.g., Ref [2]. There are a number of worldwide efforts to develop and establish new detection techniques that promise a cost-effective extension of currently available apertures to even larger dimensions. These are, for example, the observation from space of fluorescence emission by showers or the use of large arrays of radio antennas.

The detection of radio emission induced by high-energy and ultra-high-energy cosmic rays hitting the Earth's atmosphere is possible because of coherent radiation from the extensive air shower at radio frequencies. This radiation, which is emitted by secondary particles created in the air shower, can be measured with simple radio antennas, as was demonstrated first by Jelley in 1965 [3]. Recently, improved technology has led to a revival of this technique. Radio detectors, like LOPES and CODALEMA produce promising results at energies beyond $10^{17}$ eV [4], [5].

With its nearly 100% duty cycle, a signal-to-noise ratio scaling with the square of the cosmic-ray energy, its high angular resolution, and its sensitivity to the longitudinal air-shower evolution, the radio technique is particularly well-suited for detection of UHE CRs in large-scale arrays. Therefore, we are performing an R&D project to study UHE CRs using the detection of coherent radio emission from air showers in the Earth's atmosphere. The project, called AERA (Auger Engineering Radio Array), will have a dimension of about 20 km$^2$. For such an area we expect an UHE CR rate of about 5000 identified radio events per year. This data set will be used to address both scientific and technological questions. At the same time, the scale of such an array is large enough to test concepts for the deployment of hardware, for the operation of hardware and software, and to monitor the sustainability of critical parts of the whole system for a much larger array.

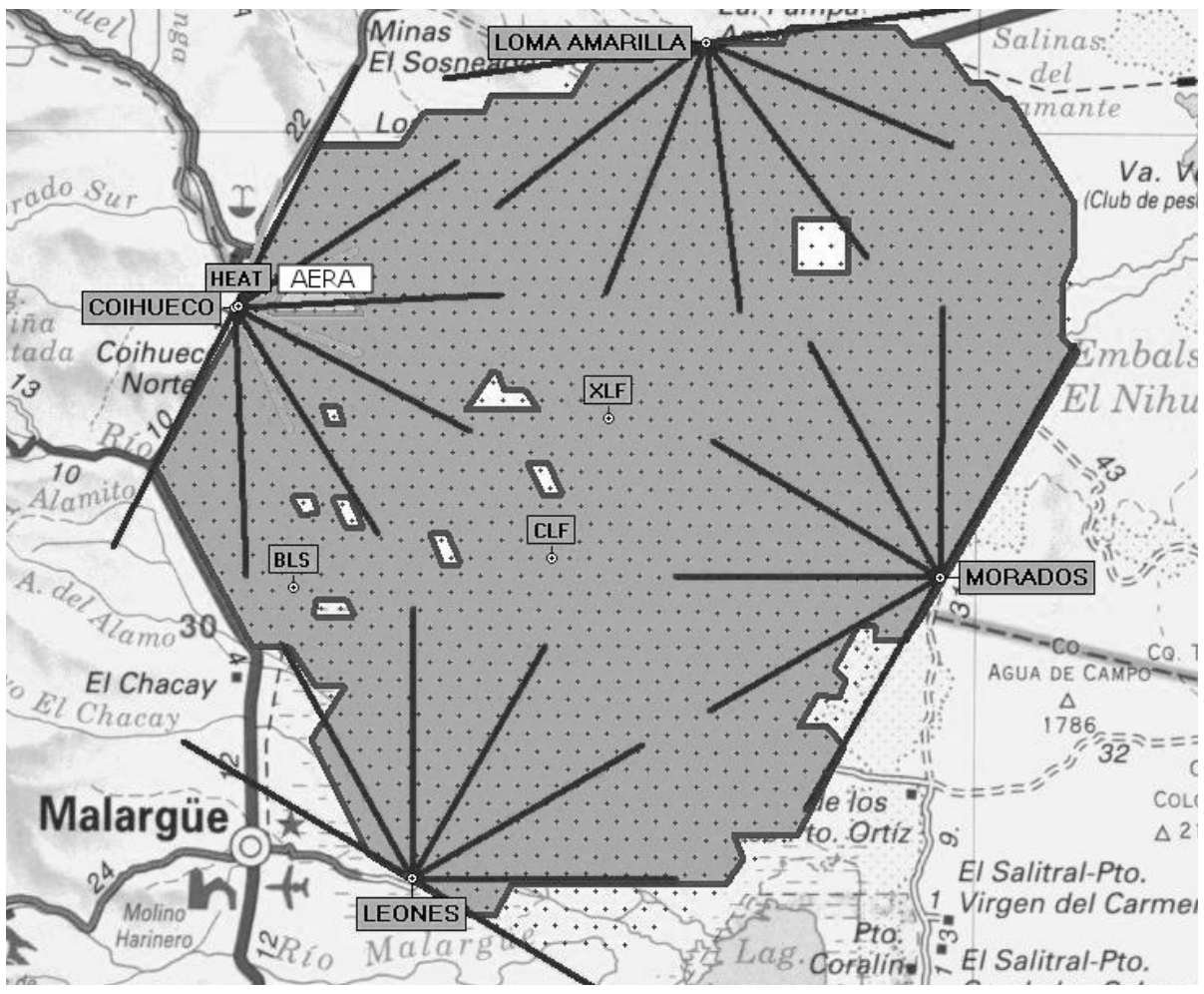

Fig. 1. Locations of the radio setups at the southern Observatory. Existing setups are located near the BLS and the CLF; the future AERA is near Coihueco.

AERA will fulfill three science goals which are not independent from each other, but have to be tackled sequentially in the time of operation and analysis and in close conjunction with data from the baseline detectors of the Observatory [6] and the enhancements AMIGA[7] and HEAT [8]. They are all located in the north-eastern part of the Observatory near the Coihueco fluorescence building, see Fig. 1. The three science goals are listed below.

1) Thorough investigation of the radio emission from an air shower at the highest energies. This includes the understanding of all the dependencies of the radio signal on general shower parameters and on the shower geometry. By this a better insight into the underlying emission mechanism will be possible; the question has to be answered which is the theory of choice and to which level one has to consider additional effects [9].
2) Exploration of the capability of the radio-detection technique. Determination of the extend and accu-

racy of stand-alone radio-detection to provide information on the most important physics quantities of UHE CRs: primary energy, primary mass, and arrival direction.

3) Composition measurements between $10^{17.4}$ and $10^{18.7}$ eV where we expect the transition from galactic to extragalactic origin of cosmic rays. With super-hybrid measurements at the AMIGA site [7], AERA will contribute with a worldwide unrivaled precision to the study of the cosmic-ray energy spectrum and composition in the energy range of the transition .

AERA will be co-located with AMIGA, which is overlooked by the HEAT detector. These enhancements of the Observatory provide worldwide the only possibility to study the details of radio emission from air showers in a timely manner.

## II. Present status of the project

Initial measurements have been made showing that indeed radio detection of UHE CRs can be performed at the southern site of the Auger Observatory with a setup near the BLS, another one near the CLF (see Fig. 1). Still the number of radio events in coincidence with its Surface Detector Array (SD) is relatively small, mainly because of the relatively small scale of the radio detector arrays used. A description of the setups used is presented in Ref. [10]. Here we list some of the initial results and present a short description of these setups. One setup consists of three dual active fat dipole antennas, mounted near the CLF in a triangular configuration with a baseline of 139 m. With this setup, self-triggered events are being recorded and using GPS time stamps, they are compared to events registered with the SD Array of the Observatory. The other setup, located near the BLS, uses dual logarithmic periodic dipole antennas and is triggered externally using a set of two particle detectors. Also in this case the Radio-Detection Stations are mounted on a triangular grid; here the baseline is 100 m. Both setups measure the electric field strength in two polarization directions: east-west and north-south.

Several coincident events between SD and externally triggered Radio-Detection Stations located near the BLS have been recorded [11]. In 27 cases we have recorded simultaneously in three different Radio-Detection Stations a coincidence with the SD. This allowed us to compare in these cases the arrival direction as determined by our Radio-Detection Stations with that from SD. The histogram in Fig. 2 displays for 27 events the measured angular difference between the arrival direction as determined with SD and as determined by our three Radio-Detection Stations. Using a Rayleigh function, the 68%-quantile of the distribution of the angular difference has been determined to be $(8.8 \pm 1.0)°$. In most events, the number of SD stations used in the event reconstruction was 3 or 4. The angular resolution for these type of SD events is about 2.0° and can thus be neglected. The major contribution to the angular uncertainty is the relative small distance of 100 m between the Radio-Detection Stations compared to the timing accuracy obtained (about 3 ns).

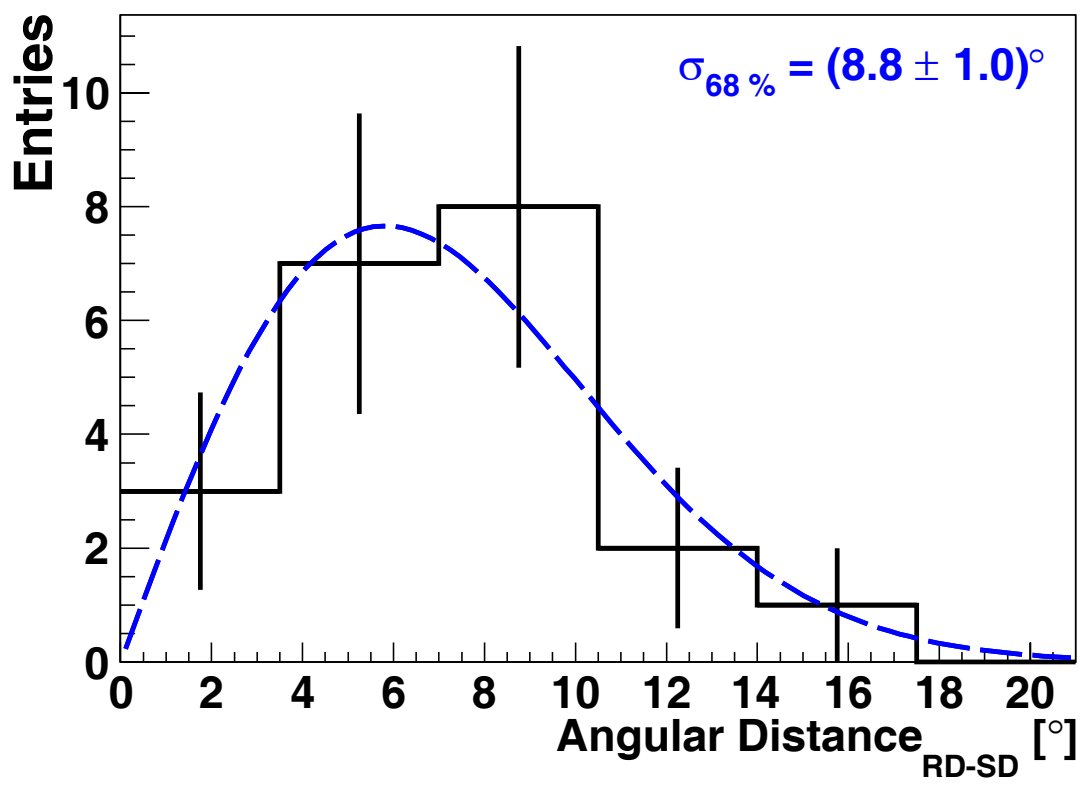

Fig. 2. Plot of the angular difference between the arrival direction of observed air showers using data from the Surface Detector Array and from the Radio-Detection Stations near the BLS. The histogram indicates the data obtained for 27 events; the dashed line is a fit through these data using a Rayleigh function; see text for details.

The sky distribution of radio events observed with the setup near the CLF and in coincidence with the SD Array is presented in Fig. 3. As expected we observe that the efficiency for the observation of cosmic rays with the SD Array does not depend on the azimuth angle. The sky plot of the observed radio events, however, is highly asymmetric with a large excess of events arriving from the south: this represents more than 70% of all events. This observed asymmetry provides further support for the geomagnetic origin of radio emission by air showers and has been observed before in the northern hemisphere [12]. The Earth's magnetic field vector at the site of the Pierre Auger Observatory makes an angle of about 60° with respect to the zenith and its azimuthal angle is 90° (i.e. north). This is indicated as the cross in Fig. 3. In the geomagnetic model, electric pulses will be strongest if the shower axis is perpendicular to the magnetic field vector. And for these observed events the trigger for radio detection was a simple pulse-height threshold on the radio signal, which explains why we detected more events coming from the south than from the north.

Studies with solitary systems have been performed. These systems are powered by solar energy, have a wireless connection to a central DAQ system, and operate in self-trigger and/or externally triggered modes. Special attention has been paid to reduce self-induced noise caused by DC-DC converters and by digital electronics. With this system self-triggered events have been obtained. However, for these events there were no data from more than two radio stations and thus the arrival direction could not be compared with that from SD. Nevertheless this is the first time ever that such an independent detection has been achieved using Radio-Detection Stations [13].

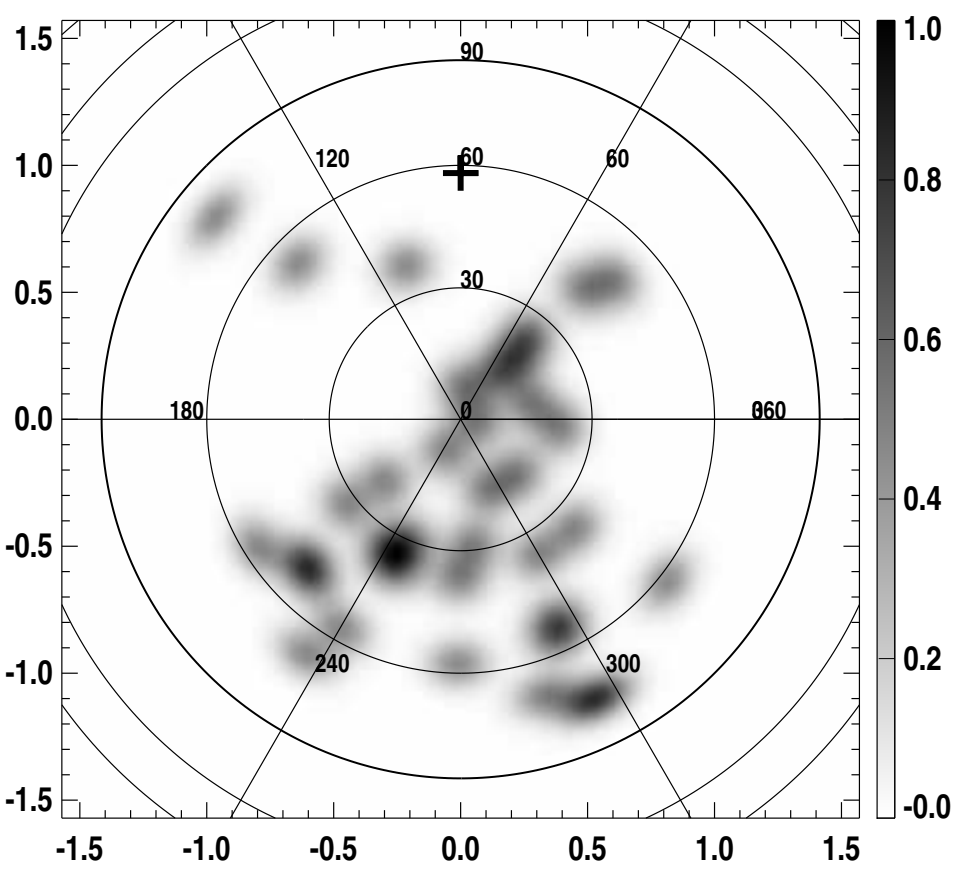

Fig. 3. Sky map of 36 radio events registered with the autonomous system in coincidence with the SD Array in local spherical coordinates. The zenith is at the center and the azimuths are oriented as follows: 0° east, 90° north. The geomagnetic field direction in Malargüe is indicated by the cross.

Noise levels have been studied in detail. For this analysis we have defined the root mean square (RMS) value for 750 consecutive samples of each time trace with have a sampling rate of 400 MS $s^{-1}$. Using a digital filter the analyzed signals were limited to the frequency band between 50 and 55 MHz. The distribution of these RMS values, measured for a period of almost one year, are plotted in Fig. 4. This figure displays for two polarizations the distribution of all the calculated RMS values as a function of the local sidereal time (LST) for one of the Radio-Detection Stations. For both directions of the polarization, we recognize the same trend: an increase in the RMS value around 18:00 LST (see also Ref. [11]). This time coincides with the passage over the Radio-Detection Station of an area near the galactic center with an increased brightness temperature in the radio frequency band [14]. These studies enable us to monitor the overall detection threshold and the gain pattern of antennas used.

The signal development of radio pulses induced by UHE CRs in the atmosphere has been modeled using extensive Monte-Carlo simulations, tracking electron- and positron density distributions, and by using macroscopic calculations [9]. In addition, a modular detector simulation package has been written, where various important components of the signal development as it passes through the electronics chain can be simulated and compared with data [15]. A conceptual design has been made for the offline analysis and visualization codes [16]. This package will be merged with the standard offline analysis which is being used for the analysis of data obtained with the other detector systems used in the Observatory.

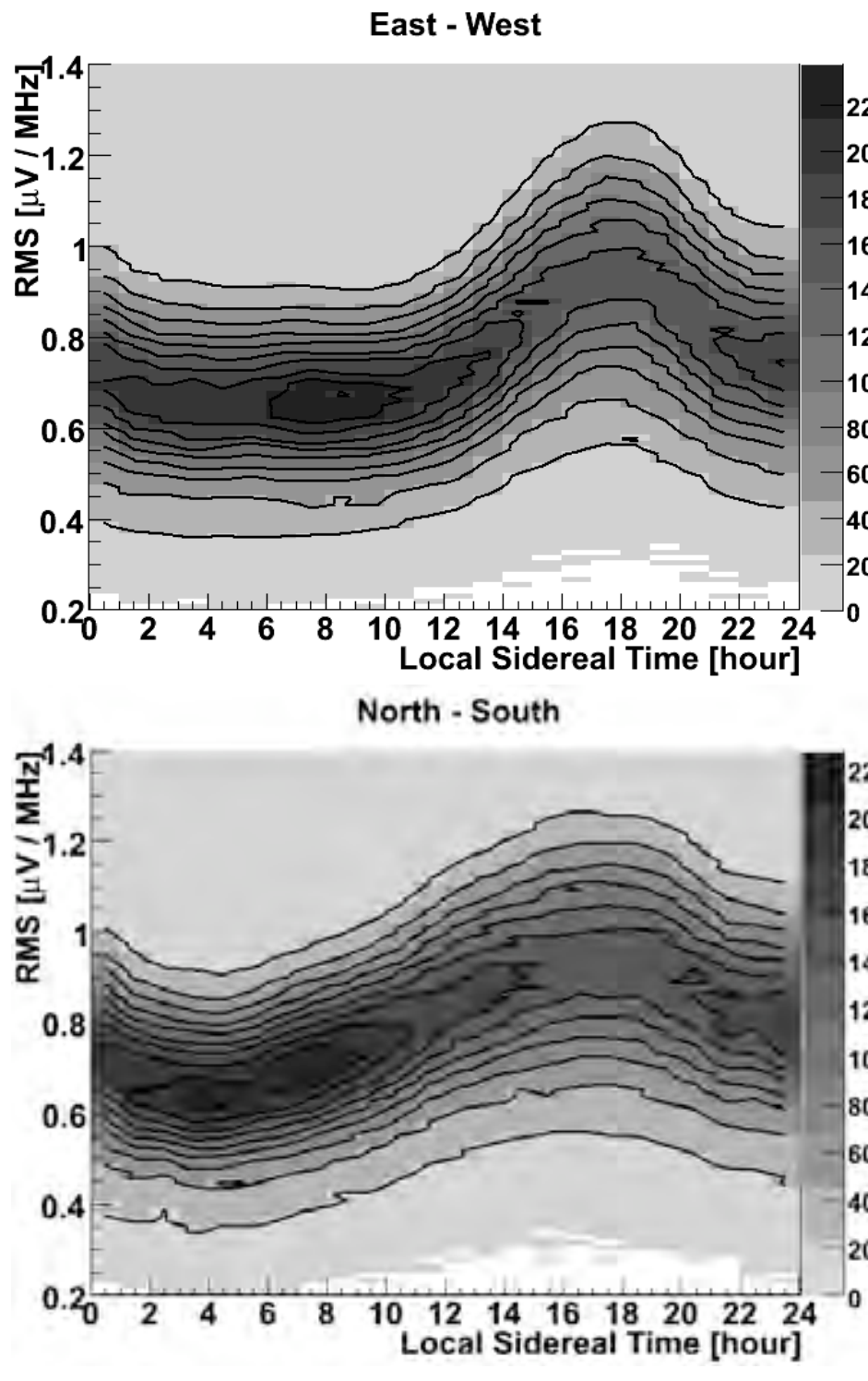

Fig. 4. RMS value of the noise level measured with one of the Radio-Detection Stations as a function of the LST filtered for the frequency band between 50 and 55 MHz. The plotted RMS values have been corrected for cable losses and amplification between the output of the logarithmic periodic dipole antenna and the receiver unit, but not for the antenna gain. The grey scale indicates number of counts per hour and per 0.012 $\mu$V $MHz^{-1}$. The data were collected between May 2007 and April 2008.

## III. AERA: THE 20 $KM^2$ ARRAY

The baseline parameters for AERA are about 150 Radio-Detection Stations distributed over an area of approximately 20 $km^2$. Figure 5 shows an areal view of the site. AERA will have a core of 24 stations deployed on a triangular grid with a baseline of 150 m. This core is about 4 km east of the Coihueco fluorescence telescope; it provides an excellent overlap for events which will be observed with both detection systems. Around this core, there will be 60 stations on a triangular grid with a pitch size of 250 m. Finally, the outer region of AERA will have 72 stations with a mutual distance of 375 m. Each Radio-Detection Station will operate on solar power and has its local data-acquisition. Like the SD of the Observatory, event definition will be based on timing information sent through wireless communication by the Radio-Detection Stations to a central data-acquisition system, located near the center of AERA. The design of the antennas and of the electronics will be optimized to have a high sensitivity in the frequency band between 30 and

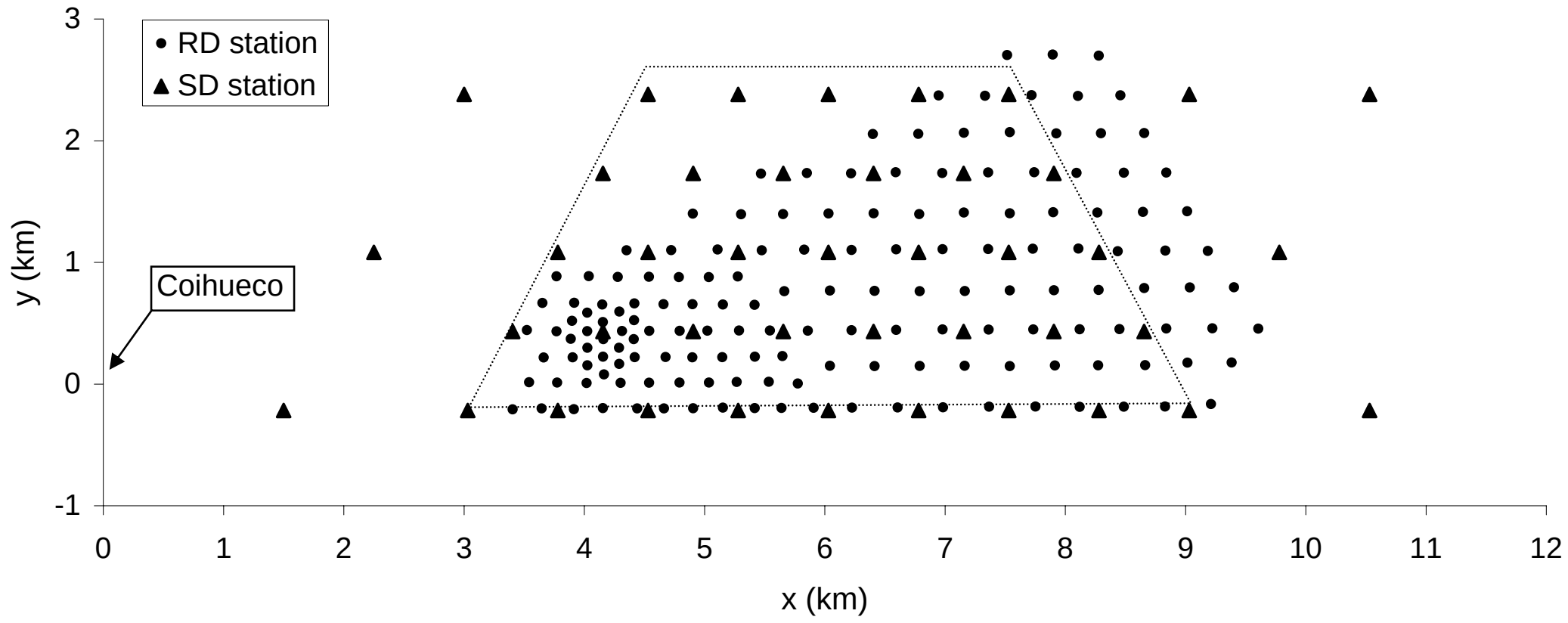

Fig. 5. Areal view of AERA at the western part of the Pierre Auger Observatory. Radio-Detection Stations are denoted as dots, detectors of the SD Array as triangles. All coordinates are relative to the Coihueco fluorescence building; see also Fig. 1.

80 MHz. Every Radio-Detection Station will have a ring buffer which can contain streaming data (4 channels, 200 MS $s^{-1}$, 12 bits per sample) for a period of 3 s.

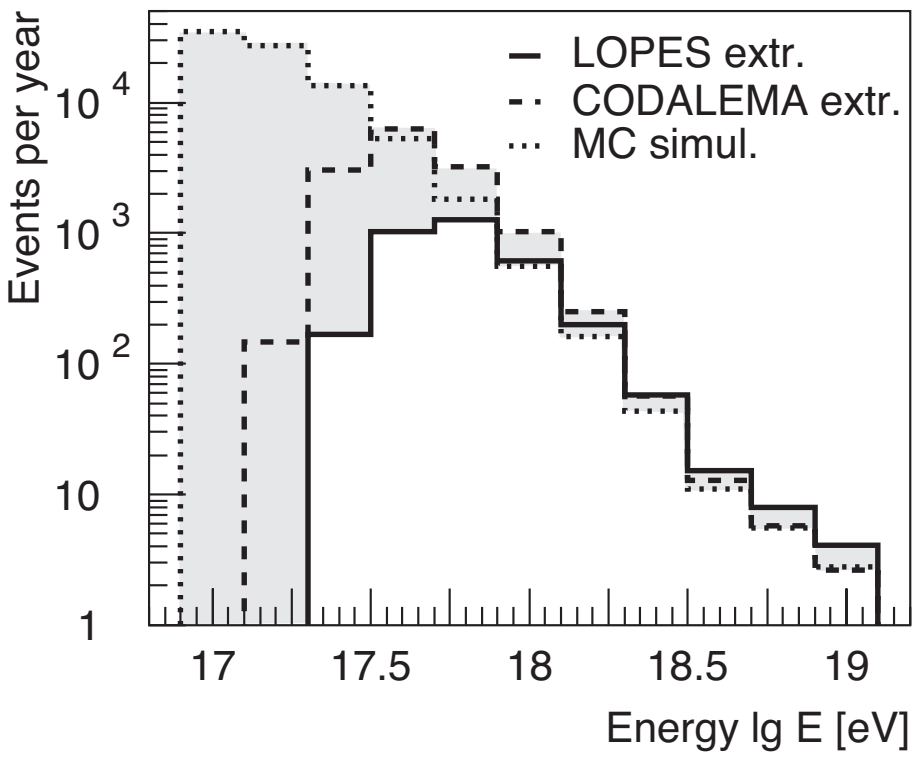

Fig. 6. Expected number of events per year as function of the shower energy according to extrapolations of LOPES and CODALEMA measurements as well as Monte-Carlo simulations, based on the REAS2 code and with zenith angle $\Theta < 60°$; see text for details.

The expected performance of the array has been calculated applying different methods. Measurements of the lateral distribution of the radio signal as obtained by the LOPES [17] and CODALEMA [18] experiments have been taken into account as well as shower simulations with the REAS2 code [19]. To estimate the expected event rates, the effective area is multiplied with the cosmic-ray flux. The present estimates are based on the spectrum measured by the Pierre Auger Collaboration [20]. The flux below $10^{18.45}$ eV has been extrapolated, assuming a spectral index of $-3.3$, which is typical for the energy region between the second knee and the ankle, $10^{17.6}$ - $10^{18.6}$ eV. The expected number of events per year is depicted in Fig. 6 for the three different approaches. The main difference between the approaches is in the threshold region, where the largest uncertainty comes from the extrapolation of the flux spectrum to lower energies. Taking a conservative approach and using the CODALEMA and LOPES extrapolations, the threshold for the radio array is around $\approx 10^{17.2}$ eV.

Summarizing, AERA will be the first large radio-detection array for the observation of UHE CRs. As it will be co-located with the other detector systems of Auger it will provide additional and complementary information on air showers which can be used, e.g., for the determination of the composition of UHE CRs.

# Hardware Developments for the AMIGA Enhancement at the Pierre Auger Observatory

**P. Buchholz*** **for the Pierre Auger Collaboration**[†]**.**

*Fachbereich Physik, Universität Siegen, Walter-Flex-Str. 3, D57068 Siegen, Germany
† Observatorio Pierre Auger, Av. San Martín Norte 304 (5613) Malargüe, Prov. Mendoza, Argentina

***Abstract.*** **To extend the energy threshold of the Auger Observatory to lower energies and to measure the number of muons in extensive air showers, the AMIGA (Auger Muons and Infill for the Ground Array) enhancement is being developed. The complete muon detector system, including the scintillation detectors, the analogue front-end, trigger and the digital readout electronics, as well as the power supplies and slow-control electronics, have been designed. Prototypes of all components have been produced and tested separately in system tests. The performances of all components and the complete system will be discussed.**

## I. Introduction

The Pierre Auger Observatory is the largest air shower detector in the world with a hybrid detection system consisting of a surface detector array (SD) and a fluorescence detector (FD). The SD of the Southern site near Mendoza, Argentina, consists of more than 1600 water-Cherenkov detectors arranged in a triangular grid with 1500 m spacing. It is overlooked by 24 fluorescence telescopes of the FD, arranged in four groups of six telescopes each [1], [2]. The Auger collaboration has been successfully taking data for several years measuring with full efficiency energies above $3 \times 10^{18}$ eV using the SD and above about $10^{18}$ eV including the FD. To reach full efficiency also for lower energies down to about $10^{17}$ eV, from which the second knee and the ankle of the energy spectrum can be fully seen and where the transition from galactic to extra galactic cosmic rays is assumed to occur, the Auger Observatory will be upgraded with two additional detector setups, AMIGA (Auger Muons and Infill for the Ground Array) and HEAT (High Elevation Auger Telescopes). HEAT will be discussed in detail elsewhere [3]. The AMIGA enhancement consists of 85 water-Cherenkov surface detectors, each with a 30 m$^2$ plastic scintillator buried about 2.5 m underground. The final depth will be decided taking the results of the BATATA detector setup into account [4]. Out of the 85 detector stations 61 are arranged in a 750 m triangular grid and 24 in a 433 m grid. They are overlooked by the FD and HEAT. The scintillators will be used to determine the muon content of the air shower, which is one of the relevant shower parameters e.g. for composition analyses. Since the Infill part of AMIGA, i.e. the additional arrays of surface detectors, is described elsewhere [5], this contribution concentrates on the muon detector part, especially on tests performed with a prototype system. For a detailed description of the muon detector system see also [5].

## II. The AMIGA muon detector

The muon detector of an AMIGA detector station consists of three independent muon scintillation counters, each divided into 64 strips and read out by a 64 channel multi anode photomultiplier (PMT). To process the PMT output signals, a modular readout system of four different electronic boards has been developed and prototypes of all modules have been produced and successfully tested. These modules are the Mother Board (MB), the Daughter Board (DB), the Digital Board (DGB) and the Power Distributor Board (PDB). By the end of 2009 a muon counter prototype will be installed in the Auger array using a 4.8 m long scintillator module (Figure 2) and the complete electronics system consisting of the existing prototypes. Following a test run of about two months with this detector, another prototype detector, consisting of a 8.8 m long scintillator module, will be added. After another test run period the first final muon counter of the AMIGA array will be installed. To equip all 85 AMIGA detector stations almost 3000 individual boards are needed, which underlines the need for an automatized test system during mass production. Each of the 85 AMIGA muon counter electronics setups will be tested and fully characterised in Siegen before being installed in the AMIGA array without any further dismounting. In the following, short descriptions of all boards and the corresponding test results are given. Figure 1 shows an overview of the complete prototype system and Figure 3 a picture of the assembled system in the laboratory.

### A. Mother Board

The MB provides a socket for the multi-channel PMT, its power supply and monitoring of the high voltage applied. Signals are distributed to 8 DBs, plugged into the MB. In all tests concerning the steering and monitoring of the high voltage of the PMT, the MB prototype performed as specified.

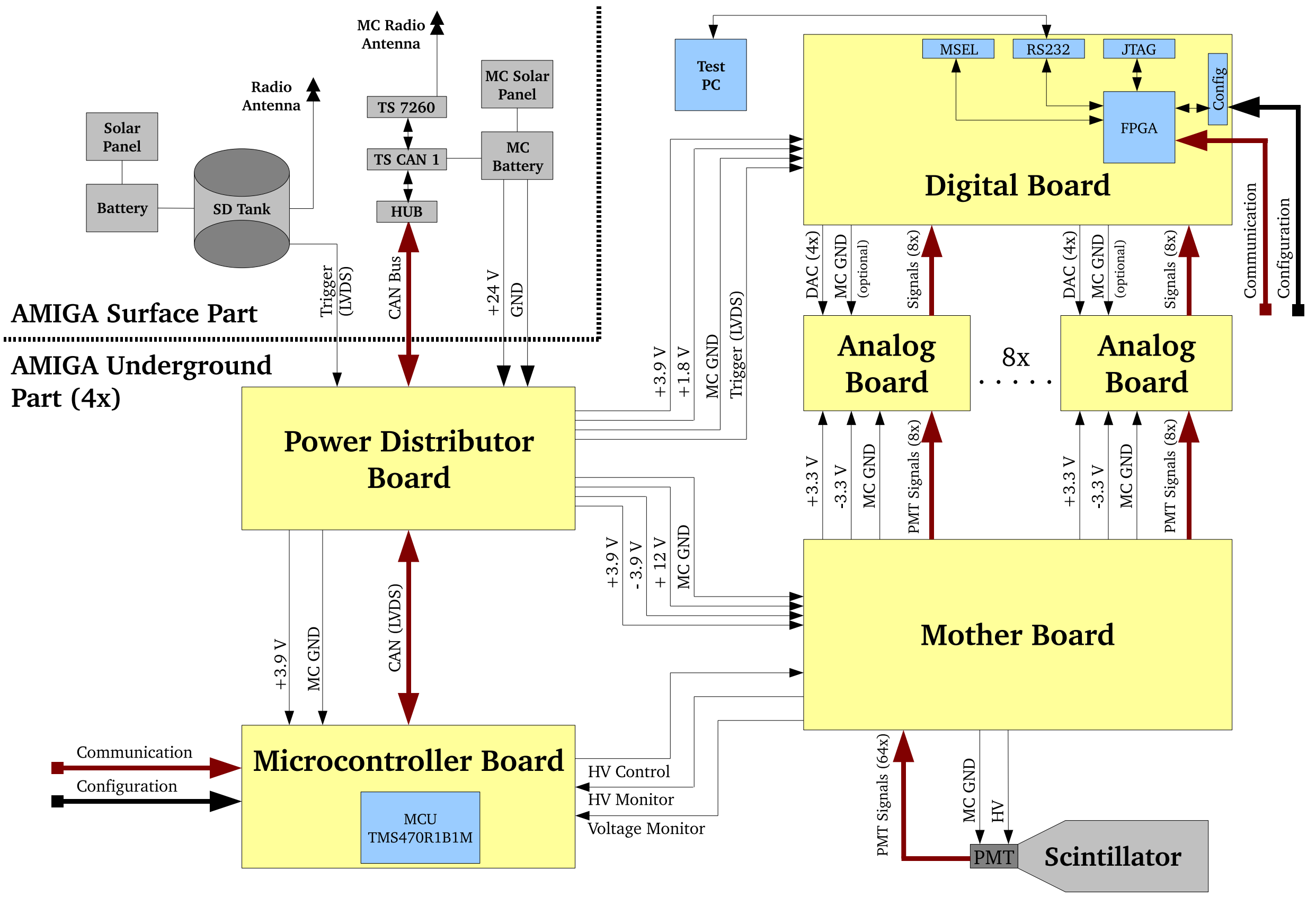

Fig. 1: Schematic overview of the AMIGA muon counter prototype system.

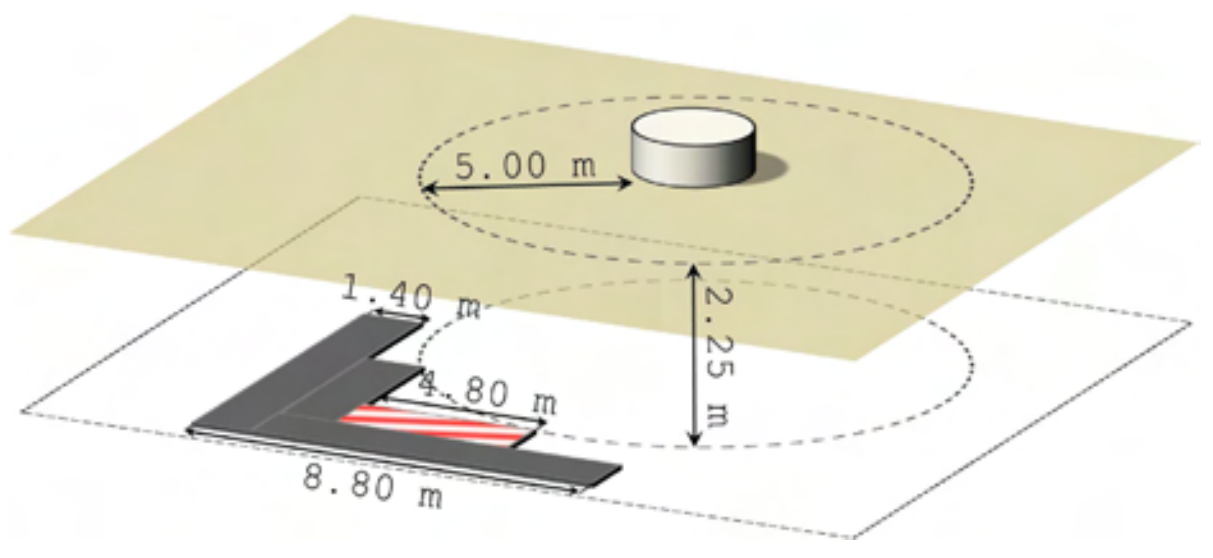

Fig. 2: Schematic view of one possible arrangement of the AMIGA detector pair consisting of one SD station and four buried muon scintillation counters. The hatched module with a length of 4.8 m will be used for the first prototype.

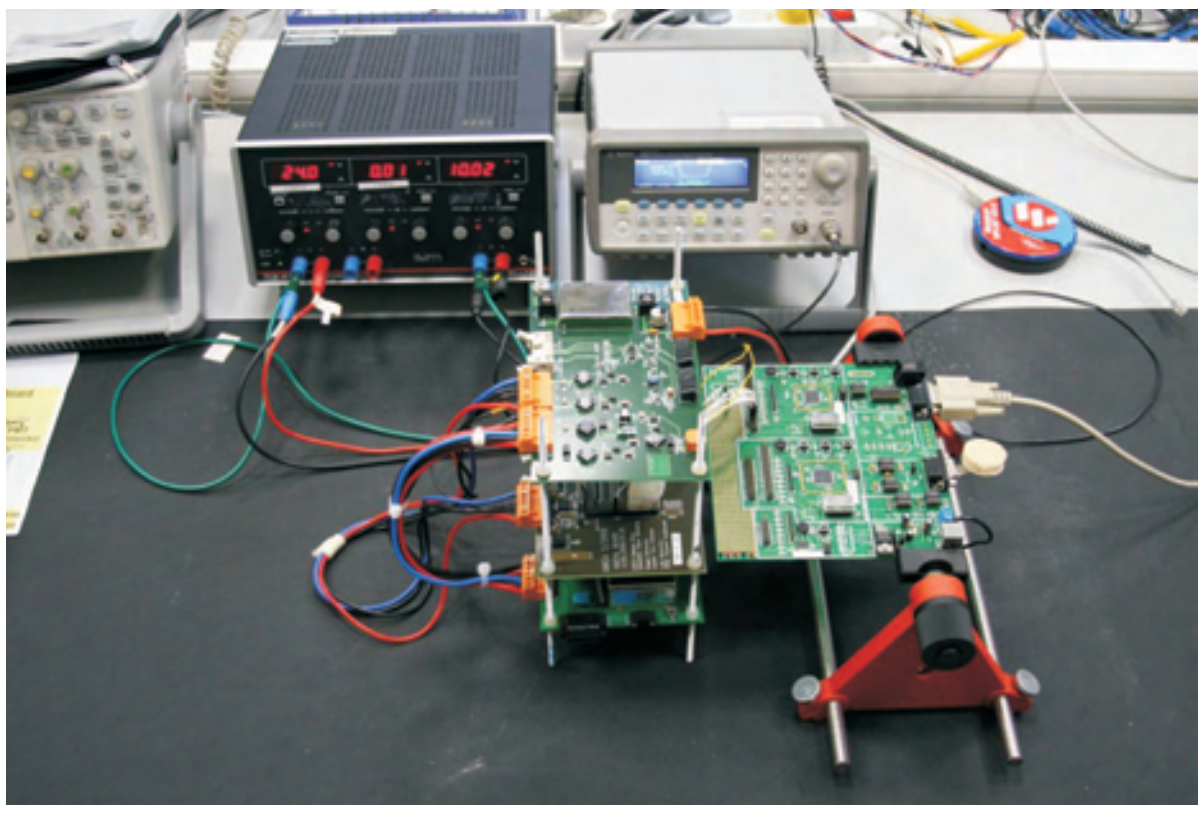

Fig. 3: Picture of the prototype system as assembled in the laboratory.

## B. Daughter Board

The DB is the signal forming module of the muon counter electronics chain. Each DB consists of eight channels, in which the analogue photomultiplier signals distributed by the MB are inverted and amplified by operational amplifiers by a factor of about 3.1. In a second step, the outputs of the operational amplifiers are compared to a given threshold. In case the amplitude of the signal in one channel exceeds the threshold voltage, the digital output signal of this channel is set to zero.

To compensate variations of the PMT gains as well as variations of the gains of the operational amplifiers, the threshold voltage can be set individually for each channel using an eight channel DAC (digital-to-analog converter) on each DB. Figure 4 shows measured single electron input pulses and the corresponding output of the DB. The bandwidth of the DB has been measured as $(134 \pm 5)$ MHz, which is more than sufficient. The typical behaviour of a single channel is shown in Figure 5.

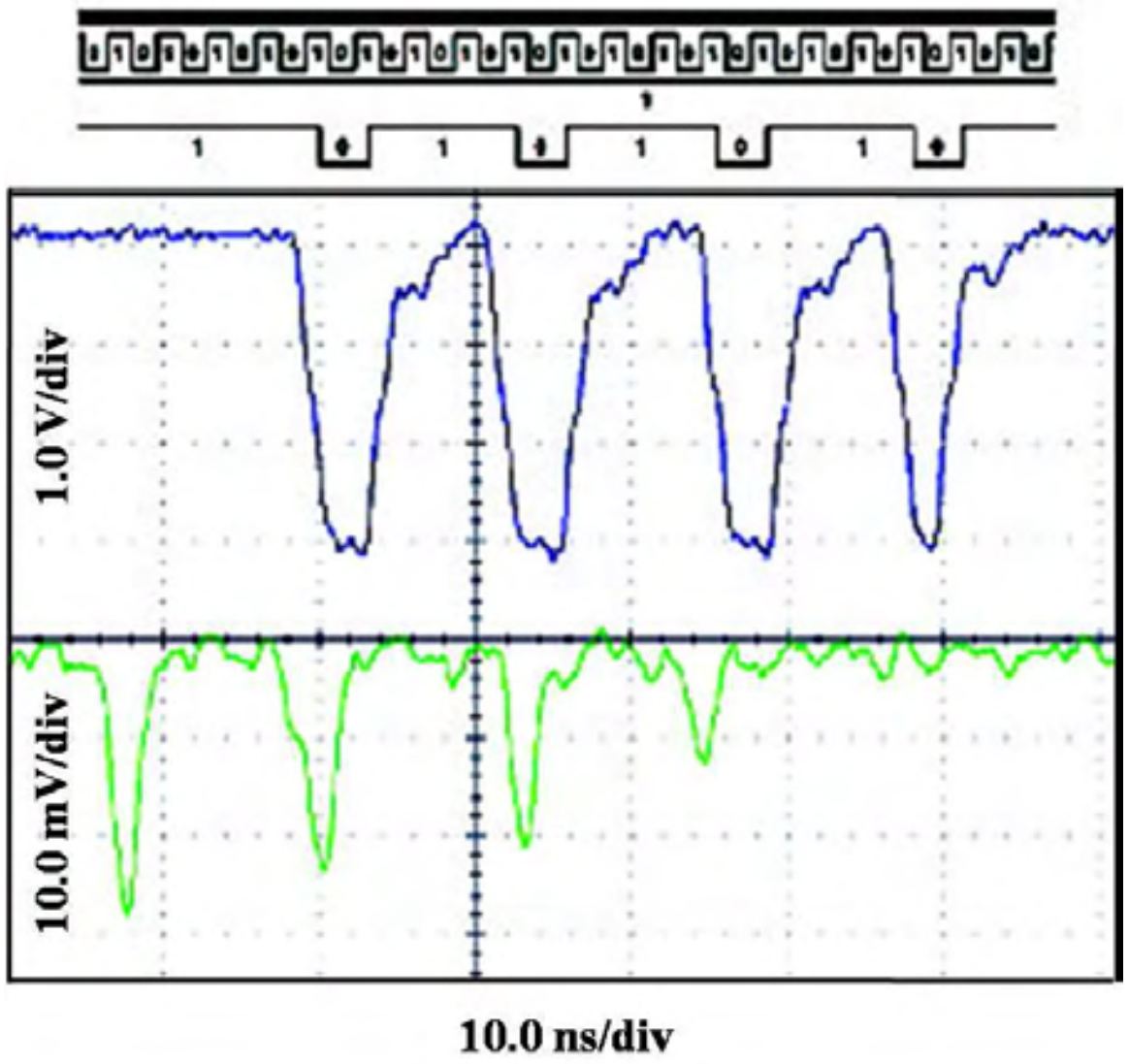

Fig. 4: Example of measured input and output pulses of the DB. The lower graph schows the amplitude of the input pulses, the middle graph the comparator output and the two upper graphs the clock and the resulting digital data in the FPGA.

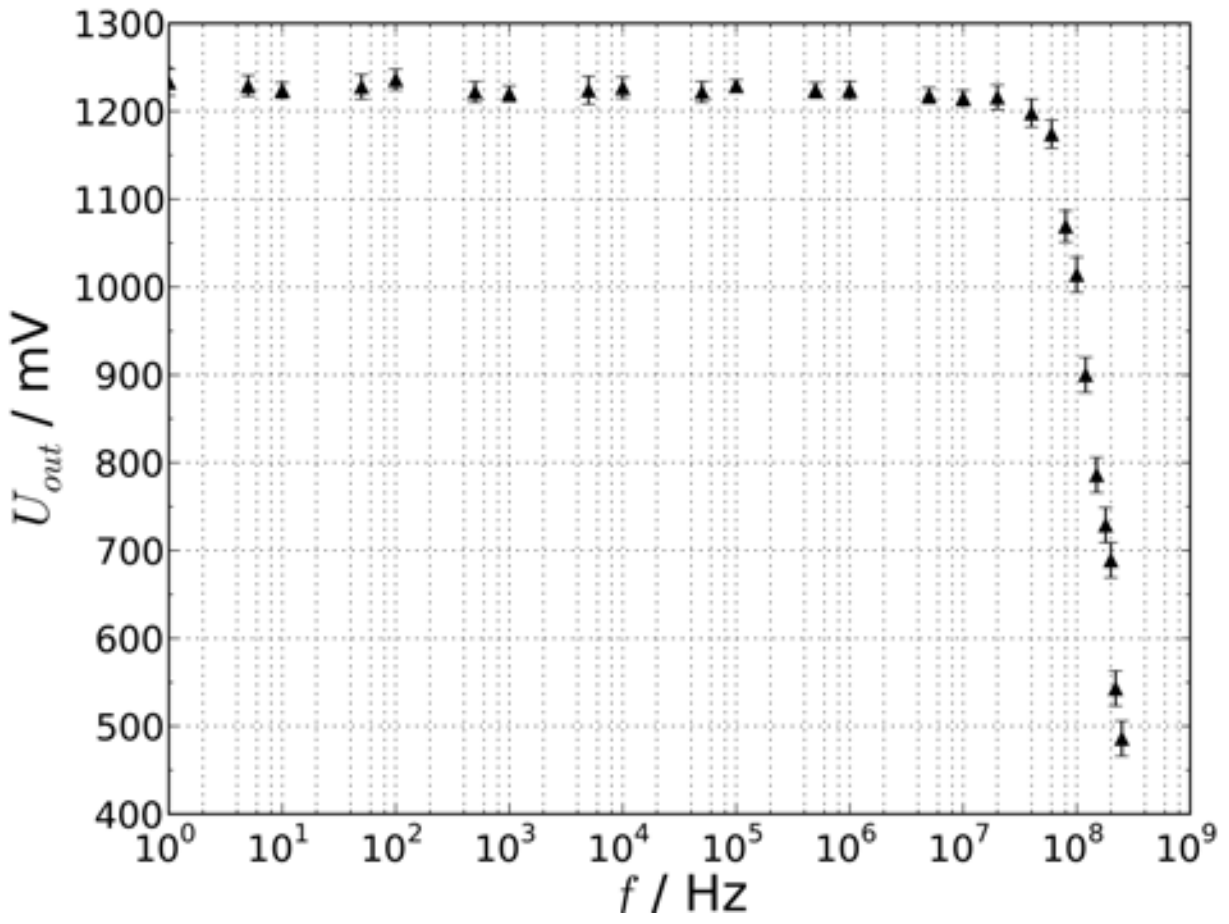

Fig. 5: Bandwidth measurement of a typical DB channel.

### C. Digital Board

The DGB continuously stores the digitised data received from the DBs into a ring buffer. Receiving a trigger signal (T1) from its corresponding SD station digitised muon data of 1.6 $\mu$s before and 3.2 $\mu$s after the trigger signal are stored in a memory. Only the muon data confirmed by a higher trigger level signal (T3) is sent via radio to the central data acquisition.

This readout is realized by using an FPGA, an external RAM, a microcontroller unit (MCU) and a single board computer (SBC) and is described in detail in [6]. The tests discussed here were performed using a prototype system consisting of two separate boards, one for the FPGA and one for the MCU. These will be merged into one DGB in the next iteration.

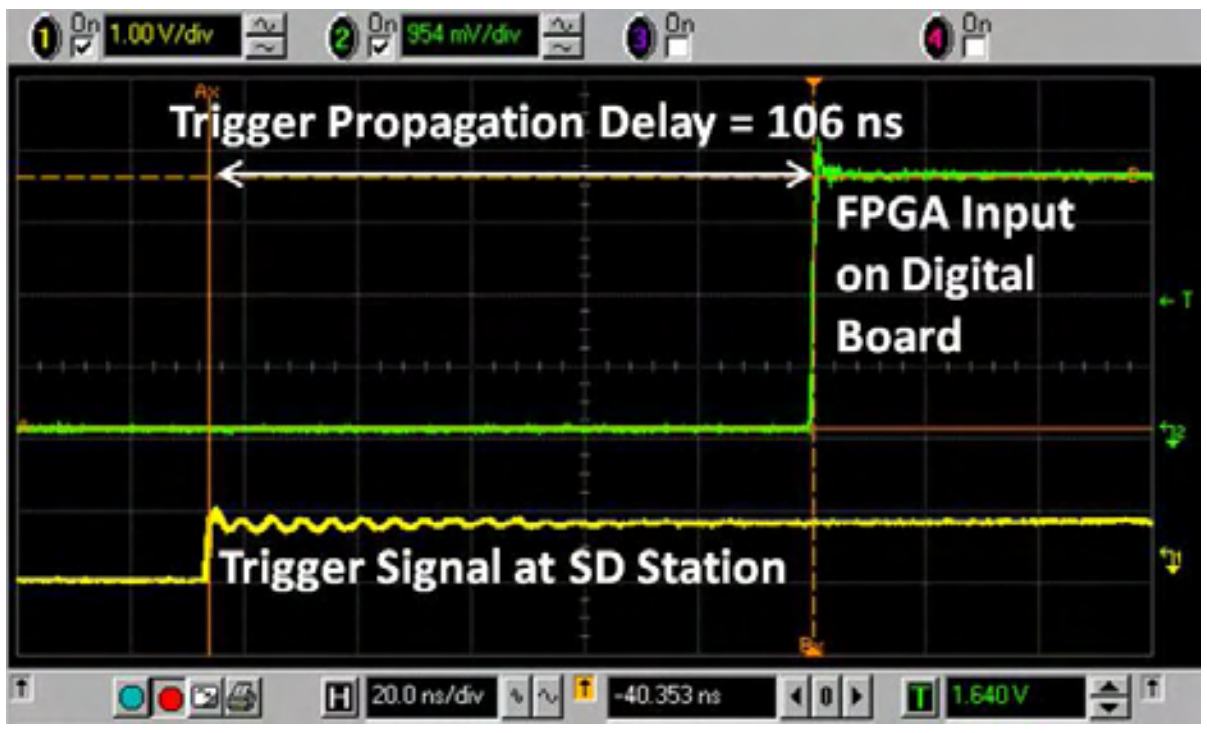

Fig. 6: Measured trigger propagation delay.

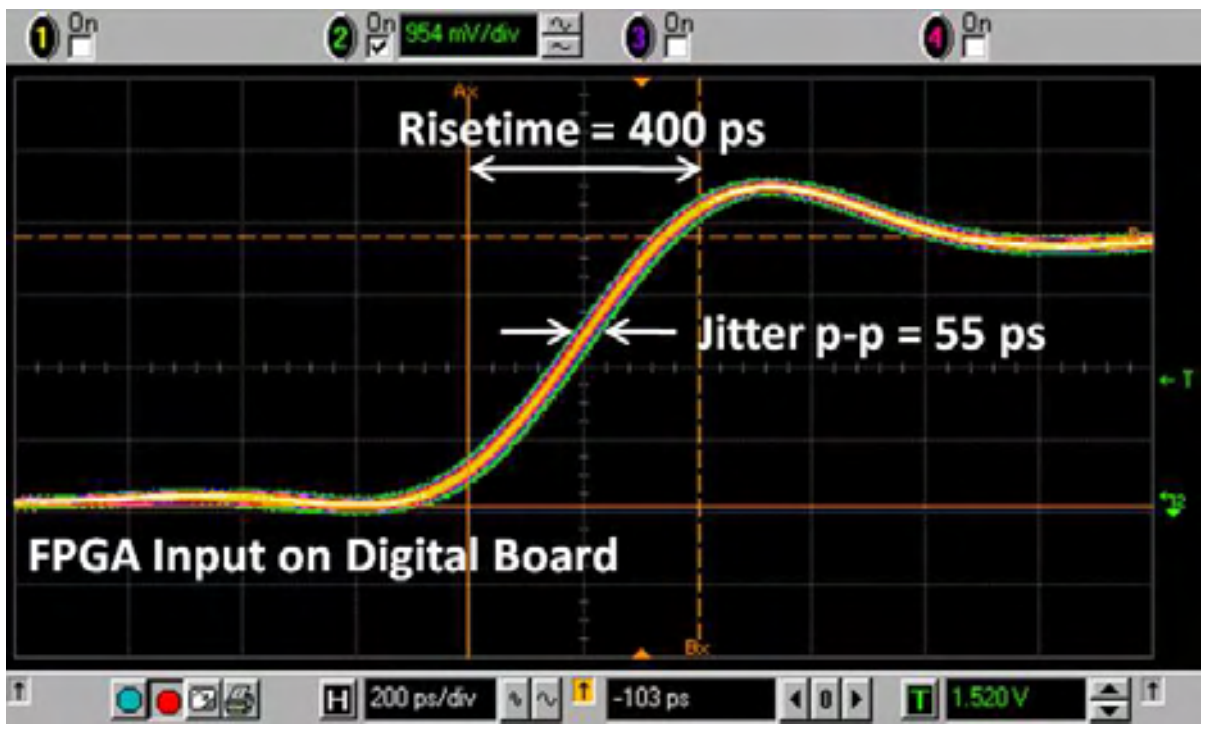

Fig. 7: Measured signal jitter.

To synchronise data from the muon detectors with events recorded by the SD, a dedicated line for the T1 trigger is used. The signal delay of this line has been measured using the prototype system. It stems from the cable length of about 15 m, the galvanic isolation in the underground electronics and the latencies of the transmitter and receiver module. The latter also adds jitter to the signal.

As can be seen in Figures 6 and 7 a delay of 106 ns and a jitter of about 55 ps have been measured. Since the digital input pulses from the DBs are sampled with a rate of 320 MHz by the FPGA, which corresponds to sampling intervals of 3.125 ns, this jitter is negligible.

### D. Power Distributor Board

The PDB derives all supply voltages needed by the electronics modules from the battery voltage of + 24 V. In addition, the ground levels of both detectors, the SD station and the muon detector, are electrically isolated from each other to reduce noise generated by ground loops. For the same reason, also the ground levels of the communication lines (trigger and CAN) are isolated by the PDB. The PDB performed in all tests within its specifications.

## III. System Tests

The full readout chain has been set up using the prototype modules described above, i.e. the performance of the complete system can be studied. As an example Figure 8 shows the efficiency of one channel of the read-out system as a function of the input signal amplitude for a given comparator threshold.

Instead of connecting a scintillator and a PMT to the electronics, a pulse generator is used. The output of the pulse generator can be fanned out with a 64-channel multiplexer to each of the 64 input channels, using a specially developed interconnection card. Square-pulses at 200 Hz, with a width of 100 ns and variable amplitudes, are used. For each amplitude, 1000 pulses are sent to the electronics.

For the FPGA, special system test software was developed. The comparator thresholds of the 64 read-out channels are programmed. The FPGA is reading the comparator output information with 160 MHz. If a digital 0 is detected on at least one channel, a trigger flag is set and the information of all comparators at trigger time and the following 4 sample times is given to the microcontroller together with a timestamp. The system test software driving the microcontroller then submits these data to a PC via Serial Port.

Thus, depending on the amplitudes of the input pulses, the electronic noise of each channel can be extracted from the efficiency curves recorded as shown in Figure 8 for a single channel.

The threshold voltage of the comparator has been set to 14.7 mV including an offset value of 9.7 mV. This offset has been measured when programming the comparator to nominal 0 V. The output voltages of the comparator can be set with a precision of 1 mV.

The data points were fitted with the function $f(u)$ given in Equation 1.

$$f(u) = 0.5 + 0.5 \cdot \frac{2}{\sqrt{\pi}} \int_0^u e^{-t^2}\, dt \qquad (1)$$

In Equation 1, $p_0$ in $u = \frac{x-p_0}{p_1}$ describes the position and $p_1$ the width of the s-curve. A $\chi^2$-fit has been performed and the threshold level has been reconstructed at $(15.1 \pm 0.1)$ mV. The electronic noise is below a level of $(0.33 \pm 0.04)$ mV, determined from the width of the fitted curve as $U_{\text{noise}} = 1.812 \cdot \frac{p_1}{2}$.

The system tests so far employ especially developed code. As soon as the original/final codes for the FPGA and the microcontroller are available, these will be used also in the system tests.

The data transferred has been checked for its correctness and no errors have been found at a CAN bus transmission rate of 1 Mbps. In future the test system will be used for several additional tests, e.g. crosstalk measurements. In parallel, the test procedures for the mass production of the final system are being developed and checked.

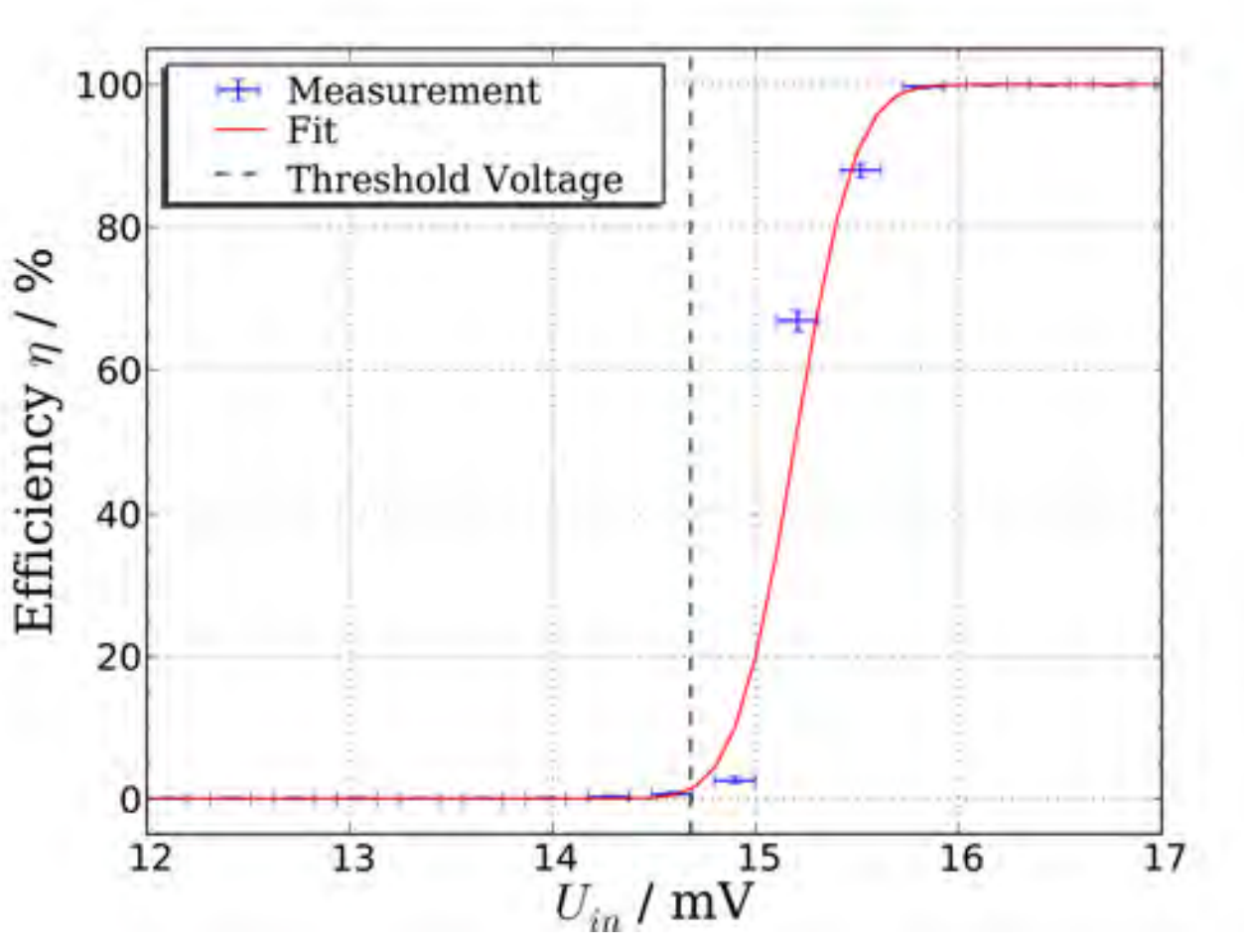

Fig. 8: Readout efficiency for a single channel using system test software. The adjusted threshold and the electronic noise level were reconstructed.

## IV. Summary

A muon detector system has been developed for the AMIGA enhancement of the Pierre Auger Observatory. This development includes a complete prototype system of the highly modular readout electronics, which has been successfully produced and tested. The performances of all components are well within their specifications. The bandwidth of the analogue front-end has been measured to be sufficiently large and the trigger propagation delay has been determined and shows negligible jitter. In addition, the readout efficiency of the complete system has been measured as a function of the input signal amplitude, which shows, that the system is fully efficient above threshold.

# A simulation of the fluorescence telescopes of the Pierre Auger Observatory using Geant4

**Pedro Assis***, **for the Pierre Auger Collaboration** †

**Laboratório de Instrumentação e Física Experimental de Partículas (LIP), Lisboa, Portugal*
†*Observatorio Pierre Auger, Av.San Martin Norte 304, (5613) Malargüe, Mendoza, Argentina*

***Abstract*. A simulation of the fluorescence telescopes of the Pierre Auger Observatory was developed, profiting from the capabilities of Geant4 to describe complex 3D geometries realistically and to allow the required optical processes to be included in the simulation. Account was taken of the description of all optical components of the telescopes. This simulation is included in the Pierre Auger offline software framework and is being used in several studies of the fluorescence detector performance. The main features of the simulation are reviewed.**

## I. Introduction

A detailed simulation of the fluorescence telescopes of the Pierre Auger Observatory was developed, taking advantage of the capabilities of Geant4 [1], [2] to describe complex 3D geometries realistically and to allow the required optical processes to be included in the simulation. The Fluorescence Detector (FD) of the Pierre Auger Observatory is composed of 24 fluorescence telescopes, located in four buildings overlooking the Surface Detector array. Each telescope features Schmidt optics consisting of a ring shaped corrector lens placed at the entrance pupil and a 11 m$^2$ spherical mirror. The incoming light is focused on to a spherical camera, containing 440 hexagonal pixels made of photo-multipliers (PMT) and light guides [3].

Geant4 is a software toolkit, developed in the C++ programming language, to simulate the interaction of particles through complex 3D geometries. In Geant4 the user defines the geometry and composition of the media, as well as a primary particle and its properties. The kernel then takes care of the tracking and interactions of the primary and secondary particles throughout the defined "world", taking into account the properties of the traversed materials and the selected physical processes.

The tracking of optical photons includes refraction and reflection at medium boundaries, Rayleigh scattering and bulk absorption. The optical properties of a medium, such as refractive index, absorption length and reflectivity coefficients, can be expressed as tabulated functions of the wavelength. In addition, specific characteristics of the optical interfaces between different media can be defined using the UNIFIED optical model [4], which provides a realistic description of surface finish and reflector coatings. Complex geometries can be defined by exploiting the Constructive Solid Geometry (CSG) functionalities in Geant4, that allows to build complex solids from simple ones using boolean operations.

This simulation is included in the Pierre Auger simulation and reconstruction framework [5]. It is being used in several performance studies of the fluorescence telescopes, in particular those related to the optical spot and the light detection efficiency.

## II. Implementation of the telescope components with Geant4

Account was taken of the description of the corrector lens profile, the details of the mirror geometry, including the parameters of each individual mirror, and the different components of the camera, including light guides and photo-multipliers (PMT). The optical properties of all materials, such as the absorption length and the refractive index, are described in the simulation.

The filter at the entrance pupil of the fluorescence telescopes is implemented as a disk made of M-UG6 [6], with a diameter of 2.2 m and a thickness of 3.25 mm. The filter is positioned perpendicular to the optical axis of the telescope at $(0, 0, -10\,\mathrm{cm})$ [1]. The refractive index is set to $n = 1.526$ and the absorption length is computed from the transmittance for the filter material with a nominal value of $0.83$ at a wavelength of 370 nm for a thickness of 3.25 mm.

The simulation of the corrector lens, made of BK7 glass, uses the Geant4 class `G4Polycone`. Given the lens axial symmetry this class enables the description of an arbitrary profile, by its decomposition into a sufficiently large number of straight segments. The segmentation of the corrector lens in $24$ petals, each with an angular width of $14.735°$ is also implemented in the simulation. A visualization of the lens profile is shown in figure 1.

The telescope mirror, composed by $64$ hexagonal elements arranged in $8$ rows, is implemented in full detail in the simulation. The mirror elements are quasi-hexagonal spherical mirrors, and their shape depends on the position in the telescope mirror. To define the mirror elements, $3$ to $5$ (depending on the type of element) trapezoids are joined and the resulting solid is intersected with a spherical shell, with inner radius equal to the radius of curvature of the mirror element. A visualisation of one mirror element is shown in figure 1. The positioning of the hexagonal elements emulates the alignment procedure performed in the real telescopes, with the centre of curvature of each element coinciding with the alignment point, which is located at the origin of the telescope reference frame. The optical properties of the mirror are implemented by defining the reflectivity in the interface surface between the air and the mirror. The curvature radius, positioning angles and reflectivity of each individual segment are read from a database.

[1] The telescope coordinate system is defined with the $z$ axis aligned with the optical axis, pointing outwards, the $y$ axis horizontal and the $x$ axis orthogonal to the $z$ and $y$ axis, pointing downwards. The origin of this coordinate system coincides with the geometrical centre of the corrector lens.

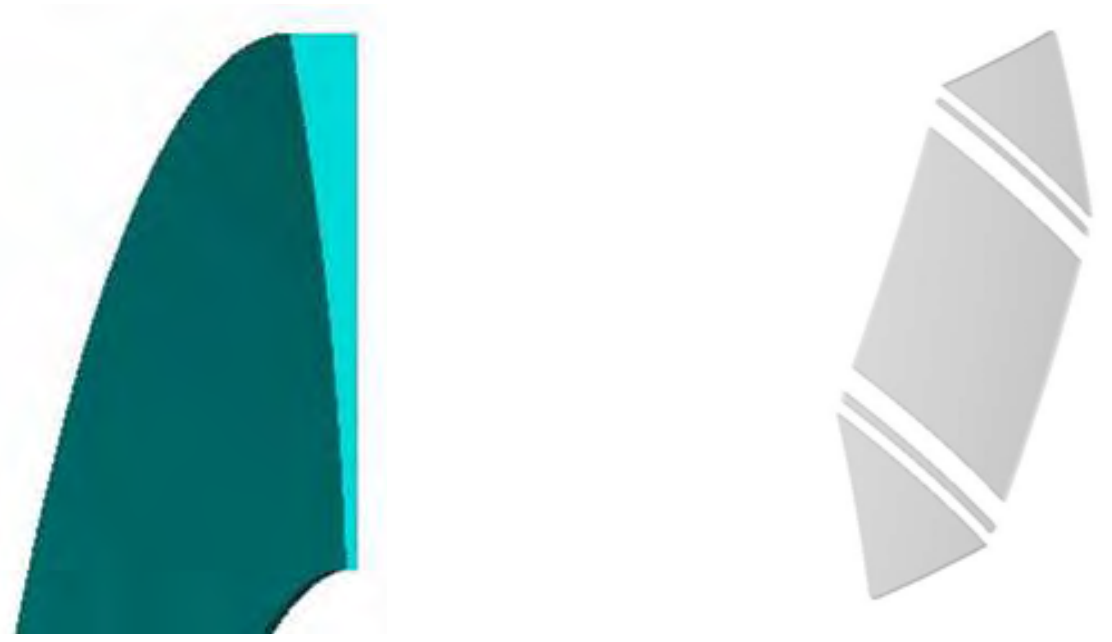

(a) (b)

Fig. 1. (a) Detail of the corrector lens in the Geant4 simulation. The curvature of the corrector ring is magnified to be visible. (b) A segment of the telescope mirror. The trapezoids joined to make the mirror element are artificially displaced for better visualization.

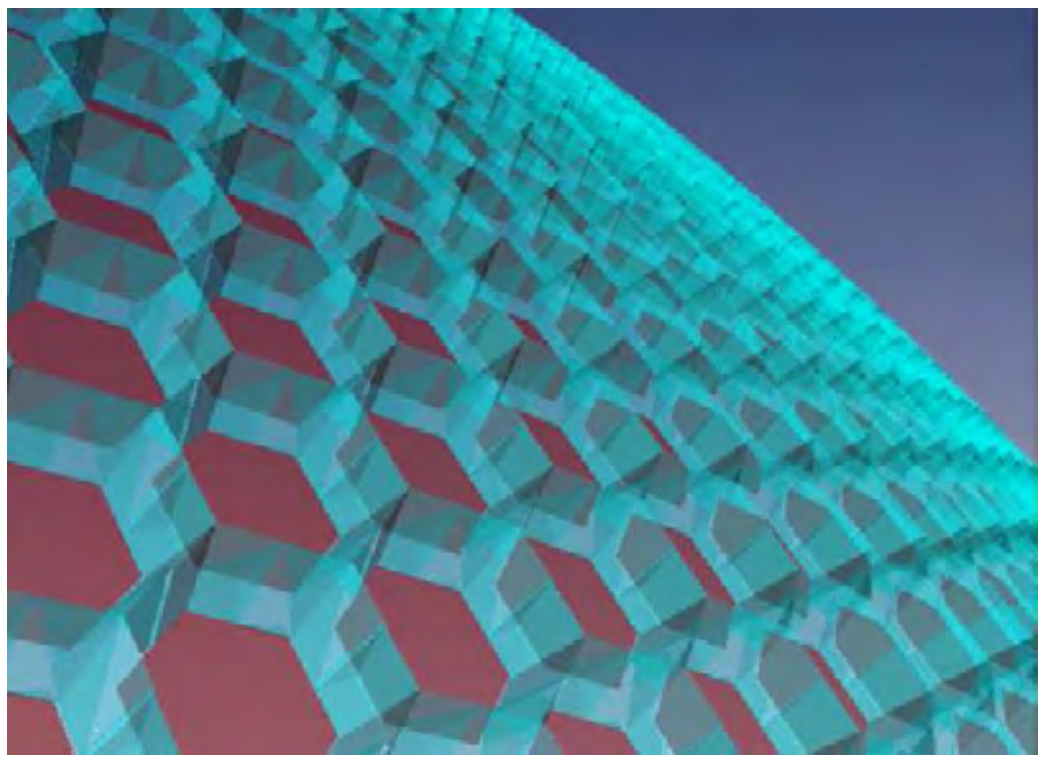

Fig. 2. View of the camera in the Geant4 simulation. The picture show the light guides in light blue and the PMT windows in red.

The cameras of the fluorescence telescopes are composed by the light guides, the PMTs and the supporting structure [7]. The light guides are placed at the vertexes of each hexagonal PMT. In the Geant4 simulation they are made by the union of three triangular prisms, each built using the Geant4 class `G4Polyhedra`. The reflectivity at the surface interface is set to $0.9$ for all wavelengths. In the simulation the PMTs include the hexagonal window and the metallic photocathode. A sensitive detector is associated to the photocathode to simulate the detection of photons. The quantum efficiency is taken into account a posteriori by applying it to the recorded signal at each pixel. Both the PMTs and the light guides are placed following the curvature of the camera. The body of the camera support and feet are also included, to correctly simulate the shadow effect. A view of the camera as implemented in the simulation is shown in figure 2.

The fully assembled telescope is visualised in figure 3 which shows photons arriving at the telescope parallel to the optical axis tracked through the various optical elements of the telescope and focused onto the camera.

## III. Simulation of the telescope performance

The simulation is being used in studies of the performance of the fluorescence telescopes, concerning the optical spot and the light detection efficiency. Some of the results of these studies are summarized below.

### A. The Optical Spot

The telescope optics was studied by characterising the image produced on the focal surface. The photon position in the focal surface is defined by the elevation angle, $\alpha = \arcsin{(x/R_{\rm FS})}$, and the azimuth angle, $\beta = \arcsin{(-y/R_{\rm FS})}$, where $R_{\rm FS}$ is the radius of curvature of the focal surface and $x$, $y$ the Cartesian coordinates of the photon at the camera. Beams of parallel photons were simulated at input angles, with respect to the telescope axis, of $\theta = 0°$, $\theta = 5°$, $\theta = 10°$ and $15°$, with $\varphi = 135°$ The spots observed at the ideal focal surface are shown in figure 4. For photons with an input angle of $\theta = 0°$ the spot presents an almost complete circular symmetry, which arises from the symmetry of the telescope with respect to the optical axis, only broken by the square camera body. For larger angles the central region of the spot is deformed along the same direction as the input direction of the photons. This is due to the photons passing through the corrector lens. Since the mirror is spherical, the contribution due to the photons crossing the hollow part of the lens still results in a symmetric spot, except for the region obscured by the camera. These spots obtained with the Geant4 simulation confirm the optics performance expected from previous ray tracing studies [8].

### B. Photon Distribution at the PMTs

Photons crossing the focal surface reach the PMTs, directly or after being reflected by the lightguides. The position of the photons on the PMT window are shown in figure 5. In the top image the spot is contained inside the pixel, with few hits on the light guide and the centre of the pixel densely illuminated. However, when the incoming photons are focused at the vertex of a light guide the three adjacent PMTs are hit, as shown in the bottom image of figure 5. In this case a large fraction of photons is reflected at the light guides, resulting in a blind circular region centred on the light guide vertex. From these example cases it is clear that the illuminated zones of the PMTs vary with the position of the spot centre. The collection efficiency of a PMT varies along the photocathode, which can yield an additional contribution to the overall telescope efficiency with the spot centre position. The non-uniformity of the PMT response can be taken into account in this simulation framework.

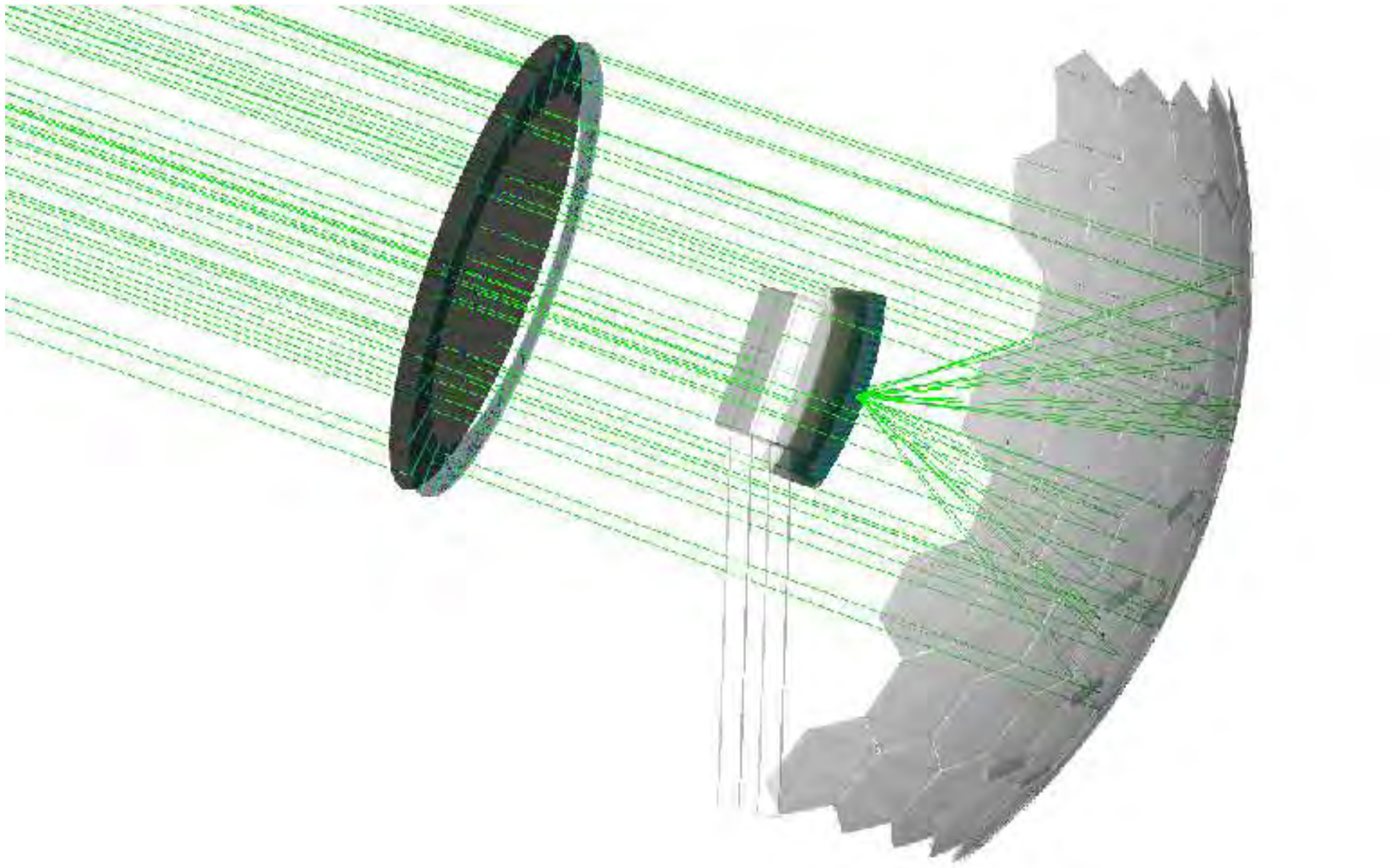

Fig. 3. View of the assembled telescope and the tracking of photons.

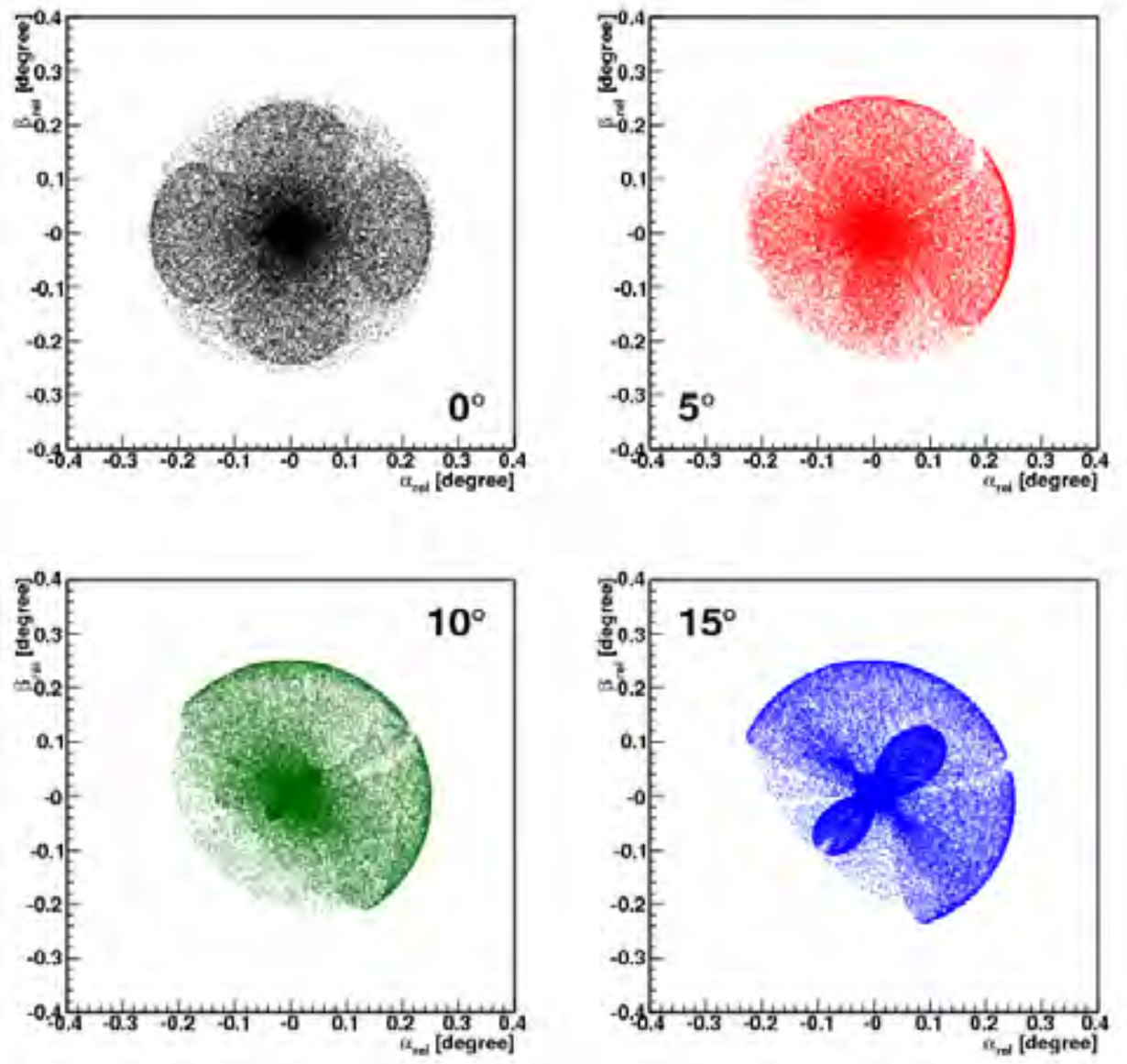

Fig. 4. Spots produced at the ideal focal surface for incident angles of 0°, 5°, 10° and 15°. For each generated photon the relative position $(\alpha_{\rm rel}, \beta_{rel})$ with respect to the expected position for an ideal optical system is shown.

### C. Comparison with laboratory data

The simulation was compared with laboratory measurements of the light detection uniformity [7] (and references therein), where a small version of the camera with seven pixels was used. Since the telescope optics is fully implemented in the Geant4 simulation and the experimental setup reproduces the spot generated by the telescope optics, the measurements can be simulated by illuminating the diaphragm with parallel light rays with varying incident directions, enabling a scan of the seven central pixels. As in the laboratory measurements two scans were performed: one through the arms of the light guides and one passing over the vertexes. The efficiency, defined as the ratio between the number of photons that arrive at the PMTs and the number of generated photons, normalised to the efficiency value in the centre of the central pixel, is shown in figure 6 for the first scan. The measurements are shown for comparison. Good agreement was found in both cases. The efficiency variations are of the order of 15% over the camera surface. The result of 0.85 for the lowest value of the efficiency obtained in the measurements is reproduced by the simulation. The steep fall in the extremes of the scan in the laboratory data is due to the fact that in the laboratory setup only seven pixels were present.

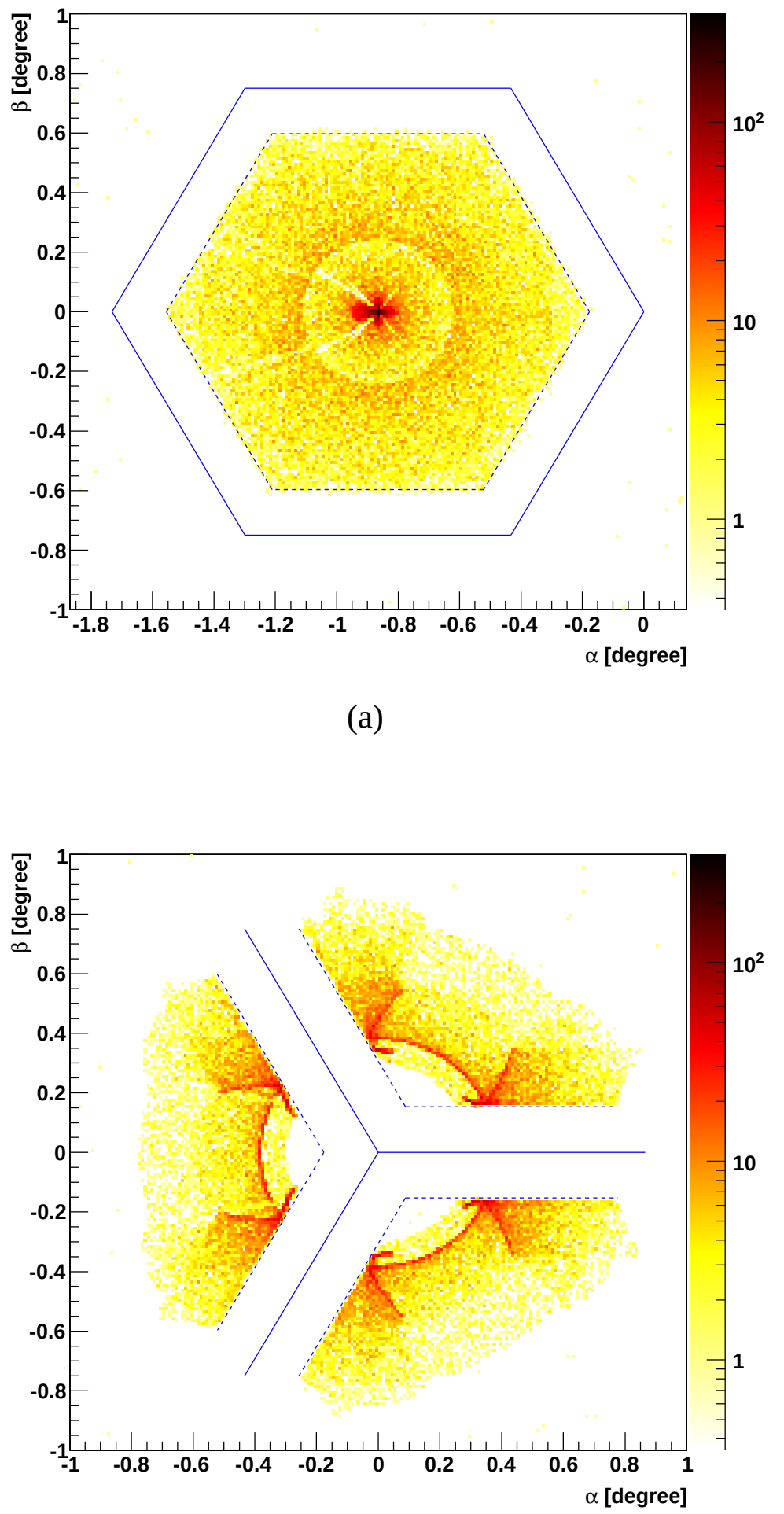

Fig. 5. The spot seen at the PMTs. Lines are drawn to represent the pixel boundaries (solid) and the light guides boundaries (dashed). Photons simulated with the expected position at the pixel centre (a) and at the vertex of the light guide (b).

## IV. Conclusions

The full simulation of the 3D geometry of the telescopes of the fluorescence detector of the Pierre Auger Observatory was implemented using the Geant4 Monte Carlo toolkit. This simulation is being used in several studies of the performance of the telescopes, in particular those related to the optical spot and the light detection efficiency. The capability of describing realistically and in detail all optical elements of the telescopes contributes to deepen the understanding of the FD performance.

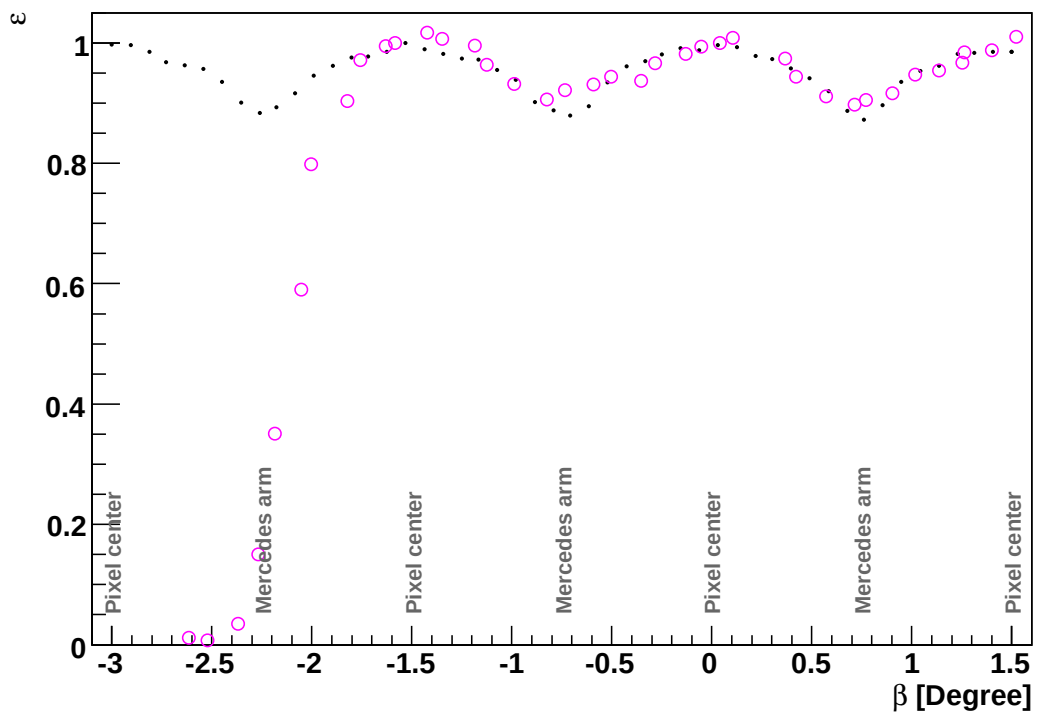

Fig. 6. Relative efficiency along a horizontal line passing through the centre of the pixels. The black dots are the simulation results and the white circles the laboratory measurements.

# Education and Public Outreach for the Pierre Auger Observatory

**G. R. Snow***, **for the Pierre Auger Collaboration**†

*University of Nebraska, Lincoln, Nebraska USA
†Observatorio Pierre Auger, Av. San Martín Norte 304, (5613) Malargüe, Mendoza, Argentina

***Abstract*. The scale and scope of the physics studied at the Auger Observatory offer significant opportunities for original outreach work. Education, outreach and public relations of the Auger collaboration are coordinated in a separate task whose goals are to encourage and support a wide range of education and outreach efforts that link schools and the public with the Auger scientists and the science of cosmic rays, particle physics, and associated technologies. The presentation will focus on the impact of the collaboration in Mendoza Province, Argentina, as: the Auger Visitor Center in Malargüe that has hosted over 40,000 visitors since 2001, a collaboration-sponsored science fair held on the Observatory campus in November 2007, the Observatory Inauguration in November 2008, public lectures, school visits, and courses for science teachers. A Google-Earth model of the Observatory and animations of extensive air showers have been created for wide public release. As the collaboration prepares its northern hemisphere site proposal, plans for an enhanced outreach program are being developed in parallel and will be described.**

## I. Introduction

Education and public outreach (EPO) have been an integral part of the Auger Observatory since its inception. The collaboration's EPO activities are organized in a separate Education and Outreach Task that was established in 1997. With the Observatory headquarters located in the remote city of Malargüe, population 20,000, early outreach activities, which included public talks, visits to schools, and courses for science teachers and students, were aimed at familiarizing the local population with the science of the Observatory and the presence of the large collaboration of international scientists in the isolated communities and countryside of Mendoza Province. The collaboration has been successful becoming part of the local culture. As an example of the Observatorys integration into local traditions, the collaboration has participated in the annual Malargüe Day parade since 2001 with collaborators marching behind a large Auger banner (see Fig. 1). The Observatory's EPO efforts have been documented in previous ICRC contributions [1]. We report here highlights of recent education, outreach, and public relations efforts.

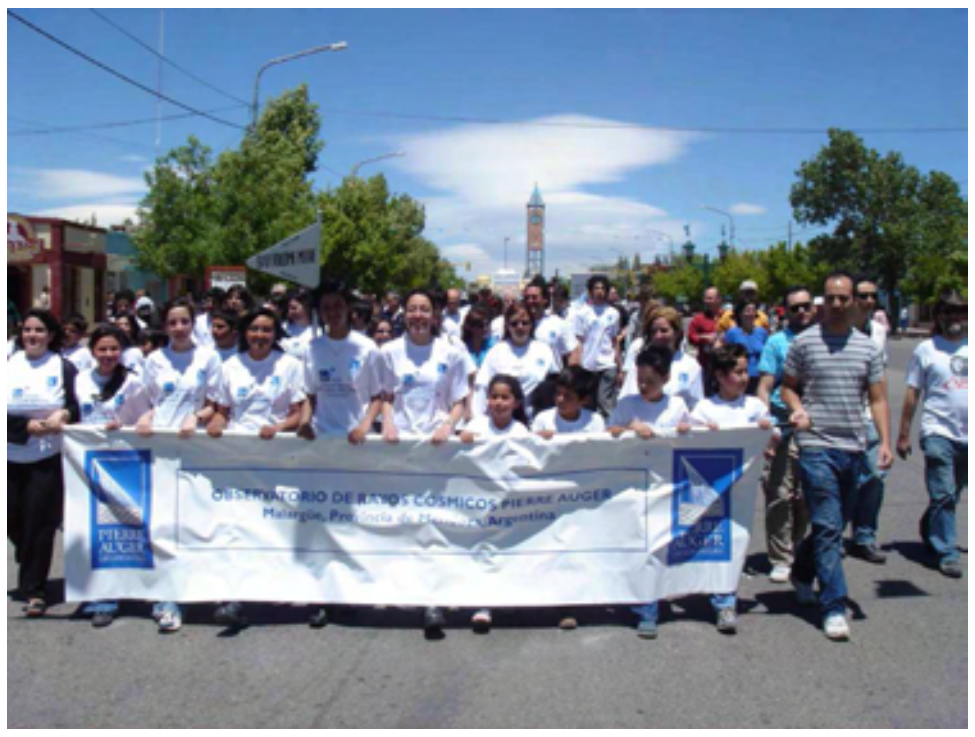

Fig. 1: The Auger collaboration and Science Fair participants in the November 2007 Malargüe Day Parade.

## II. The Auger Visitor Center in Malargüe

The Auger Visitor Center (VC), located in the central office complex and data acquisition building in Malargüe, continues to be a popular attraction. Through the end of April 2007, the VC has hosted 43,777 visitors with an average of about 6000 per year. A noticeable increase of visitors occured after the opening of a new, nearby planetarium [2] in August 2008. Fig. 2 shows the number of visitors logged per year from November 2001 through April 2009. The VC is managed by a small staff led by Observatory employee Analía Cáceres which includes local teacher Miguel Herrera and other Auger collaborators. Fig. 2 shows Auger physicist Julio Rodriguez explaining the Observatory to a visiting school group in the data acquisition center.

Recent exhibits that were field tested at the VC, notably the illuminated scale model of the Observatory developed at the Forschungscentrum Karlsruhe [1] and the Google Earth fly-over animation [3] developed by Stephane Coutu of Pennsylvania State University, have since been replicated elsewhere. As examples, copies of each display are in the interim Auger North VC at Lamar Community College in Colorado and in a new physics and astrophysics learning center called the Galileium in Teramo, Italy, whose director is Auger collaborator Aurelio Grillo.

## III. The 2007 Auger Science Fair

Following a successful Science Fair held in November 2005, the Collaboration sponsored a second Fair on November 16-17, 2007, that attracted the exhibition of 40 science projects in the areas of natural science, mathematics, and technology (see Fig. 3), in contrast to

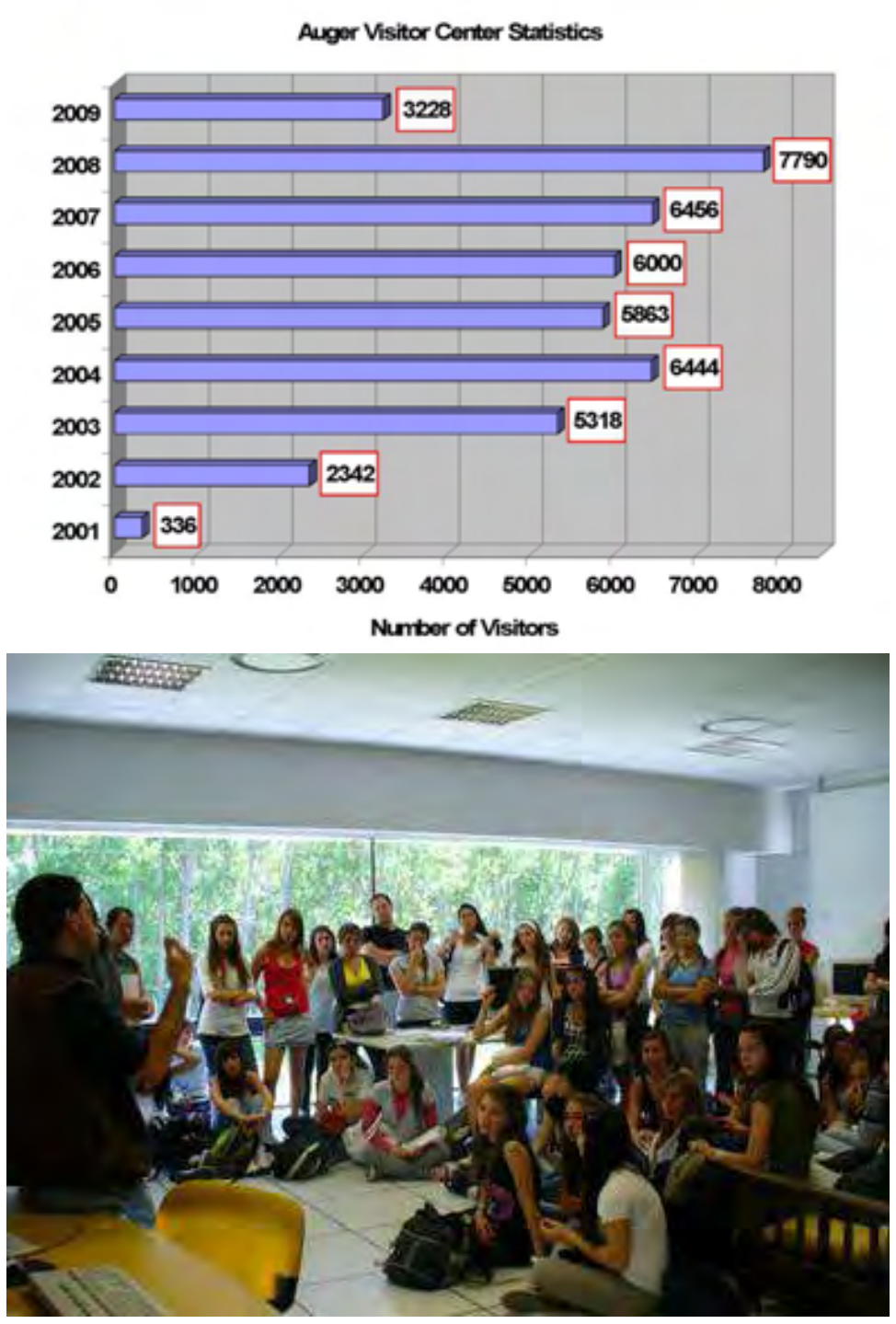

Fig. 2: Top: Number of visitors logged by year at the Auger Visitor Center. Bottom: Julio Rodriguez with a school group visiting the Observatory.

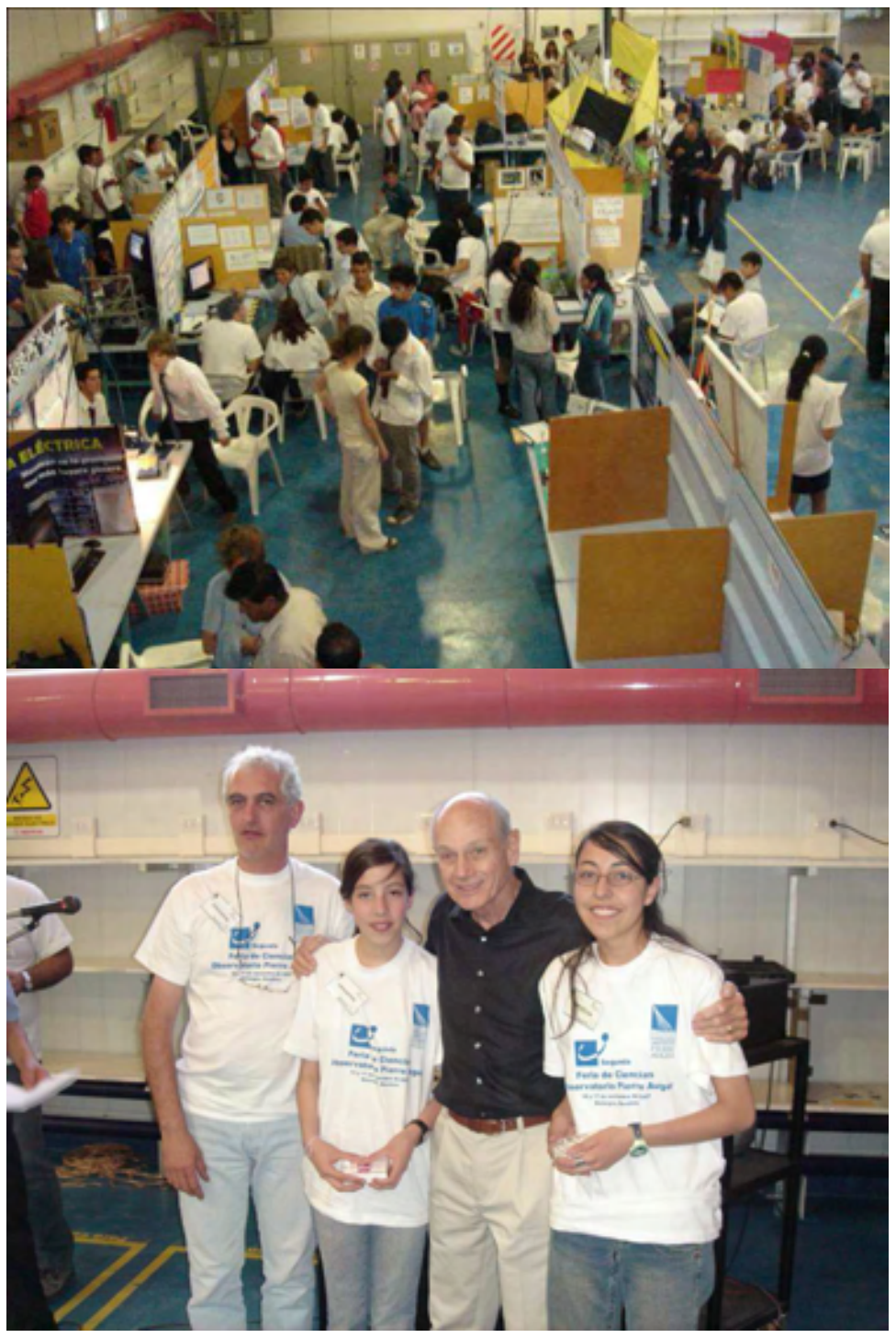

Fig. 3: Top: The 2007 Science Fair in the Assembly Building. Bottom: Jim Cronin with Science Fair award recipients.

29 projects presented at the 2005 Fair. While the 2005 Fair was open only to high school level particpants, the 2007 Fair offered awards in four age categories: grades 1-3, 4-7, 8-9, and 10-12. A team of Auger collaborators judged the projects on the basis of science content, oral and visual presentation, and the written report that accompanied each project. Although the Fair's participants came from all over Mendoza Province, a noticable number of participants and award recipients attended the James Cronin School in Malargüe. The collaboration is indebted to the Observatory staff, the local organizing team of three science teachers, and the city of Malargüe for helping to make the Science Fair a success. A third Auger Science Fair is scheduled to take place in November 2009.

## IV. The 2008 Auger Inauguration

The Collaboration held an inauguration ceremony on November 13-15, 2008, to mark the complete installation of the southern hemisphere Observatory. More than 200 guests and 100 collaborators attended. Guests included Julio Cobos, the Vice President of Argentina, Celso Jacque, the Governor of Mendoza Province, both shown in Fig. 4, the directors of Fermilab and CERN, several ambassadors, many high-level officials from funding agencies and research officers from collaborating instutions. The 3-day event featured talks on the history and status of the Observatory, the unveiling of commemorative monument, traditional folk music and dance performances during an outdoor *asado*, and opportunities for visitors to tour the vast Auger site. Many guests indicated how much they were impressed with Observatory and even enjoyed the dusty two-hour ride across the *Pampa Amarilla*.

## V. Other outreach activities

The scholarship program which brings top Malargüe students to Michigan Technical University (MTU), described in [1], has enjoyed continued success. A fifth student in the program, who would normally begin his studies in science or engineering in the fall 2009, will be delaying matriculation for one year and enroll at MTU in the fall of 2010. The first MTU student from Malargüe who enrolled in 2001 completed a Masters Degree in mechanical engineering and has embarked on an engineering career in the U.S. The second and third students have also graduated and returned to Argentina with their engineering degrees. The fourth student will begin his third year of studies at MTU in the fall 2009.

In August 2007 and 2008, Observatory employee Analía Cáceres served as a lead organizer for the National Week of Science and Technology held at the Malargüe Convention Center. These interdisciplinary symposia drew paricipants from all over Argentina and featured several Auger collaborators speaking about the Observatory and other science topics.

The public release of extensive air shower data for instructional purposes [1] has proven to be successful, and web-based information and instructions are now available in English, Spanish, French, Italian, and German [4]. The web sites register an average of 50 unique hits per day from 43 countries, although the bulk of the traffic is from the U.S., Argentina, and a few European countries. A major spike in usage occured just after the Collaboration published its notable result on the correlation of arrival directions of its highest energy events with AGN galaxies [5]. The public data served as the basis for an award-winning science fair project of two students from Lamar High School in Colorado, shown in Fig. 5.

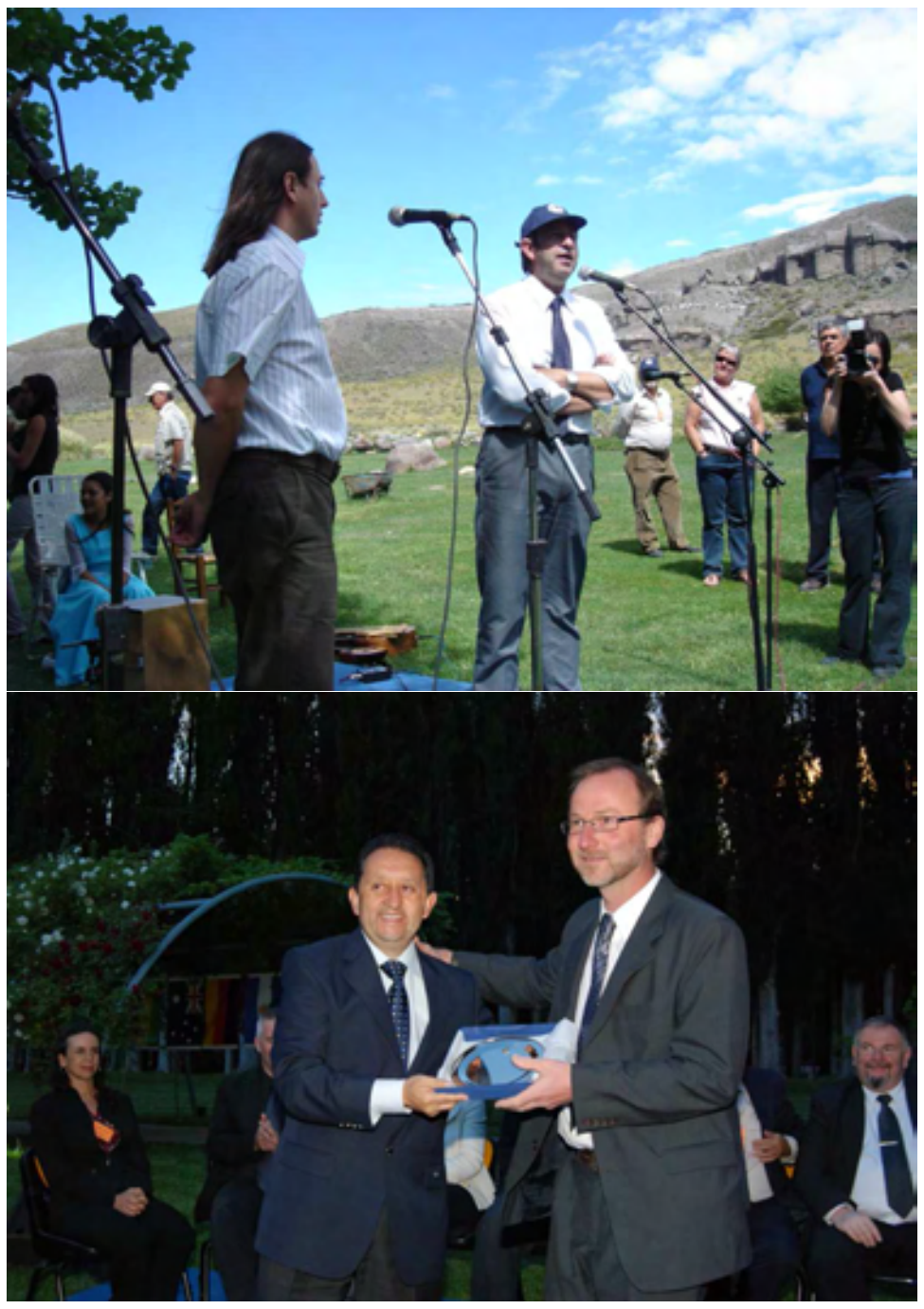

Fig. 4: Top: Auger collaborator Miguel Mostafa (left) translating the comments of Argentina Vice President, Julio Cobos, at the 2008 Inauguration. Bottom: Mendoza governor Celso Jacque presenting a plaque to Auger collaborator Ingo Allekotte at the Inauguration.

In the fall 2008, Randy Landberg, Education and Outreach Director for the Kavli Institute for Cosmological Physics, Auger collaborators from the University of Chicago, and personnel from the American Museum of Natural History (AMNH) in New York received U.S. National Scienc Foundation funding to produce a 10-minute, high-resolution video about the Observatory. The video is foreseen as one of the AMNH online Science Bulletins [6]. A AMNH film crew visited the Observatory in November 2008 to shoot footage of Observatory sites and surroundings and to film interviews with Auger collaborators. The Science Bulletin is scheduled to be released in the late spring 2009.

Many Auger collaborators have organized activities at their institutions and in local communities associated with the 2009 International Year of Astronomy.

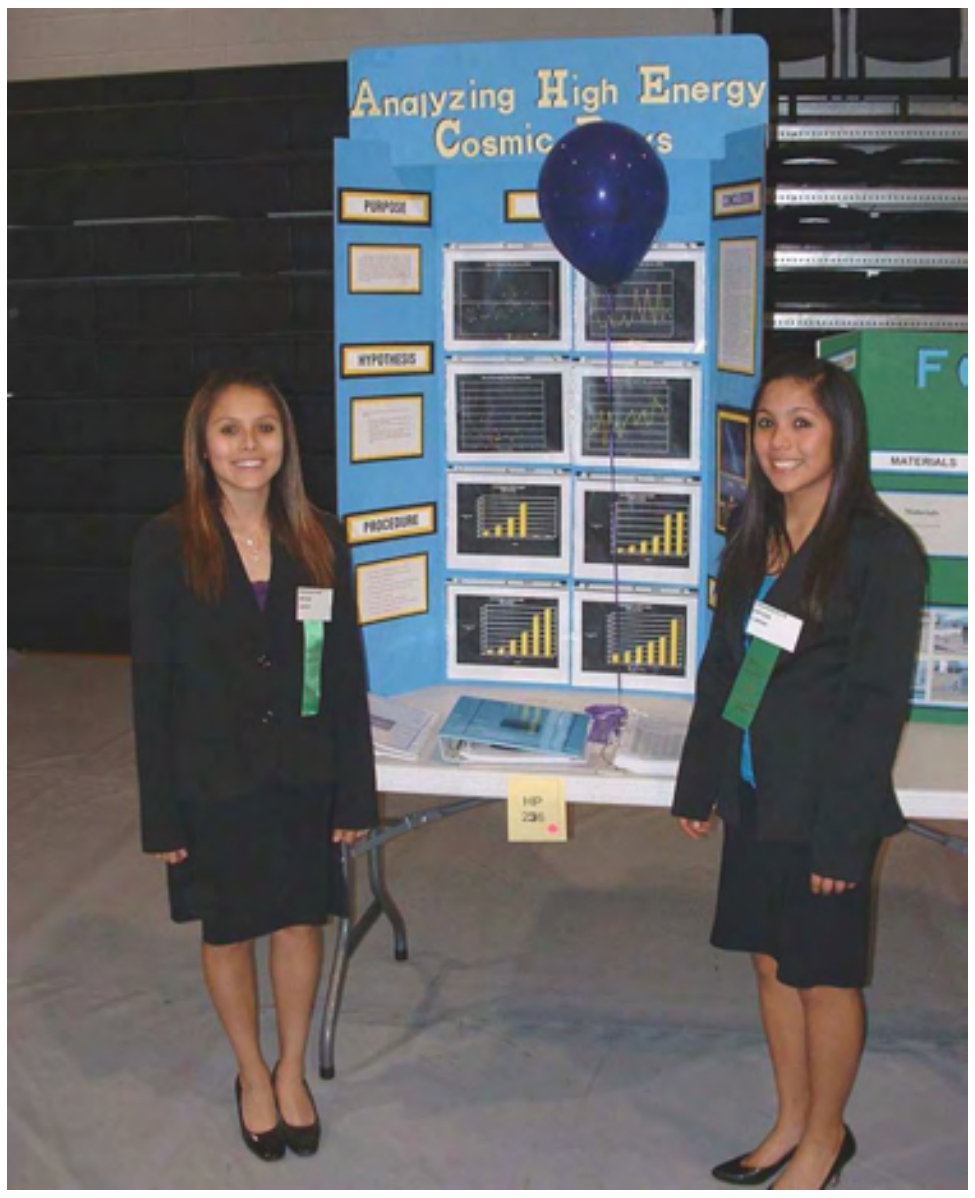

Fig. 5: Students from Lamar High School with their science fair poster based on Auger data released to the public.

## VI. Auger Northern Site Outreach

As the Auger collaboration prepares its proposal to build a northern hemisphere site in southeast Colorado, the Education and Outreach Task continues to lay the groundwork for a comprehensive outreach program that will be linked to the outreach efforts in the southern hemisphere. Primary outreach goals in this period are to promote the Observatory in southeast Colorado, provide information to people at all levels, and establish early ties with science teachers and students in the region's schools. Lamar Community College (LCC) has been named the host institution for Auger North outreach, and efforts have been led by LCC employee Brad Thompson.

An interim Visitor Center has been established (see Fig. 6) in the LCC library which features explanatory posters, take-home information brochures, a scale model of a Surface Detector (SD) station, an SD photomultiplier tube, scintillator detectors which register cosmic ray muons as viewed on an oscilloscope, and a large flat screen monitor to view the Google Earth display [3] of the Observatory. This display was recently expanded to include the layout of the proposed northern site. A larger state-of-the-art visitor center is foreseen to be included in the headquarters of Auger North at LCC.

A recent highlight was the installation of two full-sized SD detector stations in the area, one on the LCC campus, and one near the county fairgrounds in the nearby town of Las Animas. The names for these stations, Pierre's Dream and Cosmos, respectively, were determined by a contest among area primary school students. The SD stations allow area landowners to view examples of detetectors they will be asked to host on their property as Auger North proceeds.

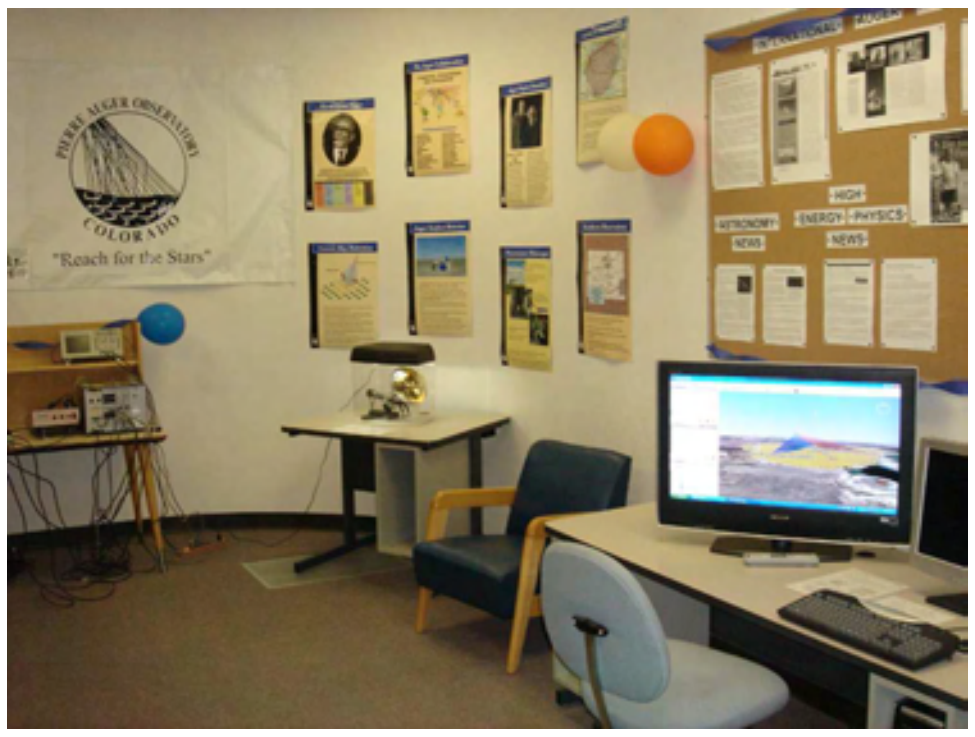

Fig. 6: The interim Visitor Center at LCC.

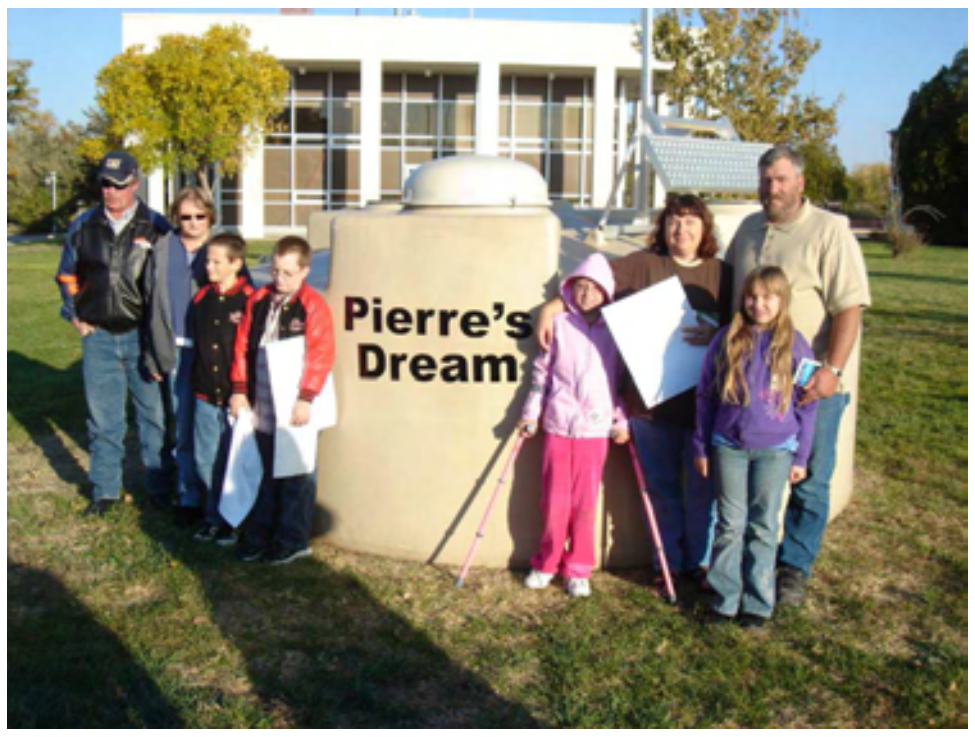

Fig. 7: Shown with the display SD station at LCC are the students and families responsible for the detector's name.

An unveiling ceremony for the detectors was held on October 27, 2007, in both Lamar and Las Animas, with the mayors of each town welcoming the Observatory to the area. The winners of the detector naming contest were also recognized at the ceremony, as shown in Fig. 7. Additional SD stations are presently being placed in visible locations in other towns in the footprint of the Auger North array, namely, in Springfield and Eads, Colorado. In addition to their role in public relations, the detectors will be filled with water and outfitted with temperature sensors for freezing studies, part of the R&D for Auger North.

# BATATA: A device to characterize the punch-through observed in underground muon detectors and to operate as a prototype for AMIGA

**Medina-Tanco Gustavo***, **for the Auger Collaboration**†

*Instituto de Ciencias Nucleares (ICN), Univ. Nacional Autónoma de México, México D.F.
† Observatorio Pierre Auger, Av. San Martín Norte 304 (5613) Malargüe, Prov. Mendoza, Argentina.

***Abstract*. BATATA is a hodoscope comprising three X-Y planes of plastic scintillation detectors. This system of buried counters is complemented by an array of 3 water-Cherenkov detectors, located at the vertices of an equilateral triangle with 200 m sides. This small surface array is triggered by extensive air showers. The BATATA detector will be installed at the centre of the AMIGA array, where it will be used to quantify the electromagnetic contamination of the muon signal as a function of depth, and so to validate, in situ, the numerical estimates made of the optimal depth for the AMIGA muon detectors. BATATA will also serves as a prototype to aid the design of these detectors.**

## I. INTRODUCTION

High energy cosmic rays are indirectly characterized by measuring the extensive air shower cascades that they trigger in the Earth atmosphere. At ground level, and at distances greater than a few tens of meters from the axis, the shower is dominated by just two components: electromagnetic (electrons, positrons and photons) and muonic. The relative weight of these two components has invaluable information about the nature of the primary cosmic ray and of the high energy hadronic processes taking place at high altitude during the first interactions. Water Cherenkov detectors used by the Pierre Auger Observatory measure the combined energy deposition of charged particles inside its volume and are not well suited for discriminating between electromagnetic and muonic components. Therefore, as part of the Auger surface low energy extension AMIGA [1], buried muon scintillators will be added to the regular surface stations of an infill region of the array. Provided enough shielding is ensured, these counters will allow the measurement of the muon signal and, in combination with their Cherenov tank companions, the estimate of the electromagnetic component. The final objective of this assemblage is to achieve high quality cosmic ray composition measurements along the ankle region [2]. Several prototyping activities are being carried out for the muon counters [3] and an additional detector is being constructed, BATATA. The main objective of the latter is to quantify the electromagnetic contamination of the muon signal as a function of depth, and so to validate in situ the numerical estimates made of the optimal depth for the AMIGA muon detectors. BATATA will be installed at the center of the future AMIGA array and will also serve as a prototype to aid the design of its detectors. The chosen site, from the point of view of soil properties and, therefore, of punch-through characteristics, is statistically equivalent to any other location of the area covered by the infill [4].

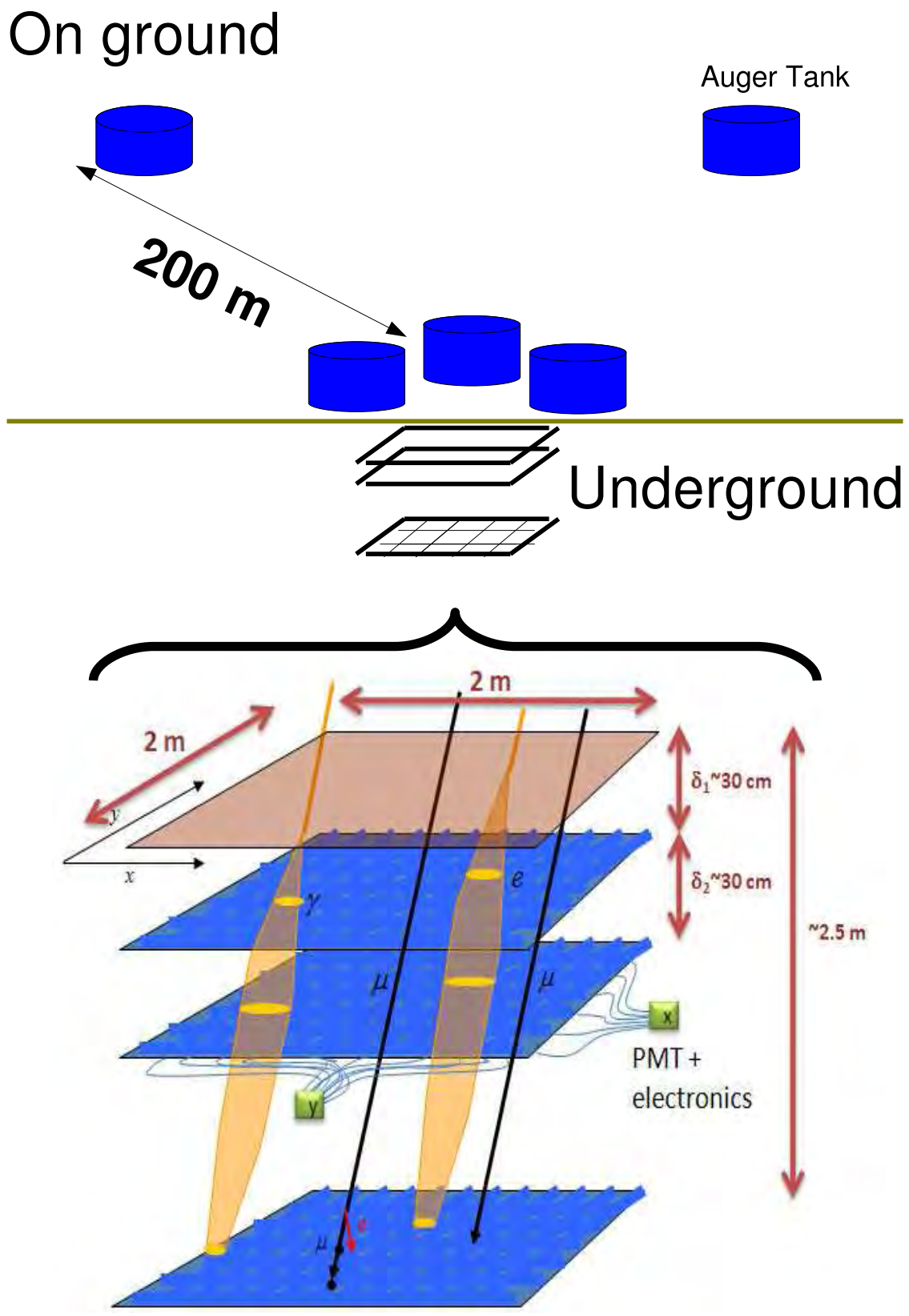

Fig. 1: Surface (top) and buried (bottom) sections of the BATATA detector and working principle.

## II. THE DETECTOR

A schematic view of BATATA is shown in Figure 1. The detector is composed by a set of three parallel dual-layer scintillator planes, buried at fix depths ranging from $0.50$ m to $2.5$ m.

Each layer in a plane is 4 $m^2$ and is composed by 49 rectangular strips of 4 cm x 2 m, oriented at a right angle with respect to its companion layer, which gives an xy-coincidencepixel of 4x4 $cm^2$. The scintillators are MINOS-type extruded polystyrene strips [5], with an embedded Bicron BC92 wavelength shifting fiber, of 1.5 mm in diameter [6]. Building and quality assessment protocols (e.g., cutting, polishing, gluing and painting) as well as the characterization procedures of the optical fibers and scintillator bars can be found in [7]. Light is collected by Hamamatsu H7546B 64 pixels multi-anode PMTs [8]. Each $x$-$y$ plane is fitted inside an individual casing, where each orthogonal direction ends in its own front-end electronic board (FE), as shown schematically in Figure 2.

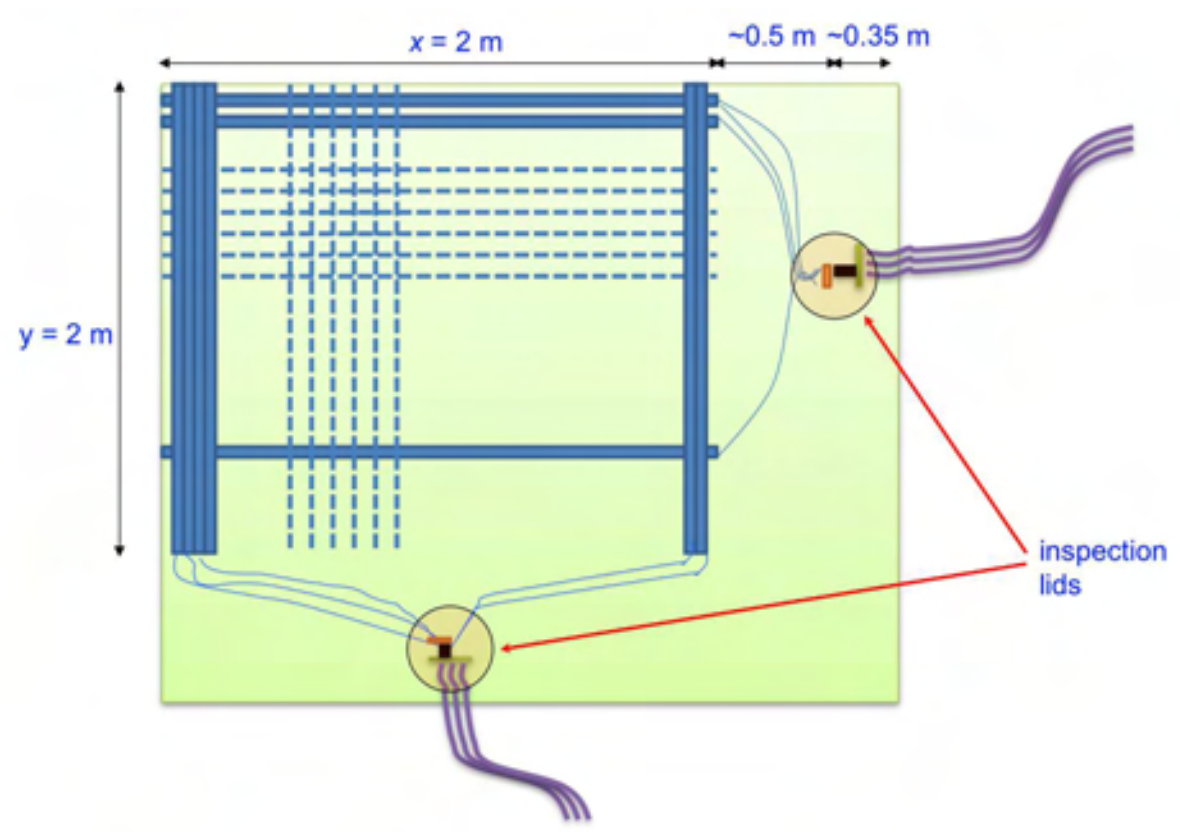

Fig. 2: Schematic view of the arrange of scintillators, optical fibers and front-end electronics inside the casing for any one of the three BATATA planes.

The casings are made of fiber glass and were specially designed to be water- and light-tight and to withstand handling during shipping and burying. They have two caps which allow for easy access to the corresponding front-end boards and optical couplings between PMTs and optical fiber cookies. The caps can be dismounted in two steps if necessary, one for inspection and servicing, and a second for cabling replacement, ensuring in both cases that water-tightness can be recovered. Additionally, the casing and sealing are versatile enough to allow for the straight forward addition of new components and cabling not specifically foreseen in the original design.

The front-end electronics works in counting mode and signals are transmitted to the surface DAQ stage using low-voltage differential signaling (LVDS). Any strip signal above threshold opens a GPS-tagged 2 $\mu s$ data collection window. Data, including signal and background, are acquired by a system of FPGA Spartan boards and a TS7800 single board computer. The code controlling the data flux at the FPGA level was written in VHDL (VHSIC -Very High Speed Integrated Circuits-hardware description language).

The front end boards are $12.3'' \times 9.3''$ and comprise 64 channels, of which only 49 are actually used at present. The multi-anode PMT and its high-voltage supply are located on the board. Each channel includes: (i) an amplification stage, which uses the AD8009 operated at amplification factor $\sim 6$, (ii) a discrimination stage, which uses a MAX9201 high-speed, low power, quad comparator with fast propagation delay (7ns typ at 5mV overdrive) connected in bipolar mode, (iii) a digital-to-analog converter TLC7226C to independently setup the discrimination voltage of each channel and, (iv) a high-speed differential line driver SN55LVDS31 to transform the discriminator output into a differential signal which is carried out to the surface and to the DAQ system some 17 m away (see, Figure 3).

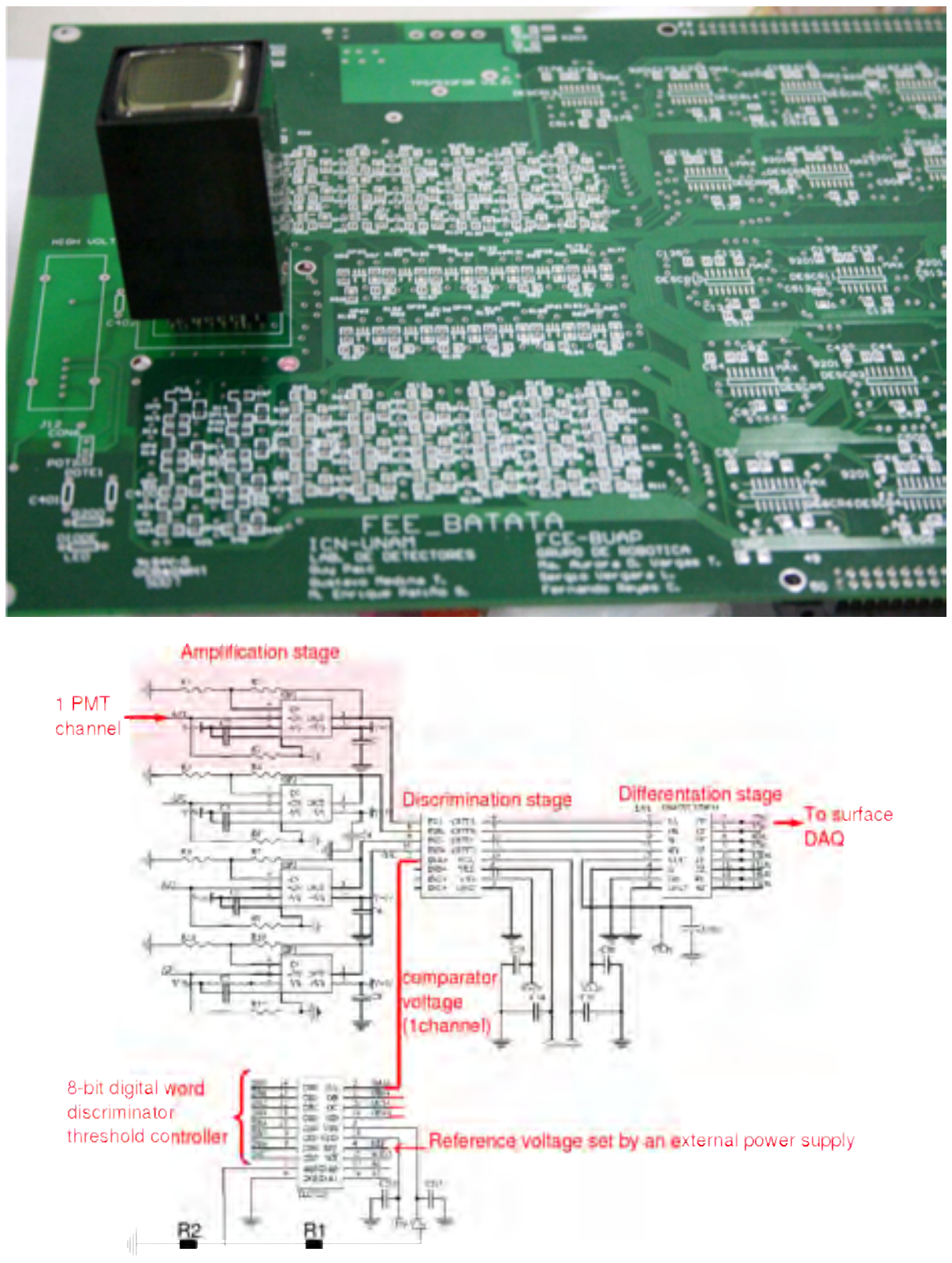

Fig. 3: Front-end board: (top) general view with multi-anode PMT, (bottom) single channel layout.

Each stage of the FE board has been tested with controlled square pulses of 50 mV height and ≈10 ns width injected at a frequency of 1 kHz. Mean outputs of the amplification and differentiation stages as well as output rates of 16 different channels are shown in Figure 4.

The front end electronics resides in an enclosure which only exchanges heat with the surrounding ground by conduction. It has being experimentally determined that, under the most unfavorable conditions, the equilibrium temperature at the surface of the board never exceeds 40 $^{o}$C, which is well inside the working range of the electronic components.

A cosmic ray EAS event triggers at least tens of channels in a time scale o a $\mu$-sec. This rate is much higher than the ∼kHz background of the detector due to low energy cosmic rays and natural radioactivity. Therefore, EAS events can be easily detected with a simple trigger scheme implemented at the FPGA level. The FPGA trigger is implemented via software and can be easily changed if appropriate. In order to ensure the stability of the trigger rate, background events are also recorded at fixed time intervals during each run and are used to re-calibrate the discriminator thresholds at the end of each run. Additionally, in order to characterize the electromagnetic punch-through, EAS inside a limited energy range and with axis at a defined distance from the detector have to be selected. In order to attain this, BATATA also counts with a small triangular surface array of 3 regular SD Cherenkov stations at a separation of 200 m. This array provides the required offline selection capability for quasi-vertical showers in the vicinity of 10 PeV.

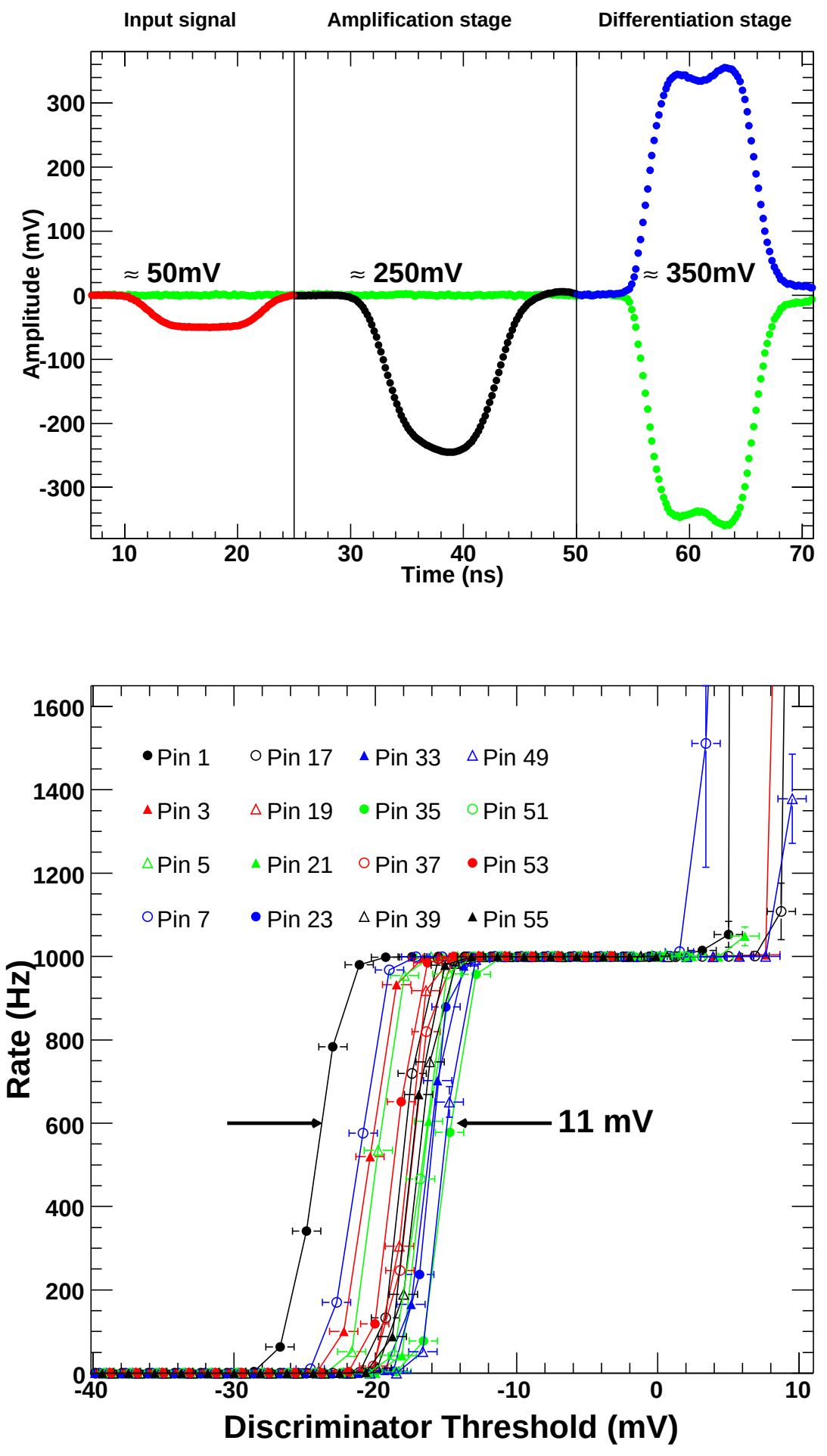

Fig. 4: Testing and characterization measurement on FE board with controlled square pulses of 50 mV height and ≈10 ns width injected at a frequency of 1 kHz.

The power consumption of BATATA is $\lesssim$ 200 W. This power is supplied by an array of 20 solar panels with their corresponding batteries, ensuring continuous operation during low insolation winter months.

## III. End-to-end simulations and data analysis

Muons are penetrating particles and propagate through the ground following a well defined track. Therefore, it is expected that, most of the times, they will only trigger one pixel at each layer. Furthermore, muons will tend to trigger $x$-$y$ pixels in coincidence inside the same plane. Electrons, positrons and photons, on the other hand, are much less penetrating and generate rapidly evolving electromagnetic showers underground which have a lateral extension. Therefore, they will tend to leave 2-dimensional footprints in the detector, specially at the depths of maximum development of the cascades between $\sim$ 30 and $\sim$ 80 cm for particles that, on the ground, have energies in the range $\sim$ 0.5–10 GeV. These differences between muon and electromagnetic signatures at the detector are used for discrimination. This is schematically shown in Figure 1, and motivates the uneven depths chosen for the three scintillator planes of BATATA.

The data analysis requires a thorough understanding of the statistical response of the detector to EAS particles impinging the ground above and in its vicinity. Therefore, a comprehensive set of numerical end-to-end simulations of the detector have been carried out. A combination of AIRES [9] and Geant4 [10] is used to simulate EAS development from the top of the atmosphere up to the ground, where particles are injected and followed using Geant4 while they propagate through the soil and into the scintillator bars. The subsequent photons produced by the dopants inside the polystyrene strips are followed while bouncing off the $TiO_2$ covering and into the wavelength shifting optical fiber. Finally, the green-shifted photons thus produced by the dopant of the fiber are transfered along up to the window of the PMT. The several parameters involved in the simulation of the scintillator are then fine-tunned to reproduce the structure of the time profiles and integrated charges of muon pulses actually measured under laboratory conditions. In parallel, the shower impinging the surface array is reconstructed using the Offline Auger reconstruction and data handling package [11].

Figure 5 shows some example of the output of the end-to-end simulations for the underground segment of BATATA. Figure 5a shows one simulated muon tracks and 2 low energy electromagnetic cascades crossing the 3 planes of the detector and the corresponding triggered strips, i.e., producing a voltage above threshold in their front-end discriminator. Figure 5b shows the simulated and the real mean pulse shape for background muons. The output of the overall simulation chain is an electronic PMT signal as shown in Figure 5c. As can be seen from Figure 5d, where a measured muon signal is shown, the simulations are able to reproduce satisfactorily the output of the detector.

## IV. Conclusions

A hodoscope, BATATA, comprising three X-Y planes of plastic scintillation detectors, complemented by an array of 3 water-Cherenkov detectors at a separation of 200 m is under construction and will be installed at the centre of the AMIGA extension to the Auger baseline design. BATATA will be used to quantify the electromagnetic contamination of the muon signal as a function of depth. and will also serve as a prototype to aid the design of the AMIGA muon detectors.

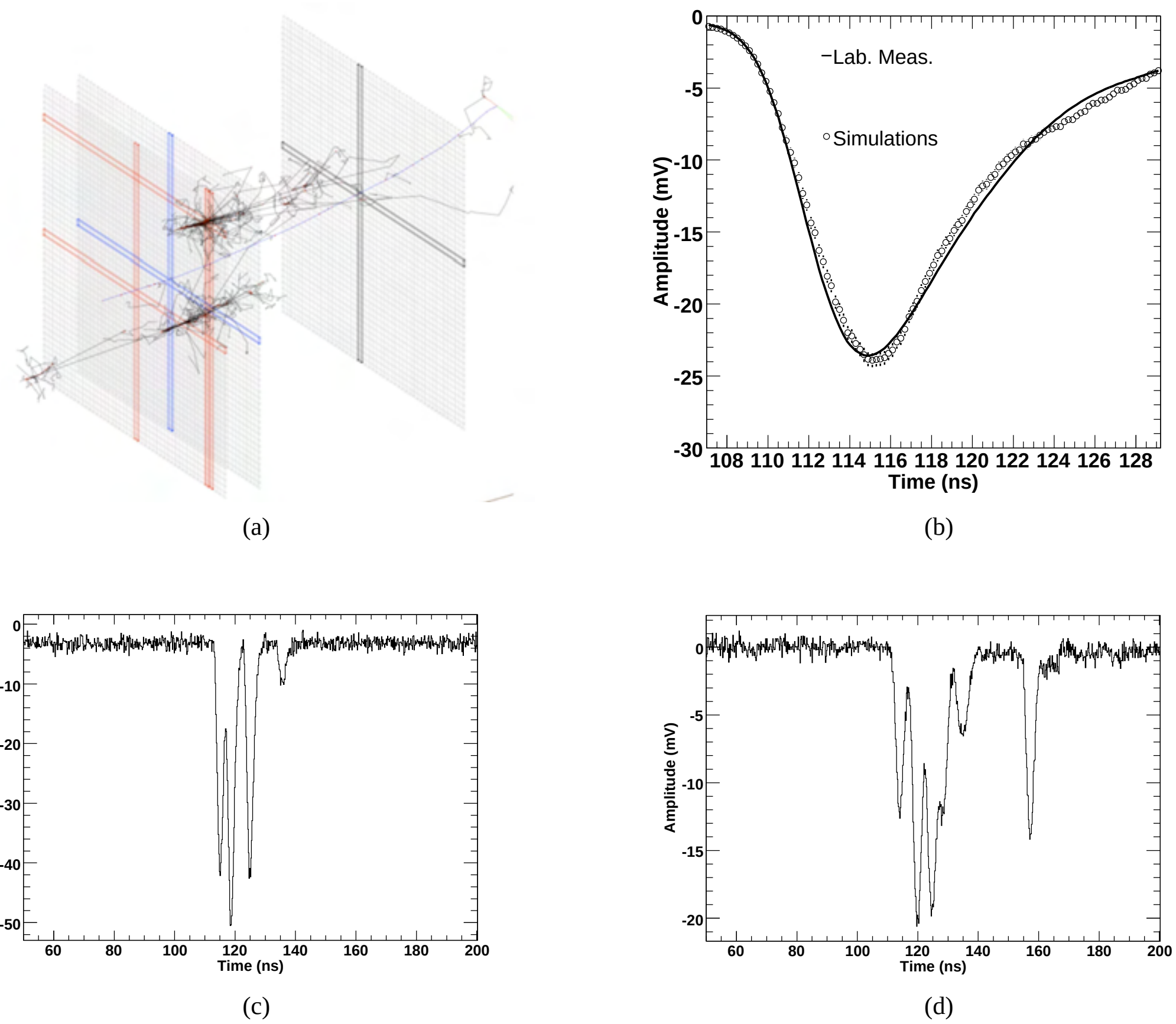

Fig. 5: End-to-end simulations. (a) Simulated muon tracks and low energy electromagnetic cascades. (b) Real and simulated mean pulse shape at 160 cm from the edge of a scintillator. (c) Simulated muon signal. (d) Actual laboratory measured muon signal.

BATATA is in its final phase of construction and its deployment will start on July 2009 with commissioning along the second semester of the year.

## ACKNOWLEDGEMENTS

This work is partially supported by the Mexican agencies CONACyT and UNAM's CIC and PAPIIT.

# Progress with the Northern Part of the Pierre Auger Observatory

**John L. Harton*** **for the Pierre Auger Collaboration**

*Colorado State University, Department of Physics, Fort Collins, CO 80523 USA*

***Abstract*****. It is planned to build the northern part of the Pierre Auger Observatory in southeast Colorado, USA. Results from the southern section of the Auger Observatory, which has recently been completed, imply a scientific imperative to create a much larger acceptance for the extremely rare cosmic rays of energy above a few times $10^{19}$ eV. The plan for Auger North is to cover an area greater than 20,000 km$^2$, seven times the area of Auger South in Argentina. The motivation for Auger North and the status of preparations will be presented including: R&D work at the Colorado site on a small surface detector array; atmospheric monitoring measurements; R&D on new electronics and communications equipment; and outreach and relations with the local community.**

## I. AUGER SOUTH REVIEW

The southern Pierre Auger Observatory near Malargüe, Argentina was completed in June 2008 and has produced science results for events with primary cosmic ray energies in the EeV range and above. The published results are based on exposures taken while the observatory was growing, and the integrated exposures are about 9,000 $km^2\ sr\ yr$. One year of data taking with the full Auger South instrument (including arrival direction zenith angles from 0 to 60 degrees) represents approximately 7,000 $km^2\ sr\ yr$. These results include the energy spectrum, with observation of the 'ankle'and GZK-like features [1], and limits on the flux of primary photons [2] [3] and tau neutrinos [4]. The photon limits are already restricting some top-down models of cosmic ray production, and the neutrino limit is the best across the energy range expected for neutrinos produced in the GZK interaction. Studies of the arrival directions of the highest energy particles show that they correlate with the matter density within about 100 Mpc, and they may correlate with relatively nearby active galactic nuclei [5] [6]. The steep drop in flux, first published by HiRes [7], occurs at about 60 EeV and generally corresponds in the Auger data to the threshold energy for the correlation with directions to nearby extragalactic objects. So the most energetic cosmic rays, those with the trans-GZK energies, appear to be of extragalactic origin. In addition data from Auger South have been used to study the average position of maximum shower development in the atmosphere, $X_{max}$, and its correlation with primary cosmic ray type. For sources further than about 20 Mpc from Earth, light nuclei, such as carbon, nitrogen, and oxygen will break up due to interactions with photons more readily than iron, so the only nuclei expected to survive these distances in significant percentages are iron nuclei and protons. The Auger South result on $X_{max}$ shows that $X_{max}$ is tending to smaller values as energies approach the GZK feature, and this can be interpreted to indicate, within current models of hadronic interactions that are extrapolated from lower energies, that the primary composition is getting heavier [8] [9]. This measurement could signal new hadronic physics if the primary cosmic rays are protons.

The Auger South instrument is composed of an array of detectors to measure particles from extensive cosmic ray showers that reach the ground and optical detectors that overlook the surface detectors. The optical detectors, which function on dark and clear nights, measure the fluorescence of atmospheric nitrogen along the axis of the particle shower. The surface array of Auger South is comprised of over 1,600 water Cherenkov detectors placed on a triangular grid with 1.5 km spacing, covering 3,000 square kilometers [10]. On the periphery of the surface array are four buildings each housing six telescopes [11] that record the nitrogen fluorescence.

The results from Auger South present compelling evidence that the sources of cosmic rays above 60 EeV are extra-galactic and that the GZK effect reduces primary energies below this level for particles from sources more distant than the nearby universe. Thus it will be possible to study the nearby sources at highest energies without a diffuse background from the entire universe. Additionally, Auger South data allow one to measure particle interactions at energies not accessible to manmade accelerators, up to $\sim$500 TeV in the center of mass system, and there may be hints of new particle physics.

## II. AUGER NORTH

The Auger South results imply an imperative to study the highest energy cosmic rays with the greatest statistical precision possible. The Auger South detector records approximately 20-25 trans-GZK events per year. The plan for Auger North is to build a ground array covering 20,500 km$^2$ with near complete fluorescence detector coverage. Auger North is proposed for southeast Colorado, USA - see Figure 1. The acceptance is seven times that of Auger South, so the full Auger Observatory, South plus North, will detect (assuming flux about the same in the north as in the south) approximately 180 of the most interesting events per year - hopefully enough

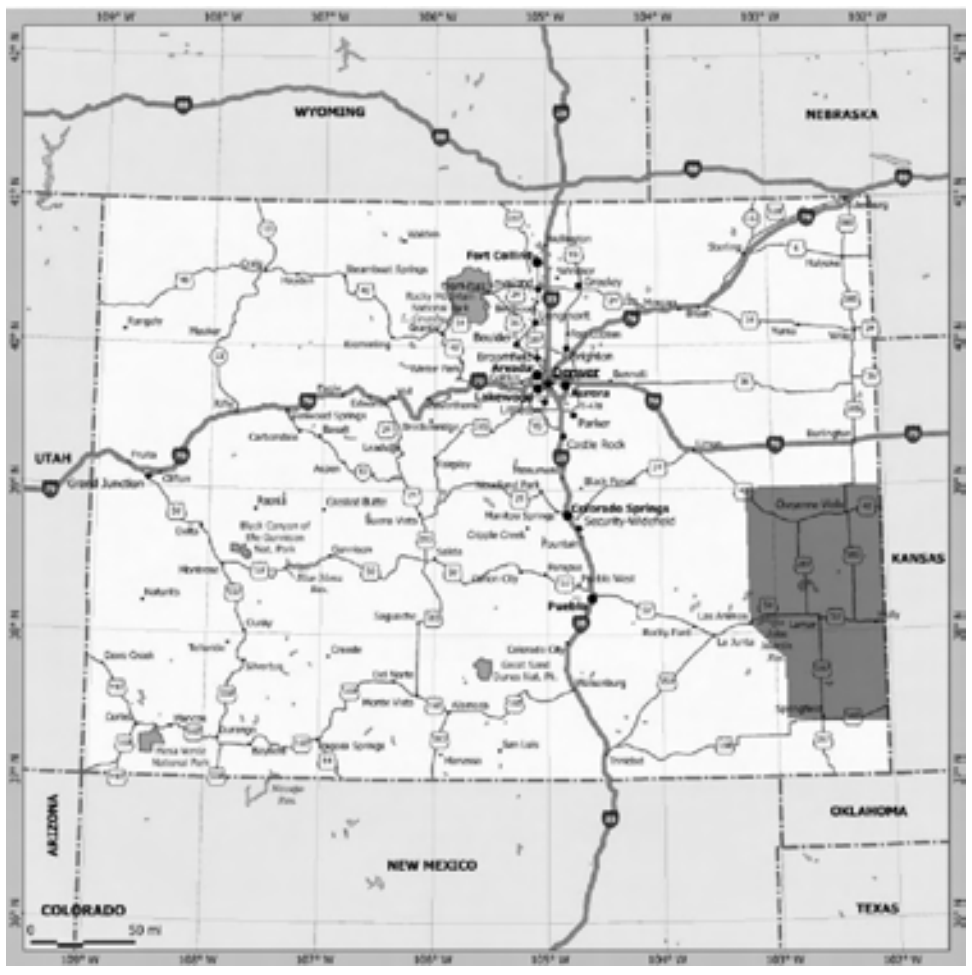

Fig. 1: Site of the northern part of the Pierre Auger Observatory. The shaded area in the southeast corner of this Colorado map shows the 8,000 square miles (20,500 square kilometers) planned for Auger North. The site could extend eastward into Kansas.

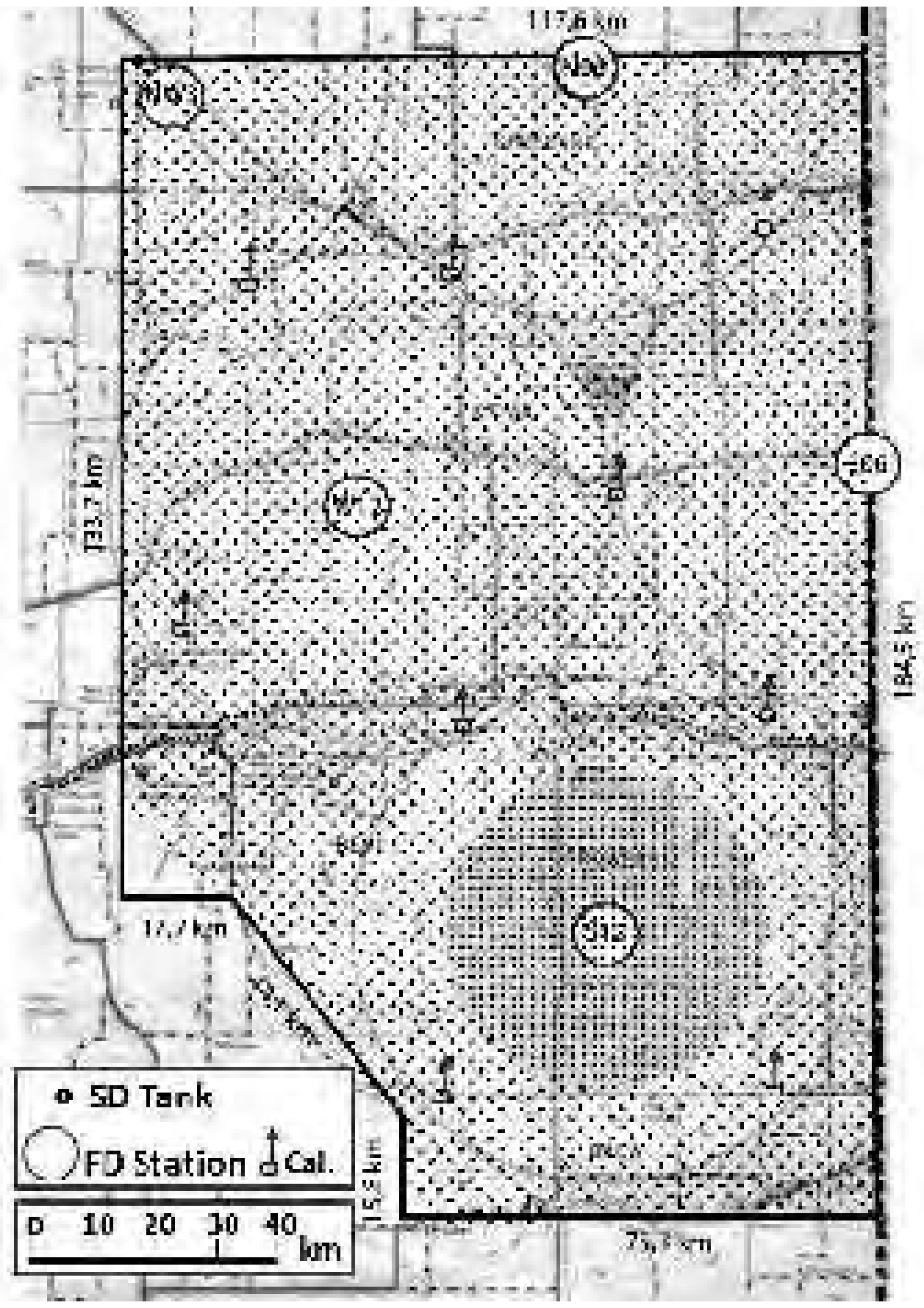

Fig. 2: Layout of the northern Auger detector. Each dot represents a Cherenkov detector. The fluorescence detector positions are given by circles, and the permanent calibration lasers are shown as small arrows. The 2000 $km^2$ infill is shaded (lower right) and has higher density of surface detectors .

to study the energy spectra from individual sources in 10-20 years of observation.

The design of Auger North focuses on the highest energies while profiting from the successful experience with the southern Auger instrument. Water Cherenkov detectors and fluorescence detectors similar to those in the south are planned. There will be a few changes. The most important is that the Auger North surface detector spacing will be $\sqrt{2}$ miles (2.28 km) on a square grid. This spacing takes advantage of the land ownership pattern and partial road grid in the western United States, which is based on 1-mile (1.6 km) squares. For most of the array only every second corner of the square mile grid will have a surface detector unit. We plan 4,000 detectors covering 8,000 square miles for the majority of the array. On 800 square miles (2000 $km^2$) we will place a detector on every grid corner. Four-hundred additional surface detectors are needed for this in-fill, which will provide a connection to the high statistics at Auger South. The energy at which Auger South approaches 100% trigger efficiency is 3 EeV. Studies using Auger South data indicate that the trigger efficiency on the $\sqrt{2}$-mile pattern will be 90% at 30 EeV and that the 800 square-mile infill will be 90% efficient at 4 EeV. We plan to build five fluorescence detector buildings with a total of 39 telescopes at Auger North. Figure 2 shows the layout of Auger North. The telescopes will be farther apart than in the south to minimize costs, but above 30 EeV Auger North will still have near full hybrid coverage.

Other important changes include the use of only one large PMT in each surface detector instead of the three at Auger South (cost savings), and the need to insulate the Cherenkov tanks (colder winters in Colorado than in Malargüe). The topography in Colorado is not as flat as at the Auger South site, and there are no hills surrounding Auger North as there are in the south. These factors make it impossible to use the communications paradigm at Auger South in which data are transmitted from individual detectors to the fluorescence buildings on the perimeter of the surface array. At Auger North the data will flow from tank to tank until reaching a collection point. At Auger South the absolute calibration of the fluorescence detector is based on periodically placing a calibrated light source at the aperture of each telescope [12] to illuminate all 440 PMTs in the detector. This calibration is checked from time to time using a mobile laser positioned a few km from the telescope [13]. The laser is fired vertically, and the calculable flux of scattered photons is used to check the calibration of a line of PMT pixels in the telescope. The method is cumbersome since it requires moving the laser in the field to illuminate different telescopes or pixels. At Auger North we plan an array of permanent lasers to complement the method using a light source at the telescope aperture.

The fluorescence detectors for Auger North will be very similar to those in Malargüe. The basic design is unchanged, but the electronics and slow control will be updated. The main testbed for the Auger North fluorescence detector is the three HEAT telescopes [14] being built at Auger South.

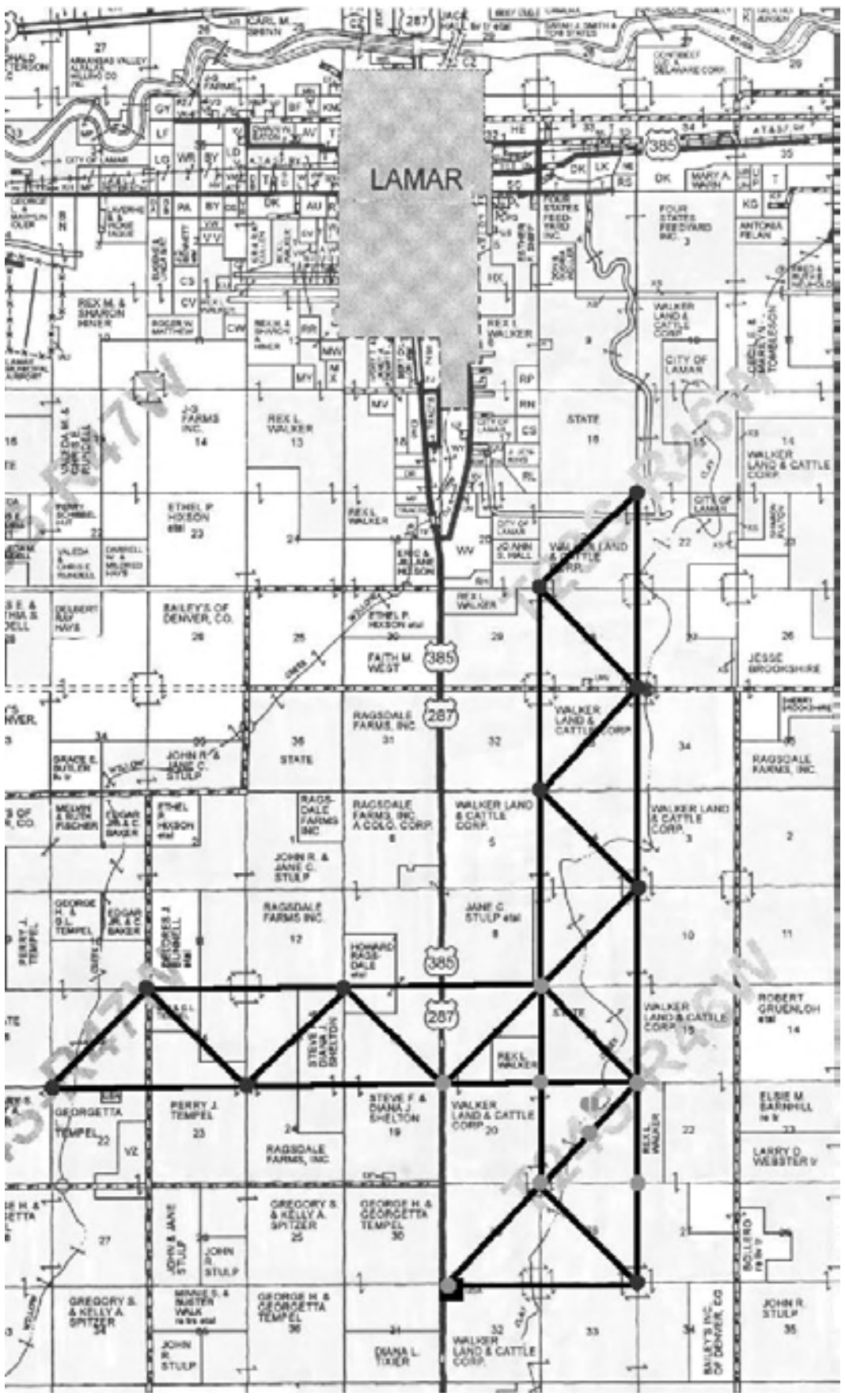

Fig. 3: Tentative layout for the Auger North R&D array. Light circles represent possible sites for fully functioning tank detectors; darker circles indicate sites for communications-only equipment. The two arms are the minimum needed to test the tank-to-tank communications system.

## III. R&D for Auger North

The main changes for Auger North require R&D to confirm the new designs and and in some cases to solve technical challenges. To this end the collaboration is building a small R&D array south of Lamar, Colorado, the main city in the site region. Lamar Community College (LCC), located a few miles north of the R&D array, is the host institution for education and outreach for Auger north. LCC is operating an interim visitors' center at the college, and puts on lectures for local residents at schools. Plans call for the main Auger North campus to be built at LCC.

Figure 3 shows the layout of the R&D array. We plan to deploy 6 to 10 fully functioning Auger North tanks and 10 to 14 stations that have only communications equipment. We plan to start the deployment in 2009. The layout of the R&D array is driven by the new communications system, which uses the long and short arms to funnel the data to the central data acquisition center. In summer 2008 Auger collaborators made a survey of the signal transmission from all points on the R&D array to the neighboring stations.

The Auger South tanks are made of roto-molded polyethylene without insulation. One approach to insulate the Auger North tanks is to use a chemically foaming resin inside the tank as a manufacturing step. The technique is promising, and a few insulated tanks have been produced, but it is not yet at a production stage. We have made studies over the past few winters with non-insulated and hand-insulated Auger tanks in Colorado and Argentina with temperature probes in and around the tanks. We have measured significant heat flow from the ground into the tank in the winter, which indicates that the bottom of the tank should not be insulated. Our thermal models generally predict the water temperature as a function of ambient temperature and incident sunlight as the tanks approach freezing and ice starts to form. Work continues with the newly manufactured tanks insulated using the foaming resin. The goal is to optimize the thickness of the insulation (for cost) in the face of a simulated historical period of extreme cold.

Atmospheric studies at the Colorado site are moving forward with a system based on successful work at Auger South. We plan to install a vertically firing laser that will be viewed by two detection systems: a Raman LIDAR receiver installed at the laser site and directed along the laser beam to record the backscattered signal; and a system with a mirror and PMT camera similar to those used at HiRes (but with only a narrow stripe of PMT's installed) which we call the AMT (Atmospheric Monitoring Telescope). The AMT will be placed 30-40 km from the vertical laser beam, and so it is very similar to the CLF[15]/XLF technique used at Auger South to measure the vertical aerosol optical depth up to approximately 6 km. The two systems will make independent measurements of the aerosols while providing a test bed for Auger North atmospheric monitoring. In addition, we plan to place a nitrogen laser approximately 4 km from the AMT to test the concept of the permanent absolute calibration laser. The AMT will use the electronics developed for the HEAT telescopes.

Other R&D includes work on surface detector electronics to extend the dynamic range of the Cherenkov detectors, and studies with new GPS receivers. We will also exercise deployment techniques and in general learn about working in Colorado to build Auger North. This R&D array is similar to the Engineering Array [16] of Auger South in that the goals are to fully reconstruct extensive air showers at each site - separately with the surface detector in Colorado and the HEAT fluorescence detectors in Argentina. The aim is to bring as many components as possible to the pre-production stage and to validate cost estimates for Auger North.

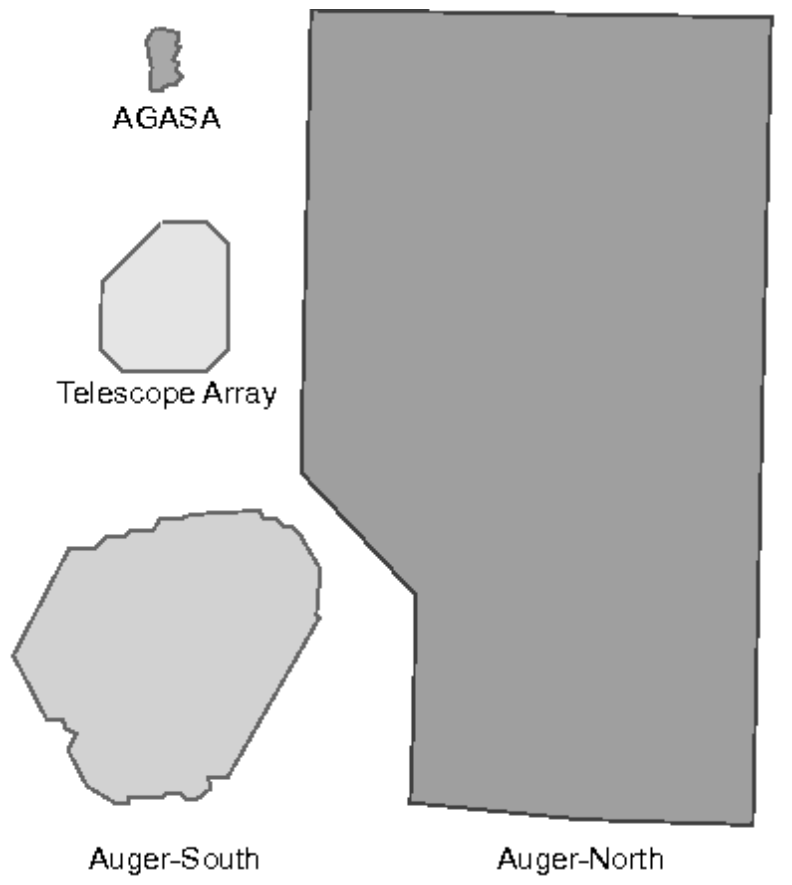

(a) Array areas to scale.

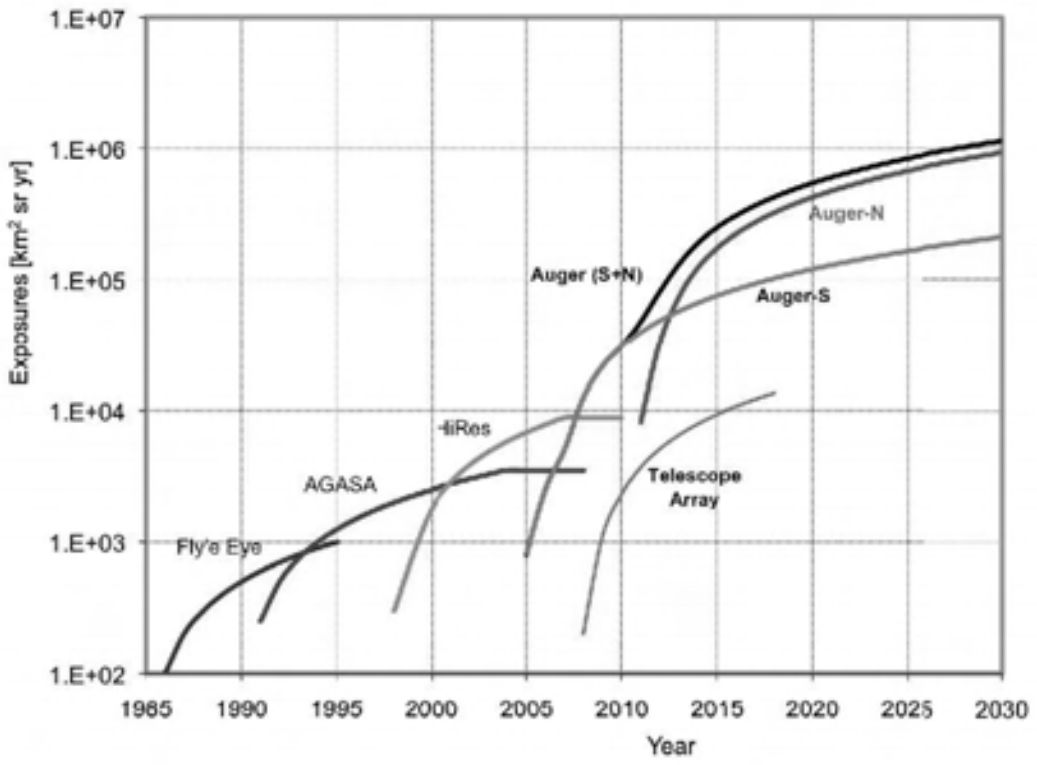

(b) Exposures

Fig. 4: Array areas and recorded or planned exposures for ground arrays of UHECR detectors: AGASA; Telescope Array; the southern part of the Auger Observatory; and the northern part of the Auger Observatory. The exposure plot is on a log scale.

## IV. Outlook

The goals of the Auger Observatory have been, since the inception of the project, to *"discover and understand the source or sources of cosmic rays with energies exceeding* $10^{19}$ *eV.*". Results from the southern section of the Observatory in Argentina have shown that the highest energy particles are likely of extra-galactic origin and that the flux of these particles drops significantly at the energy where a correlation with matter in the nearby universe sets it. The highest energy cosmic rays do not arrive isotropically at 99% confidence level. The data from Auger South indicate that the primaries are not photons and that they are tending toward heavier nuclei as the energy approaches the energy of the flux drop and the correlation with extra-galactic matter. The correlation and the tendency toward shallow $X_{max}$ pose a challenge to our understanding from the astrophysical point of view. The solution lies in gaining more data at energies at and above the energy of the flux drop and correlation, which is about 60 EeV. The Auger South detector is large enough to detect about 20-25 of these events a year. Figure 4a shows scale outlines of four cosmic ray detector arrays, and 4b plots corresponding exposures (assuming a certain start date for Auger North). The left half of Figure 4a shows AGASA, which is no longer taking data, the Telescope Array near Delta, Utah, USA, and Auger South at Malargüe, Argentina. The proposed northern part of the Auger Observatory is also shown; it is about seven times the area of Auger South. Assuming the flux of trans-GZK events in the northern hemisphere is about the same as that measured in the south, Auger North and Auger South together will measure approximately 180 trans-GZK events per year.

# Acknowledgements

The successful installation and commissioning of the Pierre Auger Observatory would not have been possible without the strong commitment and effort from the technical and administrative staff in Malargüe.

We are very grateful to the following agencies and organizations for financial support:

Comisión Nacional de Energía Atómica, Fundación Antorchas, Gobierno De La Provincia de Mendoza, Municipalidad de Malargüe, NDM Holdings and Valle Las Leñas, in gratitude for their continuing cooperation over land access, Argentina; the Australian Research Council; Conselho Nacional de Desenvolvimento Científico e Tecnológico (CNPq), Financiadora de Estudos e Projetos (FINEP), Fundação de Amparo à Pesquisa do Estado de Rio de Janeiro (FAPERJ), Fundação de Amparo à Pesquisa do Estado de São Paulo (FAPESP), Ministério de Ciência e Tecnologia (MCT), Brazil; AVCR AV0Z10100502 and AV0Z10100522, GAAV KJB300100801 and KJB100100904, MSMT-CR LA08016, LC527, 1M06002, and MSM0021620859, Czech Republic; Centre de Calcul IN2P3/CNRS, Centre National de la Recherche Scientifique (CNRS), Conseil Régional Ile-de-France, Département Physique Nucléaire et Corpusculaire (PNC-IN2P3/CNRS), Département Sciences de l'Univers (SDU-INSU/CNRS), France; Bundesministerium für Bildung und Forschung (BMBF), Deutsche Forschungsgemeinschaft (DFG), Finanzministerium Baden-Württemberg, Helmholtz-Gemeinschaft Deutscher Forschungszentren (HGF), Ministerium für Wissenschaft und Forschung, Nordrhein-Westfalen, Ministerium für Wissenschaft, Forschung und Kunst, Baden-Württemberg, Germany; Istituto Nazionale di Fisica Nucleare (INFN), Ministero dell'Istruzione, dell'Università e della Ricerca (MIUR), Italy; Consejo Nacional de Ciencia y Tecnología (CONACYT), Mexico; Ministerie van Onderwijs, Cultuur en Wetenschap, Nederlandse Organisatie voor Wetenschappelijk Onderzoek (NWO), Stichting voor Fundamenteel Onderzoek der Materie (FOM), Netherlands; Ministry of Science and Higher Education, Grant Nos. 1 P03 D 014 30, N202 090 31/0623, and PAP/218/2006, Poland; Fundação para a Ciência e a Tecnologia, Portugal; Ministry for Higher Education, Science, and Technology, Slovenian Research Agency, Slovenia; Comunidad de Madrid, Consejería de Educación de la Comunidad de Castilla La Mancha, FEDER funds, Ministerio de Ciencia e Innovación, Xunta de Galicia, Spain; Science and Technology Facilities Council, United Kingdom; Department of Energy, Contract No. DE-AC02-07CH11359, National Science Foundation, Grant No. 0450696, The Grainger Foundation USA; ALFA-EC / HELEN, European Union 6th Framework Program, Grant No. MEIF-CT-2005-025057, European Union 7th Framework Program, Grant No. PIEF-GA-2008-220240, and UNESCO.